\documentclass[longauth]{aa}   

\usepackage{graphicx}
\usepackage{txfonts}
\usepackage{lscape}             
\usepackage{placeins}           
                                
\usepackage{hyperref}

\usepackage{xcolor}

\usepackage{booktabs}

\usepackage{etoolbox}
\AtBeginDocument{\let\linenumbers\relax \let\runninglinenumbers\relax }
\AtBeginDocument{\nolinenumbers}

\begin{document}

\title{SN 2019qem: decades of extreme mass loss before the explosion of a luminous, long-lived Type IIn supernova}

\titlerunning{SN 2019qem: a luminous, long-lived Type IIn supernova}
\authorrunning{J. Majumdar et al.}

\author{Joysankar Majumdar\inst{1}\fnmsep\thanks{Corresponding author: jmajumdar@astrouw.edu.pl}
\and Mariusz Gromadzki\inst{1}
\and Takashi J. Moriya \inst{2,3,4}
\and David A.H. Buckley \inst{5,6}
\and Priscila J. Pessi \inst{7}
\and Nada Ihanec \inst{8,1}
\and Szymon Kozłowski \inst{1}
\and Łukasz Wyrzykowski \inst{1,7}
\and Callum Aster \inst{9}
\and Panos Charalampopoulos \inst{10,11}
\and Ting-Wan Chen \inst{12}
\and Massimo Della Valle \inst{13}
\and Lluís Galbany \inst{10,14}
\and Avishay Gal-Yam \inst{15}
\and Claudia P. Guti\'errez \inst{10,14}
\and Cosimo Inserra \inst{9}
\and Erkki Kankare \inst{16}
\and Hanindyo Kuncarayakti \inst{16,11}
\and Tom\'as E. M{\"u}ller-Bravo \inst{17,18}
\and Matt Nicholl \inst{19}
\and Jacco H. Terwel \inst{17}
\and David R. Young \inst{19}
\and Andrzej Udalski \inst{1}
\and Michał K. Szymański \inst{1}
\and Igor Soszyński \inst{1}
\and Paweł Pietrukowicz \inst{1}
\and Radosław Poleski \inst{1}
\and Jan Skowron \inst{1}
\and Krzysztof Ulaczyk \inst{20,1}
\and Przemek Mróz \inst{1}
\and Krzysztof Rybicki \inst{1}
\and Patryk Iwanek \inst{1}
\and Marcin Wrona \inst{21,1}
\and Mateusz J. Mróz \inst{1}
\and Moira Andrews \inst{22,23}
\and Craig Pellegrino \inst{24}
\and D. Andrew Howell \inst{22,23}
\and Curtis McCully \inst{22}
\and Daichi Hiramatsu \inst{25}
\and Megan Newsome \inst{26}
\and Giacomo Terreran \inst{27}
\and Estefania Padilla Gonzalez \inst{28}
\and Joseph R. Farah \inst{29,30,31,32}
\and Mauritz Wicker \inst{1}
\and Michał Małkowski \inst{1}
\and Andrew Fraser Gillan \inst{7}
}

\institute{Astronomical Observatory, University of Warsaw, Al. Ujazdowskie 4, 00-478 Warsaw, Poland
\and
National Astronomical Observatory of Japan, National Institutes of Natural Sciences, Tokyo, Japan
\and
Graduate Institute for Advanced Studies, SOKENDAI, Tokyo, Japan
\and
School of Physics and Astronomy, Monash University, Clayton, VIC, Australia
\and
South African Astronomical Observatory, PO Box 9, Observatory, Cape Town 7935, South Africa
\and
Department of Astronomy, University of Cape Town, Private Bag X3, Rondebosch 7701, South Africa
\and
Astrophysics Division, National Centre for Nuclear Research, Pasteura 7, 02-093 Warsaw, Poland
\and
Isaac Newton Group of Telescopes, Apartado 321, E-38700 Santa Cruz de La Palma, Spain
\and
Cardiff Hub for Astrophysics Research and Technology, School of Physics \& Astronomy, Cardiff University, Queens Buildings, The Parade, Cardiff, CF24 3AA, UK
\and
Institute of Space Sciences (ICE-CSIC), Campus UAB, Carrer de Can Magrans, s/n, E-08193 Barcelona, Spain
\and
Finnish Centre for Astronomy with ESO (FINCA), FI-20014 University of Turku, Finland
\and
Graduate Institute of Astronomy, National Central University, 300 Jhongda Road, 32001 Jhongli, Taiwan
\and
Observatorio Astronomico, INAF-Padova, Vicolo Osservatorio 5, 35122, Padova, Italy
\and
Institut d'Estudis Espacials de Catalunya (IEEC), 08860 Castelldefels (Barcelona), Spain
\and
Department of Particle Physics and Astrophysics, Weizmann Institute of Science, Rehovot, Israel
\and
Tuorla Observatory, Department of Physics and Astronomy, FI-20014 University of Turku, Finland
\and
School of Physics, Trinity College Dublin, The University of Dublin, Dublin 2, Ireland
\and
Instituto de Ciencias Exactas y Naturales (ICEN), Universidad Arturo Prat, Chile
\and
Astrophysics Research Centre, School of Mathematics and Physics, Queen's University Belfast, Belfast BT7 1NN, UK
\and
Department of Physics, University of Warwick, Coventry CV4 7AL, UK
\and
Department of Astrophysics and Planetary Sciences, Villanova University, 800 Lancaster Ave., Villanova, PA 19085, USA
\and
Las Cumbres Observatory, 6740 Cortona Drive, Suite 102, Goleta, CA 93117-5575, USA
\and
Department of Physics, University of California, Santa Barbara, CA 93106-9530, USA
\and
NASA Goddard Space Flight Center, 8800 Greenbelt Road, Greenbelt, MD 20771, USA
\and
Department of Astronomy, University of Florida, Bryant Space Science Center, Gainesville, FL 32611-2055, USA
\and
University of Texas at Austin, 1 University Station C1400, Austin, TX 78712-0259, USA
\and
Adler Planetarium, 1300 S DuSable Lake Shore Dr, Chicago, IL 60605, USA
\and
Space Telescope Science Institute, 3700 San Martin Dr, Baltimore, MD 21218, USA
\and
Center for Astrophysics | Harvard \& Smithsonian, 60 Garden St, Cambridge, MA 02138, USA
\and
Black Hole Initiative, Harvard University, 20 Garden St, Cambridge, MA 02138, USA
\and
Miller Institute for Basic Research in Science, 206B Stanley Hall, Berkeley, CA 94720, USA
\and
Department of Astronomy, University of California, Berkeley, CA 94720-3411, USA
}

\date{Received October 25, 2026}

 
  \abstract
   {How massive stars lose huge amounts of mass in the final decades before core collapse is still debated. Type IIn supernovae (SNe) are powered by interaction of fast-moving ejecta with dense circumstellar material (CSM) and therefore probe this mass loss.}
   {We present the luminous, long-lived Type IIn SN 2019qem and use it to reconstruct how much mass its progenitor lost, and when.}
   {We combine host-subtracted optical photometry, 26 optical spectra from $-22$ to $+699$\,d (rest-frame days from the $I$-band maximum), NEOWISE mid-infrared photometry up to $+1553$\,d and an analysis of the host galaxy. We model the bolometric light curve with thin-shell CSM-interaction models in \texttt{REDBACK}, with the shock velocity constrained by the H$\alpha$ line widths, and compare them with an analytic interaction model.}
   {SN 2019qem rose in $\approx30$\,d to $M_I = -20.5$\,mag and a bolometric peak of $7.0\times10^{43}$\,erg\,s$^{-1}$, and radiated $6.9\times10^{50}$\,erg in the optical within 575\,d, closely resembling luminous, strongly interacting SNe IIn such as SN 2010jl and SN 2015da. A \ion{He}{ii}/\ion{N}{iii} flash feature at $-22$\,d and unresolved narrow Balmer lines reveal dense, slow CSM. From $+78$ to $+445$\,d the H$\alpha$ wings are shaped by electron scattering, and the line stays blueshifted by 400-700\,km\,s$^{-1}$, which points to an optically thick, asymmetric interaction region rather than to newly formed dust. In the preferred model, $\approx7\,M_\odot$ of ejecta run into $\approx9\,M_\odot$ of gas ejected $\approx2$\,yr before explosion and then into a wind of $\approx0.6\,M_\odot$\,yr$^{-1}$ (for $v_w = 100$\,km\,s$^{-1}$), sweeping up $\approx32\,M_\odot$ by $+544$\,d. An analytic model matches the light curve with a 30 times thinner wind, but only with a shock three times faster than the line widths allow. The mid-infrared excess brightens for two years while the optical fades, tracing dust heated by the ongoing interaction. The host is a low-mass, star-forming galaxy, $\log(M_*/M_\odot) = 9.19$, with a sub-solar metallicity at the explosion site.}
   {In its final decades the progenitor of SN 2019qem lost mass far faster than any steady stellar wind, ending with an eruption of $\approx9\,M_\odot$ two years before core collapse, consistent with luminous-blue-variable-like eruptive mass loss. Light curves alone cannot separate the CSM density from the shock velocity, so spectroscopic line widths are essential for measuring the mass-loss rates of interacting supernovae. Taken at face value, the mass budget implies an initial progenitor mass of $\gtrsim40\,M_\odot$.}


\makeatletter
\global\aa@nbauthors=1
\makeatother

\maketitle

\section{Introduction}
Within the hydrogen-rich supernovae (Type II SNe) family, the presence of strong and narrow hydrogen Balmer emission lines in optical spectra forms a distinctive subclass, Type IIn supernovae \citep{1990MNRAS.244..269S,1997ARA&A..35..309F,2017hsn..book..195G}. These narrow emission lines do not originate in the fast-moving supernova ejecta but they are produced in slow, dense CSM that is mostly gas expelled by the progenitor star in the years to centuries before the explosion. Like other supernova classes since \citet{1941PASP...53..224M}, the IIn class is defined from the spectrum, so it describes the circumstellar gas rather than the type of explosion.  This distinction matters because the bright light from the interaction can hide the explosion itself, so the nature of the underlying event is not always clear \citep{2017hsn..book..403S}. Most Type IIn supernovae are thought to come from the core collapse of massive stars, but in some cases, a thermonuclear explosion of a white dwarf inside dense CSM can produce a similar spectrum \citep{2017hsn..book..403S}. For this reason, Type IIn supernovae are best understood as tracers of extreme mass loss and dense circumstellar gas, and they are useful probes of the late mass-loss history of their progenitors \citep{2020A&A...637A..73N}.

The intense radiation from the supernova ionises this gas, and later the expanding ejecta crashes into it, producing the narrow emission lines \citep{1994MNRAS.268..173C, 2017hsn..book..403S}. As the high-velocity ejecta overtake the dense CSM, a forward shock propagates into the circumstellar gas and a reverse shock is driven back into the ejecta. The strongly radiative region between them, the cold dense shell (CDS), efficiently converts kinetic energy into radiation \citep{1982ApJ...258..790C,2017hsn..book..875C}. Due to the large kinetic energy and efficient conversion, the interaction can sustain a high luminosity for months to years, far longer than the few months over which ordinary hydrogen-rich (Type II-P/L) supernovae radiate the energy deposited by the shock, with $^{56}$Ni decay powering mainly the later tail \citep{2016ApJ...823..127N}.

The spectrum of a Type IIn supernova changes in a characteristic way with time. At early times, the gas around the star is still optically thick. The spectrum then shows a smooth blue continuum with narrow emission lines that have broad wings. These wings are produced by electron scattering in the dense CSM and not by the fast motion of the gas \citep[e.g.][]{2001MNRAS.326.1448C,2009MNRAS.394...21D,2017hsn..book..403S,2018MNRAS.475.1261H}. In the first days after the explosion, the ultraviolet flash from the shock can ionise the closest gas and produce narrow high-ionization lines of helium, carbon, and nitrogen. These features are called flash features, and they fade within days to weeks \citep{2014Natur.509..471G,2016ApJ...818....3K,2021ApJ...912...46B,2023ApJ...952..119B,2024ApJ...970..189J}. Near the time of peak brightness, the photosphere moves inward, and intermediate-width lines of a few thousand kilometres per second appear. These lines come from the post-shock gas in the cool dense shell \citep{2017hsn..book..403S}.

A key question is how the progenitor produced such dense gas so close to the explosion. Normal steady winds of massive stars are not dense enough to make a Type IIn supernova \citep{2014ARA&A..52..487S}. Among single stars, only the giant eruptions of Luminous Blue Variable (LBV) stars and the strong winds of extreme red supergiants and yellow hypergiants can reach the required density \citep[e.g.][]{2007ApJ...656..372G,2009Natur.458..865G,2014ARA&A..52..487S,2017hsn..book..403S}. This points to eruptive or episodic mass loss in the last phase of the star's life and direct support for this picture comes from outbursts seen before some Type IIn explosions, often called precursors \citep{2007Natur.447..829P,2014ApJ...789..104O,2021ApJ...907...99S}. The clearest case is SN 2009ip, whose LBV progenitor was seen to erupt for years and then explode as a supernova \citep{2011ApJ...732...32F,2013MNRAS.430.1801M}. Binary interaction may also produce such dense CSM, because in binary population models the stars that reach core collapse during or shortly after mass transfer make up about 5\% of all core-collapse supernovae, close to the observed fraction of Type IIn and Ibn events, and a disc-like CSM reproduces the observed spread in their radiated energies \citep{2026A&A...706A.169E}.

Type IIn supernovae show a wide range of peak brightness, rise time, and decline rate. In a large untargeted sample, the mean peak absolute magnitude is about -19.2 with a spread of more than one magnitude, and the rise times split into fast and slow risers at around 20 and 50 days, respectively \citep{2020A&A...637A..73N,2026ApJ..1005...82H}. Some objects fade slowly and keep interacting for years, such as the enduring events SN 1988Z and SN 2005ip, which require several to about 20 $M_\odot$ of CSM \citep{2009ApJ...695.1334S,2012ApJ...756..173S}. At the bright end are the superluminous Type IIn supernovae ($M\lesssim-21$\,mag; \citealt{2012Sci...337..927G}), such as SN 2006gy and SN 2006tf \citep{2007ApJ...666.1116S,2008ApJ...686..467S}, and just below them luminous events such as SN 2010jl; all of them need a very massive and dense shell \citep{2007ApJ...666.1116S,2012Sci...337..927G}. Some light curves exhibit bumps or rebrightening episodes \citep{2017A&A...605A...6N}, and others, such as the strongly interacting SN 2013L and the hydrogen-rich superluminous SN 2021aaev, combine early flash features with long-lived interactions \citep{2020A&A...638A..92T,2026ApJ...999..176H}. This variety shows that the CSM and the progenitors are also varied, so well-observed single objects remain valuable for linking the observed properties of an event to the structure of its CSM and the mass-loss history of its progenitor.

Type IIn supernovae are also strong sources of infrared light, because they form and heat dust; in mid-infrared surveys they are the most luminous and longest-lived emitters across all supernovae, because their dense CSM both holds dust and keeps heating it \citep{2011ApJ...741....7F,2013AJ....146....2F,2019ApJS..241...38S}. Dust grains absorb optical and ultraviolet light and re-emit it in the infrared. This infrared excess can have two origins. In the first case, new dust forms in the cool dense shell behind the shock, or later in the ejecta. This is seen as a rising dust mass, a falling dust temperature, and emission lines that become more blueshifted with time, because the dust hides the receding side of the gas \citep{2012AJ....143...17S,2011A&ARv..19...43G,2018A&A...611A..67B,2022MNRAS.511.2021L}. In the second case, dust that was already present around the star is heated by the bright peak of the supernova, and its infrared emission reaches us with a delay set by the longer light-travel path via the dust. This is called an infrared echo \citep{2013AJ....146....2F}. The enduring events SN 2005ip and SN 2006jd showed long-lived infrared excesses of this kind \citep{2011ApJ...741....7F,2012ApJ...756..173S}. Telling these origins apart requires mid-infrared monitoring over several years together with the evolution of the line profiles, as available for SN 2019qem.

In this paper, we present a study of the Type IIn supernova SN 2019qem\footnote{\url{https://www.wis-tns.org/object/2019qem}}, at $\alpha = 06^{\rm h}03^{\rm m}45.01^{\rm s}$, $\delta = -64^\circ22'35.9''$ (J2000), with an internal name of OGLE19qem. 
The dataset covers a wide range of wavelengths, including ultraviolet photometry from Swift, optical photometry from OGLE, Las Cumbres Observatory and Gaia, near-infrared photometry from the NTT, and mid-infrared photometry from NEOWISE, which follows the source for more than four years after the explosion. \citet{2019TNSCR2193....1G} classified SN 2019qem as a Type IIn SN based on the spectrum obtained on the night of 20 September 2019 from the Southern African Large Telescope (SALT) and reported the redshift of 0.042.

Here we present the ground-based and space-based observations and analysis in Section \ref{sec:observations}. The results are presented in Section \ref{sec:results}, the nature of the SN in Section \ref{sec:discussion}, and our conclusions in Section~\ref{sec:conclusion}.

\section{Observations} \label{sec:observations}
\subsection{Photometry}
The Optical Gravitational Lensing Experiment (OGLE; \citealt{2015AcA....65....1U}) is a long-term survey whose main goal is the study of sky variability, including searches for variable stars, transients and gravitational microlensing events. OGLE-IV detected SN 2019qem through its Transient Detection System\footnote{\url{https://ogle.astrouw.edu.pl/ogle4/transients/}} \citep{2014AcA....64..197W} with the 1.3-m Warsaw University Telescope at Las Campanas Observatory, Chile, in the \textit{I} band, which closely matches the standard Cousins $I$ band \citep{2015AcA....65....1U}. After a non-detection of SN 2019qem from 15 January 2019 (MJD=58\,498) to 4 September 2019 (MJD=58\,730) with $I>$ 22 mag, it was observed with $I$ = 20.98$\pm$0.28 mag on 12 September 2019 (MJD=58\,738) \citep{2019TNSTR1836....1G}, a $3.9\sigma$ detection confirmed by the brightening four days later. On 16 September 2019 (MJD=58\,742), it reached $I$ = 17.49$\pm$0.01 mag and the brightest OGLE point, $I$ = 16.03$\pm$0.01 mag on 17 October 2019 (MJD=58\,773), follows an 11-day gap in the OGLE coverage; the epoch of maximum, MJD 58\,769.0, is derived in Sect. \ref{subsec:lightcurve}.

The Black Hole Target and Observation Manager (BHTOM\footnote{\url{https://bhtom.space/}}; \citealt{2024lsst.confE...4W,2024eas..conf.2399W}) is an automated platform that archives photometry of transients and reduces optical images. From it we retrieved the archival Gaia alert \citep{2021A&A...652A..76H}, NEOWISE \citep{2014ApJ...792...30M}, AllWISE and Swift/UVOT \citep{2005SSRv..120...95R} photometry of SN 2019qem\footnote{\url{https://bhtom.space/public/targets/Gaia19exe}}.

Las Cumbres Observatory \citep{2013PASP..125.1031B} is a global network of robotic telescopes. The optical follow-up of SN 2019qem was conducted with its 1-m telescope network, including Siding Spring Observatory (COJ), South African Astronomical Observatory (CPT), and Cerro Tololo Interamerican Observatory (LSC). The optical images were obtained through the Supernova Exchange (SNEx)\footnote{\url{https://supernova.exchange/public/}} platform via the Global Supernova Project. SN 2019qem was also observed in the near-infrared (NIR) with SofI \citep{1998Msngr..91....9M} on the New Technology Telescope (NTT) through the extended Public ESO (European Southern Observatory) Spectroscopic Survey for Transient Objects (ePESSTO+) collaboration \citep{2015A&A...579A..40S}. The Las Cumbres Observatory and SofI images were reduced using BHTOM with the CCDPhot suite of image processing along with CPCS photometry calibration tools \citep{2019CoSka..49..125Z,2020past.conf..190Z} and calibrated to standard magnitudes based on Gaia Synthetic Photometry \citep{2023A&A...674A..33G}. The Gaia, Swift/UVOT and near-infrared photometry and the Las Cumbres Observatory $B$ and $V$ photometry are not host-subtracted. We show them in Fig.~\ref{fig:lightcurve} for completeness but do not use them in the analysis, apart from the UVOT epoch at $-12.9$\,d, which we use to test the blackbody extrapolation (Sect.~\ref{subsec:lightcurve}).

\subsection{Spectroscopy}
The Southern African Large Telescope (SALT; \citealt{Buckley2006}) obtained three spectra of SN 2019qem with the Robert Stobie Spectrograph (RSS; \citealt{Burgh2003}), the first on the night of 20 September 2019, about 9 days after the discovery, and the last on 10 October 2021. To reduce them we used the PyRAF-based PySALT package \citep{Crawford2010}\footnote{\url{https://astronomers.salt.ac.za/software/}}, which includes corrections for gain and cross-talk and performs bias subtraction. We extracted the spectra with standard IRAF\footnote{\url{https://iraf-community.github.io/}} tasks, including wavelength calibration, background subtraction and one-dimensional extraction. Because of the SALT design, absolute flux calibration is not possible\footnote{The pupil (i.e. the view of the mirror from the tracker) moves during all SALT observations, so the effective area of the telescope changes during the exposures. See \citet{Buckley2006} and \citet{Crawford2010} for details.}. However, spectrophotometric standards observed during twilight give a relative flux calibration, which recovers the correct spectral shape and relative line strengths. Another 19 spectra were obtained with the NTT and the ESO Faint Object Spectrograph and Camera (EFOSC2) through the ePESSTO+ collaboration \citep{2015A&A...579A..40S} between 26 October 2019 and 6 October 2021. Another optical spectrum was obtained with VLT/FORS2 on 19 September 2020 within the FOSSIL (FORS+ Survey of Supernovae in Late Times) programme \citep{2022ApJ...941L..32K} and reduced with standard procedures, including a correction for telluric absorption. The EFOSC2 spectra were reduced using the PESSTO Pipeline for NTT\footnote{\url{https://github.com/svalenti/pessto}}. The spectra taken on the same night were merged to provide broader wavelength coverage. Telluric absorption was removed from most spectra. For the ePESSTO+ spectra, a telluric model was fitted to the spectrophotometric standard star observed on the same night and removed from the spectrum of the supernova. The regions of the strongest telluric bands are shaded in Fig. ~\ref{fig:all_spec}. Three optical spectra were obtained with the FLOYDS spectrograph on the Faulkes Telescope South of Las Cumbres Observatory and are publicly available in its archive\footnote{\url{https://archive.lco.global}}. The FLOYDS raw data were processed with the ORAC pipeline \citep{2008AN....329..295C}. No spectrophotometric standard star was observed on the nights of the $-4$ and $+4$\,d FLOYDS spectra, so their calibration in the blue is less reliable, and we use them only for H$\alpha$. Finally, each spectrum was corrected with a single multiplicative factor, chosen so that its synthetic $g$, $r$ and $i$ magnitudes match the host-subtracted Las Cumbres Observatory photometry interpolated to the epoch of the spectrum using the \texttt{Light Curve Fitting}\footnote{\url{https://zenodo.org/records/23271467}} package \citep{2023zndo...7872772H}.

\section{Results} \label{sec:results}

\subsection{Host galaxy} \label{subsec:host}
There is no catalogued redshift for the galaxy WISEA J060344.96-642234.2, which is the host for SN 2019qem. The initial redshift of the supernova has been reported to be 0.042 by \citet{2019TNSCR2193....1G}. To determine the redshift more precisely, we fitted the narrow host-galaxy emission lines (H$\beta$, \ion{O}{iii}, H$\alpha$, \ion{N}{ii} and \ion{S}{ii}) in the last SALT/RSS spectrum taken on 10 October 2021, requiring all lines to share a single redshift. We adopt a heliocentric redshift of $z = 0.0427 \pm 0.0001$ for SN 2019qem, where the uncertainty is dominated by the wavelength calibration. Assuming a Hubble constant, $H_0 = 67.66$ km s$^{-1}$ Mpc$^{-1}$ \citep{2020A&A...641A...6P}, this redshift corresponds to a luminosity distance $d_{\rm L} = 195.3$\,Mpc and a distance modulus $\mu = 36.453$\,mag. We correct for Galactic foreground reddening using $E(B-V) = 0.049$ mag \citep{2011ApJ...737..103S} and the extinction law of \cite{1999PASP..111...63F} with $R_V$ = 3.1. We do not correct for extinction in the host galaxy. Its SED fit gives a low mean attenuation, $A_V = 0.10\pm0.03$\,mag (see below), and the SALT spectra show no clear narrow \ion{Na}{i}\,D absorption at the host redshift, although \ion{He}{i}\,$\lambda$5876 emission lies close to it. Any host extinction would make SN 2019qem even more luminous.

\begin{figure}
    \centering
    \includegraphics[width=0.99\linewidth]{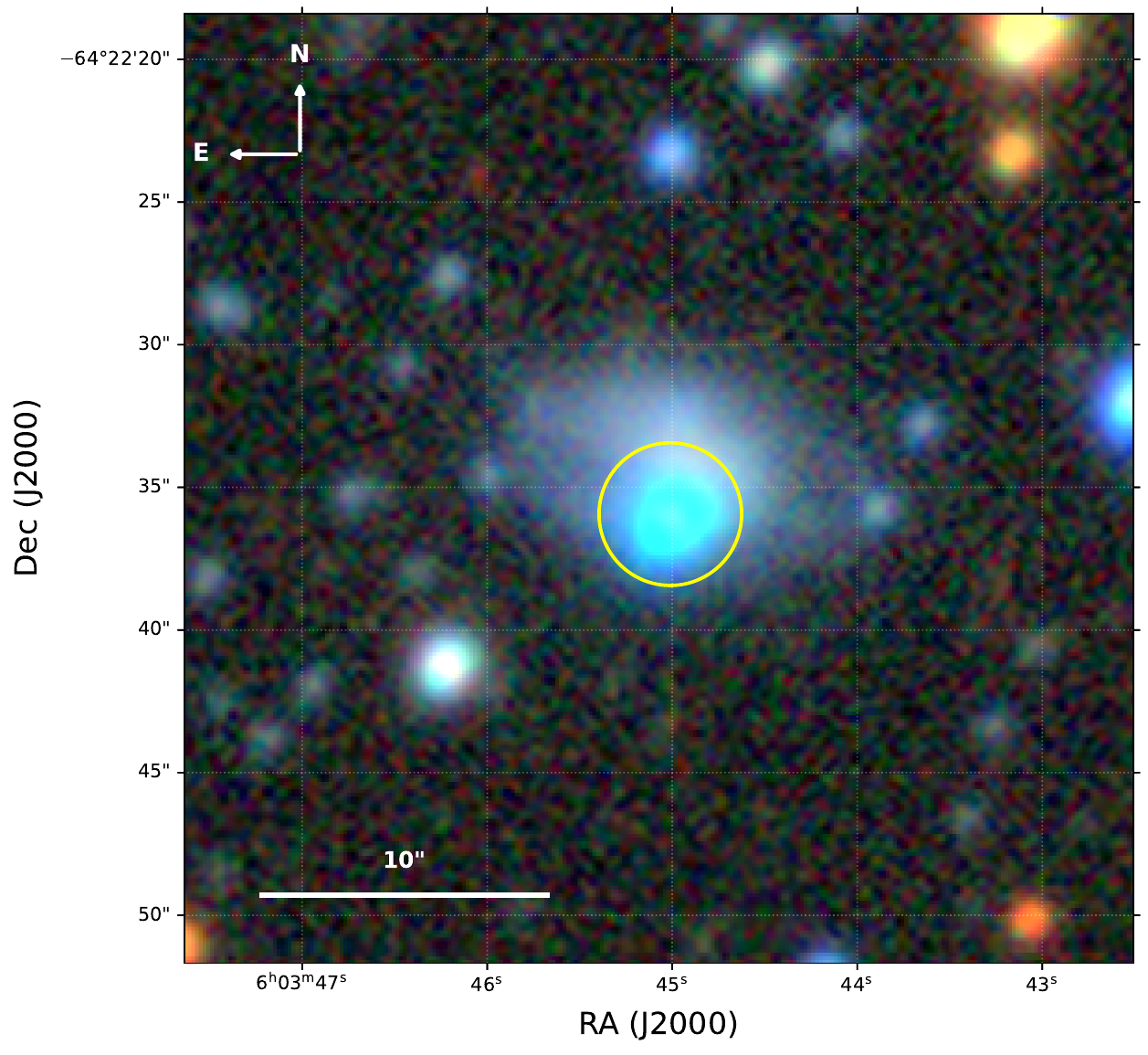}
    \caption{SN 2019qem in DESI Legacy Surveys DR10 \citep{2019AJ....157..168D}. The supernova on top of the host galaxy is marked with a yellow circle, near the centre of the image. Image courtesy of Legacy Surveys / D. Lang (Perimeter Institute).}
    \label{fig:rgb_image}
\end{figure}

Fig. \ref{fig:rgb_image} shows the field of the SN. The host galaxy is a smooth and diffuse disc galaxy with no clear spiral structure at the resolution of these images. The NOIRLab Source Catalog DR2 \citep{2021AJ....161..192N} resolves WISEA J060344.96-642234.2 as a single extended object, measured on DECam images taken between December 2012 and February 2013. Its centroid lies at $\alpha = 06^{\rm h}03^{\rm m}44.98^{\rm s}$, $\delta = -64^\circ22'34.1''$ (J2000). The SN is $1.80''$ from this centroid, and the spread between catalogue centroids given below indicates an uncertainty of about $0.2''$. At $z = 0.0427$ in the \citet{2020A&A...641A...6P} cosmology, where $1''$ corresponds to 0.871 kpc, this is a projected offset of 1.57 kpc.

To measure the size and shape of the host, we modelled the OGLE deep $V$- and $I$-band images with \texttt{GALFIT} \citep{2002AJ....124..266P,2010AJ....139.2097P}. The fits give an effective radius $R_{\rm e} = 3.12''$ (2.72 kpc) in $V$ and $2.76''$ (2.40 kpc) in $I$, a Sérsic index $n = 0.75 \pm 0.02$ and $0.79 \pm 0.01$, and an axis ratio $b/a = 0.72 \pm 0.01$ and $0.73 \pm 0.01$, respectively. The host therefore has a disc-like light profile ($n < 1$) and is slightly more extended in $V$ than in $I$. With the $I$-band radius, the host-normalised offset is $r/R_{\rm e} = 0.65$ (0.58 with the $V$-band radius). Allowing for the range of catalogue centroids ($1.59''$–$1.80''$, i.e. 1.38-1.57 kpc) and both radii, it lies between 0.51 and 0.65. SN 2019qem therefore exploded in the inner disc of its host, inside the effective radius but clearly away from the nucleus.

We assembled the spectral energy distribution (SED) of the host from ultraviolet to mid-infrared catalogue photometry obtained before the explosion. We fitted it with \texttt{CIGALE} \citep{2019A&A...622A.103B}, using a delayed-$\tau$ star-formation history with an optional recent burst, the stellar populations of \citet{2003MNRAS.344.1000B} with a \citet{2003PASP..115..763C} initial mass function, nebular emission, a modified \citet{2000ApJ...533..682C} attenuation law and the dust emission models of \citet{2014ApJ...780..172D}. Table~\ref{tab:cigale} lists the full parameter grid. The fit (Fig.~\ref{fig:host_sed}) gives $\log(M_*/M_\odot) = 9.19 \pm 0.14$, ${\rm SFR} = 0.19 \pm 0.04\,M_\odot\,{\rm yr}^{-1}$, $\log({\rm sSFR/yr^{-1}}) = -9.92 \pm 0.16$ and $A_V = 0.10 \pm 0.03$ mag. Dropping the near-infrared points or their aperture offset, or adopting a higher Galactic reddening, changes these by less than their uncertainties ($\log M_* = 9.12$–$9.19$), whereas fixing the stellar metallicity at $Z = 0.008$ ($\approx0.4\,Z_\odot$, close to the gas-phase value at the explosion site derived below) lowers the mass to $\log M_* = 8.83$ at a clearly worse $\chi^2$. A mock test recovers $M_*$, SFR, and $A_V$; the mass-weighted age ($4.7 \pm 1.9$ Gyr) is only indicative, and the stellar metallicity is unconstrained. At its mass, the host lies 0.24 dex below the star-forming main sequence of \citet{2014ApJS..214...15S}, comparable to the $\approx 0.2$ dex scatter of the relation, so it is an ordinary star-forming galaxy. Independently, the \texttt{GALFIT}'s total host magnitudes, $V = 17.82$ and $I = 17.09$, agree with the NSC DR2 photometry to within 0.05 mag and give $M_V = -18.79$ and $M_I = -19.45$ after correction for Galactic extinction.

The narrow host lines in the $+699$\,d SALT spectrum, whose slit was centred on the supernova, give the gas-phase metallicity at the explosion site. Fitting the narrow lines together with the residual broad H$\alpha$ of the supernova, we measure $\log([\ion{N}{ii}]\,\lambda6583/{\rm H}\alpha) = -0.86\pm0.04$ and $\log[([\ion{O}{iii}]\,\lambda5007/{\rm H}\beta)/([\ion{N}{ii}]\,\lambda6583/{\rm H}\alpha)] = 1.13\pm0.05$. The N2 and O3N2 calibrations of \citet{2004MNRAS.348L..59P} give $12+\log({\rm O/H}) = 8.41$ and 8.37, with systematic uncertainties of 0.18 and 0.14\,dex, and the calibration of \citet{2016Ap&SS.361...61D}, which relies mainly on [\ion{N}{ii}]/[\ion{S}{ii}], gives 8.27. The [\ion{S}{ii}]\,$\lambda6716/\lambda6731$ ratio of 1.48 places the gas in the low-density limit, as expected for an \ion{H}{ii} region. The narrow Balmer decrement, H$\alpha$/H$\beta\approx6$, is high, and part of the narrow H$\alpha$ may still come from the supernova. Replacing H$\alpha$ by $2.86\,{\rm H}\beta$ raises the N2 and O3N2 values by at most 0.2\,dex. The gas at the explosion site is therefore sub-solar, at about 0.3-0.8 of the solar oxygen abundance \citep[$12+\log({\rm O/H})_\odot = 8.69$;][]{2009ARA&A..47..481A}.

\subsection{Light curves} \label{subsec:lightcurve}

\begin{figure*}
\centering
\includegraphics[width=0.9\textwidth]{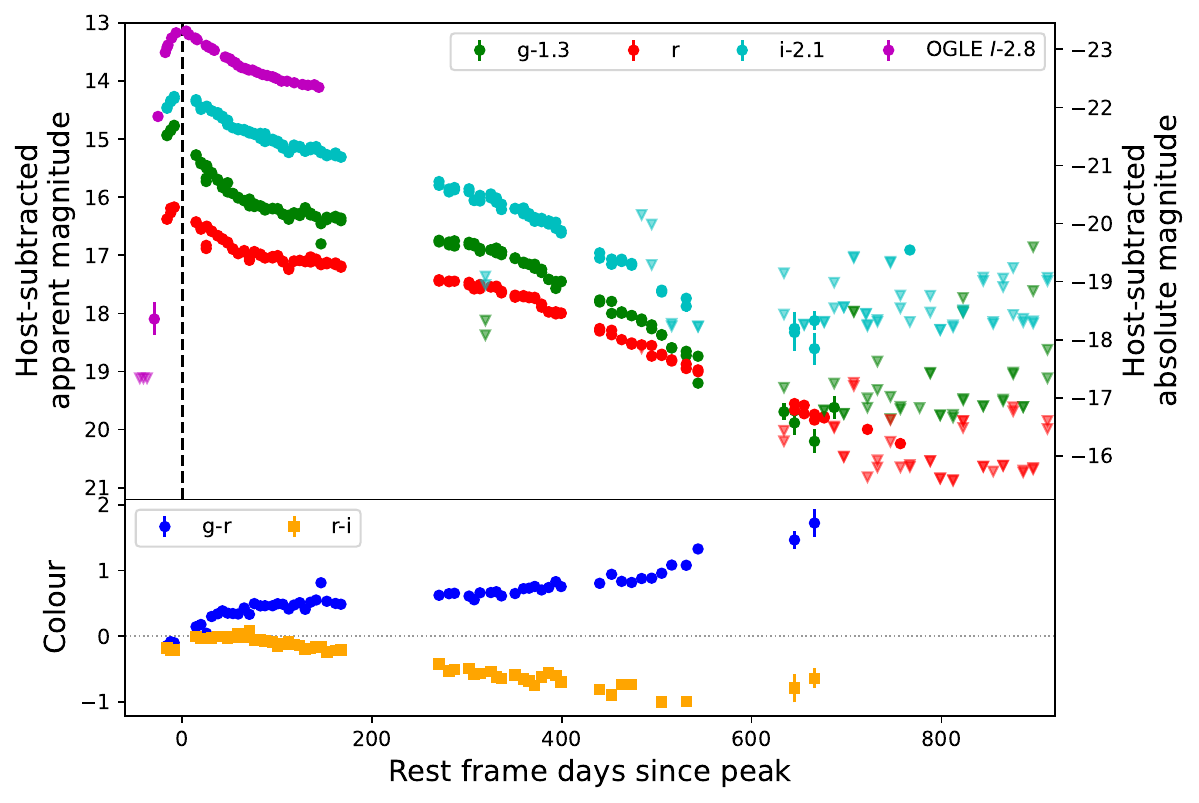} 
\caption{Host subtracted light curve of SN 2019qem in $g$, $r$, $i$ and $I$ bands.}
\label{fig:diff_lc}
\end{figure*}

Fig. \ref{fig:lightcurve} shows the observed, non-host-subtracted multi-band light curves of SN 2019qem. The top panel shows the long-term mid-infrared variability along with the optical light curve, whereas the bottom panel shows the zoomed light curve for the whole supernova evolution. We can see a large bump in $W1$ and $W2$, reaching a maximum about 2 years after the optical maximum (Sect.~\ref{subsec:mir}). We have not included OGLE $I$-band light curve data in Fig. \ref{fig:lightcurve} because it was already host-subtracted. At late times the optical points in Fig.~\ref{fig:lightcurve} are dominated by the host galaxy. After host subtraction (Fig.~\ref{fig:diff_lc}) the supernova is detected in $r$ until $\approx+760$\,d and only upper limits remain afterwards, so the late scatter of the Las Cumbres Observatory points in Fig.~\ref{fig:lightcurve} reflects the host and the photometric errors, not variability of the supernova.

We interpolate the OGLE $I$-band light curve with Gaussian Processes (GP) using \texttt{GaussianProcessRegressor} from the \texttt{scikit-learn} package \citep{2011JMLR...12.2825P} with a Matern-5/2 kernel. We used interpolation by converting the magnitude of OGLE $I$-band to flux as shown in Fig. \ref{fig:ogle_gp_fit}. We estimate the peak epoch and absolute magnitude of SN 2019qem to be $t_{\text{peak}}^{I}$ = 58\,769.0$^{+1.9}_{-2.0}$ MJD and $M_{\text{peak}}^{I}$ = -20.49$^{+0.01}_{-0.03}$ mag, respectively. We do not apply K-corrections, and synthetic photometry on our spectra gives $|K_I|\lesssim0.1$\,mag at this redshift. All phases in this paper are reported as rest-frame days since $t_{\text{peak}}^{I}$, which is better determined than the explosion epoch (about 30\,d earlier, see below). We take the explosion epoch as the time at which the GP median, conditioned on the pre-explosion non-detections as a zero-flux baseline, starts to rise. This gives MJD 58\,738.03$^{+0.09}_{-0.10}$, only 0.3\,d before the first detection, and the estimate is driven by that faint first point; without it the same procedure gives MJD 58\,730.3. We therefore adopt $t_{\text{exp}}^{I}$ = 58\,738.0$^{+0.1}_{-7.6}$ MJD, where the lower uncertainty reaches the last non-detection (MJD 58\,730.4). The rise is sampled by only two detections before an 11-day gap, too few for a power-law fit to constrain the explosion epoch better. We also used GP interpolation to estimate the rise time of the supernova, considering these as 10\% and 1/e flux fractions with respect to the peak. The estimated values are $t_{\text{rise},1/e}$ = 23.9$^{+1.8}_{-2.0}$ days and $t_{\text{rise},10\%}$ = 27.8$^{+1.8}_{-2.0}$ days. Given the time of explosion, the rise time computed as the time elapsed since the explosion is $t_{\text{rise}}$ =  ($t_{\text{peak}}^{I}$ - $t_{\text{exp}}^{I}$)/(1+z) = 29.7$^{+7.5}_{-1.9}$ days, where the lower uncertainty comes from the epoch of maximum.

We removed the host contribution by image subtraction with \texttt{subphot}\footnote{\url{https://github.com/joysankar-astro/subphot}}, which uses \texttt{hotpants} \citep{2015ascl.soft04004B} for image subtraction. We used recent Las Cumbres Observatory images taken on 13 March 2026 as the template for image subtraction. The template was taken $\approx2200$ rest-frame days after maximum. Extrapolating the late decline of the supernova, about 0.5\,mag per 100\,d in $r$ between $+500$ and $+760$\,d, its residual flux in the template is negligible. The first panel of Fig. \ref{fig:diff_lc} shows the host-subtracted light curves of SN 2019qem, including the OGLE $I$-band light curve. The supernova is detected in $r$ until $\approx+760$\,d. The second panel of Fig. \ref{fig:diff_lc} shows the colour evolution. Near peak, both $g-r$ and $r-i$ are blue, after which $g-r$ reddens steadily, from 0.3 mag at $\sim$50 d to 0.8 mag at $\sim$400 d and 1.3 mag by $\sim$550 d, consistent with progressive cooling of the ejecta. In contrast, $r-i$ turns increasingly blue after $\sim$100 d, reaching -1 mag by $\sim$500 d, which points to an $r$-band flux excess likely due to strong H$\alpha$ emission.

To estimate the bolometric luminosity of SN 2019qem, we fitted a blackbody to the host-subtracted $griI$ photometry, as we have host-subtracted light curves from these bands only. OGLE $I$ overlaps in wavelength with $i$, but it provides the only data on the rise and fills gaps in the Las Cumbres Observatory coverage. We used the \texttt{estimate\_bolometric\_luminosity} function from \texttt{REDBACK}\footnote{\url{https://redback.readthedocs.io/en}} \citep{2024MNRAS.531.1203S} to fit a blackbody to the spectral energy distribution in 1-d bins, producing the bolometric light curve shown in Fig.~\ref{fig:bolo_lc}, with 57 multi-band epochs between $-29$ and $+545$\,d. The input is the host-subtracted $g$, $r$ and $i$ photometry together with OGLE $I$ converted to AB, all corrected for Galactic extinction. On three earlier nights, at $-29$, $-26$ and $-18$\,d, only OGLE $I$ is available; for these we reconstructed the missing bands assuming the colours measured at $-16$\,d. The light curve peaks at $(7.0\pm0.4)\times10^{43}$\,erg\,s$^{-1}$ at $-8.5$\,d, with $T_{\rm bb}\simeq12\,100$\,K and $R_{\rm bb}\simeq2.1\times10^{15}$\,cm. Only about 35 \% of the luminosity of a 12\,000\,K blackbody falls within the observed bands, so near peak the result relies on the extrapolation to the ultraviolet. At $-12.9$\,d, where Swift/UVOT photometry is available, including the UVOT bands raises $T_{\rm bb}$ to 13\,200\,K and the luminosity by only 9 per cent, which supports the extrapolation. The light curve then declines at $\simeq0.5$\,mag per 100\,d with $T_{\rm bb}\simeq6900$\,K before flattening after $+400$\,d, and integrates to $6.9\times10^{50}$\,erg between $-29$ and $+545$\,d. After $\sim$100\,d the $r$ band is increasingly dominated by H$\alpha$ ($r-i\approx-1$\,mag by $+500$\,d). Excluding $r$ from the fits after $+300$\,d raises $T_{\rm bb}$ and lowers the luminosity by typically 10 and at most 25 per cent, so the late-time luminosities carry this systematic uncertainty, and the late temperatures and radii are only approximate. This luminosity excludes the mid-infrared emission (Sect.~\ref{subsec:mir}).

\subsection{Spectral evolution} \label{subec:spectra}

\begin{figure*}
\centering
\includegraphics[width=\textwidth]{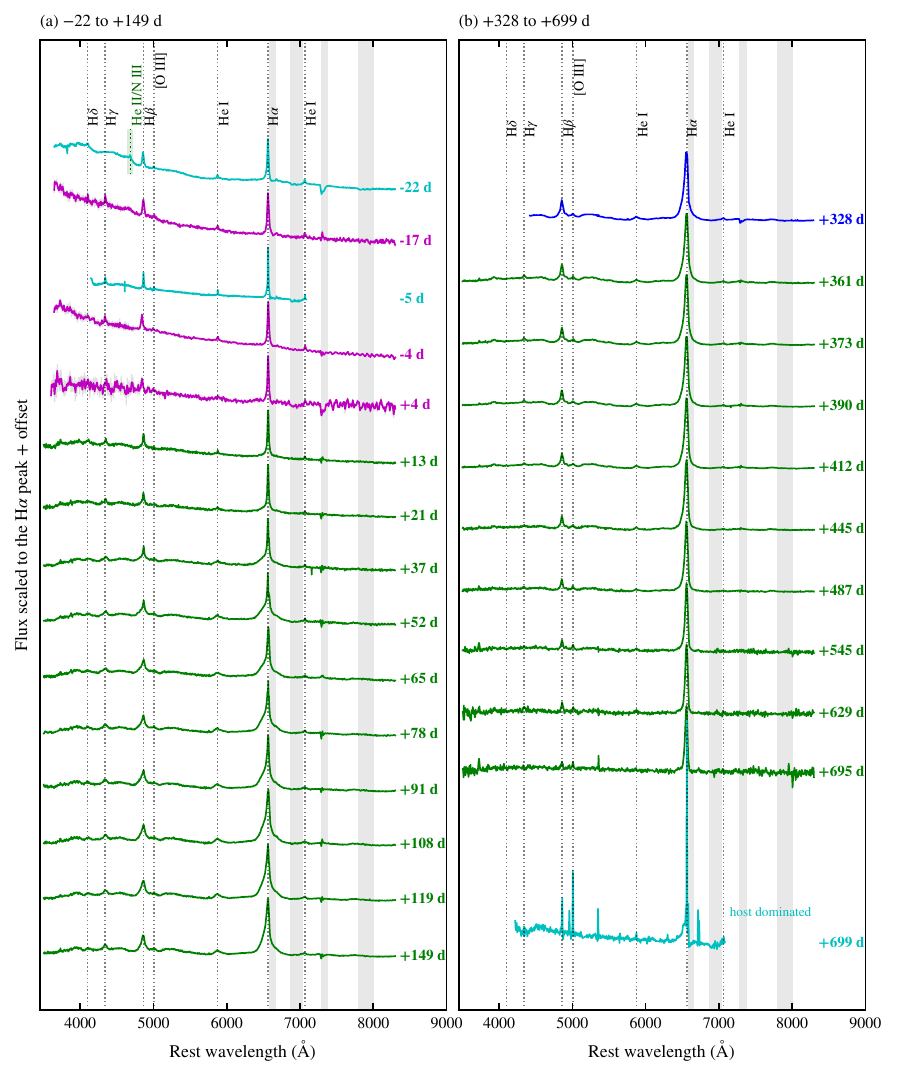}
\caption{Optical spectral evolution of SN 2019qem, (a) from $-22$ to $+149$\,d and (b) from $+328$ to $+699$\,d. Each spectrum is scaled to its H$\alpha$ peak and offset for clarity. Cyan spectra are from SALT, magenta from FLOYDS at Las Cumbres Observatory (smoothed; the unsmoothed spectra are shown in grey), green from NTT and blue from VLT/FORS2. The green band marks the \ion{He}{ii}\,$\lambda$4686/\ion{N}{iii}\,$\lambda$4640 flash feature in the $-22$\,d spectrum (cf. Fig.~\ref{fig:flash}). The $+699$\,d SALT spectrum is dominated by the host galaxy; it is scaled to the broad H$\alpha$ emission of the supernova and shifted down so that its strong narrow host lines do not overlap the $+695$\,d spectrum. The grey shaded bands are strong telluric absorption regions.}
\label{fig:all_spec}
\end{figure*}

Fig. ~\ref{fig:all_spec} shows the full optical spectral evolution for SN 2019qem. The log of the spectra has been tabulated in Table \ref{tab:log_spectra}. The spectra show the three phases that are typical of a strongly interacting Type IIn supernova. Around maximum light, from about -22 to +4 days, the spectrum has a blue continuum with narrow Balmer emission lines (H$\alpha$, H$\beta$, H$\gamma$, and H$\delta$) superimposed on it. Each Balmer line consists of a narrow core sitting on broad and smooth wings. The wings are smooth (but not symmetric; Sect.~\ref{subsec:line_evo}), and in the dense circumstellar gas of a Type IIn supernova, they are best explained by electron scattering of the line photons rather than by fast motion of the emitting gas \citep{2001MNRAS.326.1448C,2017hsn..book..403S}. Weak \ion{He}{i} $\lambda$5876 emission is also present, and the higher Balmer lines are clearly seen. During the following months, from about +13 to +149 days, the continuum becomes redder and fainter, while the emission lines grow stronger. An intermediate-width component and a broad component develop in H$\alpha$ and, more weakly, in H$\beta$. \ion{He}{i} $\lambda$5876 and a weaker \ion{He}{i} $\lambda$7065 lines are also present. In this phase emission from the interaction dominates the spectrum, as the forward shock drives into the dense CSM and a cold dense shell forms behind it \citep{1994ApJ...420..268C,2017hsn..book..403S}. At late times, from about +328 to +699 days, the continuum has almost disappeared and the spectrum is dominated by emission lines. H$\alpha$ is by far the strongest feature by the end of the sequence. The broad component of H$\alpha$ becomes weaker and narrower as the shock decelerates, while the intermediate-width component remains visible. The last NTT spectrum, at +695 days, is faint and is dominated by H$\alpha$ alone, while the SALT spectrum at $+699$\,d is dominated by the host galaxy, and about half of its H$\alpha$ flux is unresolved host emission. Its broad H$\alpha$ appears several times weaker than at $+695$\,d, which cannot be intrinsic over four days. The spectrum was scaled to the host-subtracted photometry although its continuum is mostly host light, so its absolute flux is underestimated, and we use it only for the host lines (Sect.~\ref{subsec:host}). The persistence of strong H$\alpha$ emission for about two years is direct evidence that the luminosity is powered by long-lived circumstellar interaction and not by radioactive decay. We do not detect the dense forest of narrow high-ionization coronal lines that appeared in some enduring Type IIn supernovae such as SN 2005ip \citep{2009ApJ...695.1334S}, although our resolution and signal-to-noise ratio limit the sensitivity to weak narrow lines.

\begin{figure}
    \centering
    \includegraphics[width=0.99\linewidth]{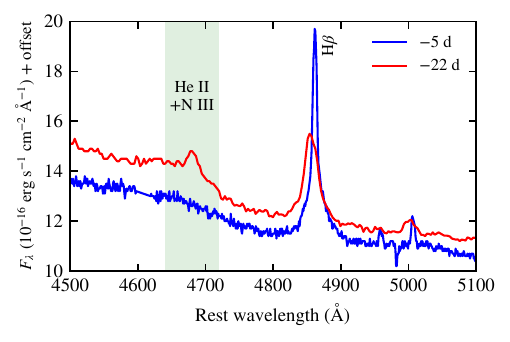}
    \caption{The SALT spectra at $-22$\,d (red) and $-5$\,d (blue) around H$\beta$. The shaded region marks the \ion{He}{ii}\,$\lambda$4686/\ion{N}{iii}\,$\lambda$4640 flash feature, present at $-22$\,d (about 8 rest-frame days after explosion) and absent at $-5$\,d.}
    \label{fig:flash}
\end{figure}

The earliest spectra of SN 2019qem show high-ionisation emission, known as flash features, that is produced when the ultraviolet flash from the explosion ionises the dense circumstellar gas closest to the star \citep{2014Natur.509..471G,2016ApJ...818....3K}. Such features are seen in a large fraction of young Type II supernovae, not only in SNe IIn \citep{2021ApJ...912...46B}. In the first SALT spectrum, taken at a phase of about -22 days (8-15 rest-frame days after explosion, depending on the explosion epoch), we identify an emission feature near 4660 $\AA$ (see Fig. \ref{fig:flash}). We attribute this feature to the blend of \ion{He}{ii} $\lambda$4686 and \ion{N}{iii} $\lambda$4640, which is the characteristic high-ionization signature seen in the early spectra of many interacting supernovae. Its presence shows that a dense and highly ionised circumstellar medium already surrounds the star at the time of explosion. This feature has disappeared in the second SALT spectrum at about -5 days and it is not seen in any of the later spectra. This behaviour is expected because flash features usually last from a few days to about two weeks after the explosion, as the ionizing flash fades and the shock overtakes the emitting gas.

\subsection{Spectral line evolution} \label{subsec:line_evo}

\begin{figure}
\centering
\includegraphics[width=\linewidth]{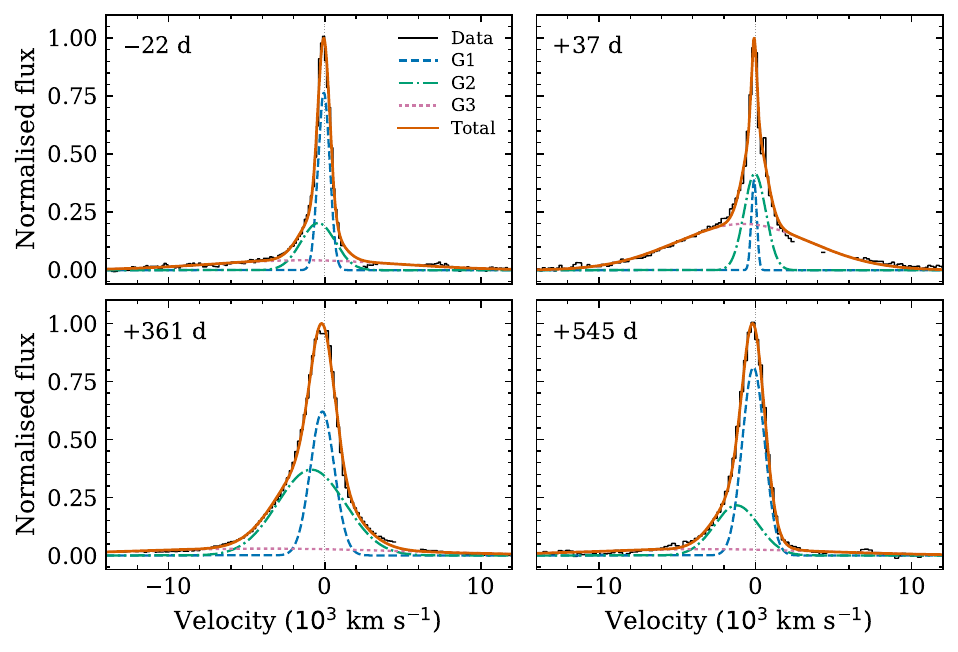}  
\caption{Three-Gaussian fits to H$\alpha$ at four epochs. The data are shown in black, the components G1-G3 (Table~\ref{tab:halpha_fits}) as broken lines and their sum as a solid line.}
\label{fig:halpha_fit}
\end{figure}

H$\alpha$ is the strongest line in all 26 spectra, from $-22$ to +699 d, and its shape changes considerably over this time (spectral sequence, Fig. ~\ref{fig:all_spec}). We followed \cite{2021MNRAS.506.4715R} to describe these changes consistently. We subtract a straight-line continuum fitted on both sides of the line, fit the profile with three components (first with Gaussians, then with Lorentzians), which represent narrow emission from the unshocked CSM, intermediate-width emission from the shocked gas, and broad emission from the fastest gas or from electron scattering, and compare the two with the Bayesian information criterion (BIC). \ion{He}{i}\,$\lambda$6678 is masked in every spectrum. Velocities are measured from the rest wavelength of H$\alpha$ at $z = 0.0427$. The instrumental resolution, about 270\,km\,s$^{-1}$ for SALT/PG0900 and several hundred km\,s$^{-1}$ for the other set-ups, matters only for the narrowest component, which we flag as unresolved where appropriate; the intermediate and broad components are several times wider. The host contributes little to the narrow component, because the host H$\alpha$ flux measured at $+699$\,d is only 2--4 per cent of the narrow-component flux near maximum. Table \ref{tab:halpha_fits} lists the Gaussian components, and Fig. \ref{fig:halpha_fit} shows four representative fits.

Near maximum ($-22$ to +21 d) H$\alpha$ has the classic 'Eiffel Tower' shape of SNe IIn \citep{2021MNRAS.506.4715R}, a narrow peak, carrying a third to a half of the line flux, on top of wings that extend several thousand km s$^{-1}$. The narrow line is unresolved in our highest-resolution spectrum (SALT/RSS at $-5$ d, resolution 270 km s$^{-1}$), so its intrinsic full width at half maximum (FWHM) is $\lesssim 200$ km s$^{-1}$ and the larger widths in the lower-resolution spectra come from blending with the broad emission. At these epochs the model with the lower BIC alternates between Gaussian and Lorentzian (Table~\ref{tab:halpha_fits}), so neither profile shape is consistently preferred. The narrow component then fades. It carries only about 5 per cent of the flux at +37 d and cannot be separated from the rest of the line after +52 d and from $+65$\,d onwards the narrowest Gaussian (G1, FWHM 1100--1800\,km\,s$^{-1}$) traces intermediate-width emission, not the narrow circumstellar line. Meanwhile, the H$\alpha$ flux rises almost fourfold up to +149 d, and at +37 to +65 d, 70–80 per cent of it is in a broad Gaussian component (FWHM $\approx 10\,000$ km s$^{-1}$), whose wings extend to higher velocities than the shocked shell and are probably broadened by electron scattering (Sect.~\ref{sec:discussion}). From +78 to +445 d, the Lorentzian model is preferred at every epoch ($\Delta{\rm BIC} = {\rm BIC_G} - {\rm BIC_L}$ between 15 and 503, where $\Delta{\rm BIC} > 10$ is conventionally strong evidence). The Gaussian fits reproduce the extended wings only by adding a weak, very broad component (FWHM $10\,000$–$21\,000$ km s$^{-1}$), which we do not consider physical. The broad Lorentzian component narrows steadily, from about 6400 km s$^{-1}$ at +78 d to 2700 km s$^{-1}$ at +487 d. After about +490 d, the Gaussian model gives the lower BIC again, because the red wing has become too weak to match a Lorentzian.

The narrowing and the asymmetry are also seen directly in the profile, independent of any model. The velocities at which H$\alpha$ falls to 10 per cent of its peak shrink from $-7200$ and +6200 km s$^{-1}$ at +78 d to $-3000$ and +1600 km s$^{-1}$ at +695 d, with the red wing shrinking faster. The line stays blueshifted throughout, and its flux-weighted centroid is at $-60$ to $-500$ km s$^{-1}$ near maximum and at $-420$ to $-690$ km s$^{-1}$ from +37 d onwards. A core with FWHM $\approx 1600$–1900 km s$^{-1}$ persists from +328 to +545 d, and intermediate-width emission is still present at +695 d, so the interaction continues to the end of our coverage. The +699 d SALT spectrum is dominated by host lines (narrow H$\alpha$ and [\ion{N}{ii}]); only its weak, blueshifted broad emission comes from the supernova.

According to the classification scheme of \citet{2021MNRAS.506.4715R}, SN 2019qem is a `gold' SN IIn as it has narrow circumstellar emission sitting on broader H$\alpha$ from $-22$ to +52 d, the supernova is offset from the nucleus of a star-forming host (Sect.~\ref{subsec:host}), and its peak absolute magnitude ($M \approx -20$) is far brighter than those of SN impostors and gap transients.

\subsection{Comparison to other supernovae} \label{subsec:comparison}

\begin{figure*}
\centering
\includegraphics[width=0.9\textwidth]{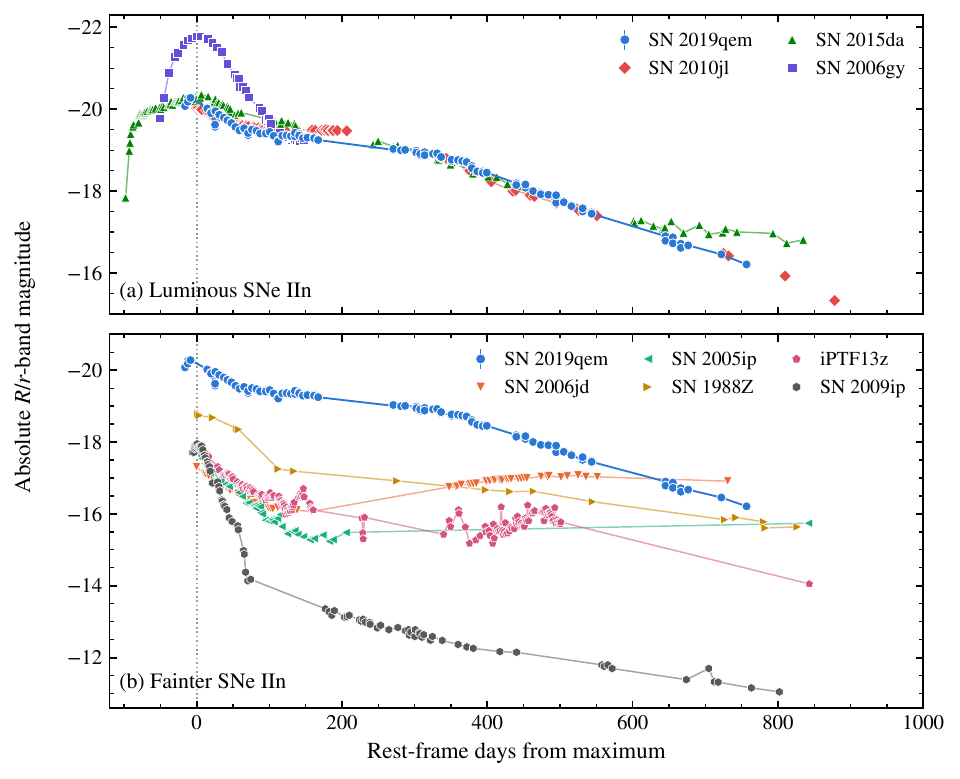} 
\caption{$R/r$-band absolute light curves of SN 2019qem (host-subtracted $r$) and of other Type IIn supernovae, (a) the luminous SN 2006gy, SN 2010jl and SN 2015da and (b) the fainter SN 1988Z, SN 2005ip, SN 2006jd, SN 2009ip and iPTF13z.}
\label{fig:lc_comp}
\end{figure*}

We compare SN 2019qem with well-studied Type IIn supernovae that span the luminosity range of the class: SN 1988Z \citep{1999MNRAS.309..343A}, SN 2005ip and SN 2006jd \citep{2009ApJ...695.1334S,2012ApJ...756..173S}, SN 2006gy \citep{2007ApJ...666.1116S}, SN 2009ip \citep{2013MNRAS.430.1801M}, SN 2010jl \citep{2014ApJ...797..118F}, SN 2015da \citep{2020A&A...635A..39T} and iPTF13z \citep{2017A&A...605A...6N}. Fig. \ref{fig:lc_comp} shows the $R/r$-band absolute light curves. With a peak of $M_r=-20.3$, SN 2019qem is a luminous Type IIn, about 1\,mag brighter than the mean peak of the untargeted sample of \citet{2020A&A...637A..73N}. At maximum it is about as bright as SN 2010jl, about 1.5\,mag fainter than SN 2006gy, and 1.5 to 3.0\,mag brighter than SN 1988Z, SN 2005ip, SN 2006jd, SN 2009ip and iPTF13z. During the first 300 days its decline closely follows that of SN 2010jl. SN 2015da is the closest match in the shape of the light curve. After a correction for a large host extinction, $E(B-V)\approx1$\,mag, it peaked at $M_R\approx-20.4$ after a slow rise of about 100\,d \citep{2020A&A...635A..39T}. Between $+250$ and $+550$\,d both supernovae fade at about 0.6\,mag per 100\,d (0.58 and 0.61\,mag, against 0.66\,mag for SN 2010jl), after which SN 2015da levels off and stays 0.3-0.7\,mag brighter than SN 2019qem. Another long-lived event, SN 2019vxm, reached $M_V\approx-20.0$, showed early flash-ionisation lines of \ion{He}{ii}, \ion{N}{iii} and \ion{C}{iii}, and developed a mid-infrared excess from seven months after maximum \citep{2026arXiv260630390C}. Fig.~\ref{fig:lc_comp} uses the host-subtracted $r$-band photometry of SN 2019qem, which fades from $M_r\approx-17.7$ at $+500$\,d to $\approx-16.2$ at $+760$\,d, close to SN 2010jl at the same phases.

\begin{figure*}[ht]
    \centering
    \includegraphics[width=\textwidth]{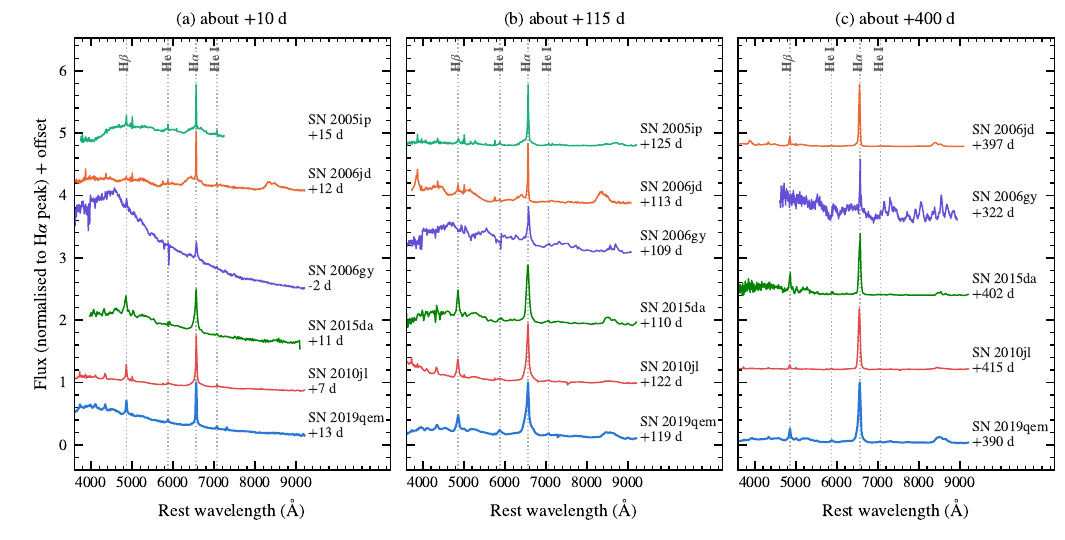}
    \caption{Comparison of the spectra of SN 2019qem (blue) with archival spectra of five well-studied Type IIn supernovae (SN 2005ip, SN 2006jd, SN 2006gy, SN 2010jl and SN 2015da) from WISeREP, at three epochs, (a) about +10 days, (b) about +115 days, and (c) about +400 days. Each spectrum is corrected for Galactic extinction and, for SN 2015da and SN 2006gy, for their large host extinction ($E(B-V)=0.97$ and, in total, 0.72\,mag; \citealt{2020A&A...635A..39T,2007ApJ...666.1116S}), shifted to the rest frame and normalized to its H$\alpha$ peak, and for each object the available epoch closest in phase to SN 2019qem is shown. SN 2019qem most closely resembles SN 2010jl and SN 2015da at all three epochs. The phase of each spectrum is given next to it.}
    \label{fig:spec_comp}
\end{figure*}

Fig. \ref{fig:spec_comp} compares the spectra at $+13$, $+119$ and $+390$\,d with archival spectra from WISeREP \citep{2012PASP..124..668Y}, originally published by \citet{2009ApJ...697..747K}, \citet{2010ApJ...709..856S}, \citet{2012AJ....143...17S}, \citet{2012ApJ...756..173S}, \citet{2016MNRAS.456.2622J} and \citet{2020A&A...635A..39T}, each corrected for extinction and normalised to the H$\alpha$ peak. Near maximum, SN 2019qem shows a blue continuum and Balmer lines with a narrow core and broad wings, very similar to SN 2010jl and SN 2015da. The H$\alpha$ equivalent width, which does not depend on reddening, is about 80\,$\AA$ in both SN 2019qem and SN 2015da, against about 180\,$\AA$ in SN 2010jl. Their continuum slopes also agree, but this is not an independent test, because the host extinction of SN 2015da was partly derived by matching its continuum to that of SN 2010jl \citep{2020A&A...635A..39T}. SN 2005ip and SN 2006jd have flatter continua and narrower H$\alpha$, and SN 2006gy has an even bluer continuum once its large extinction is removed, but weak H$\alpha$. At about $+120$\,d, SN 2019qem, SN 2010jl and SN 2015da all show strong H$\alpha$ with a broad base, while the lines of SN 2005ip and SN 2006jd remain narrower. At about $+400$\,d the spectrum of SN 2019qem is dominated by H$\alpha$ and is again closest to those of SN 2010jl and SN 2015da. Overall, SN 2019qem is most similar to SN 2010jl and SN 2015da in both its light curve and its spectral evolution. This similarity in luminosity, decline rate and line profiles suggests comparable CSM densities and shock velocities. For SN 2010jl, \citet{2014ApJ...797..118F} inferred $\dot M\approx0.1\,M_\odot$\,yr$^{-1}$ for $v_w = 100$\,km\,s$^{-1}$ and $\gtrsim3\,M_\odot$ of CSM, which we compare with our model in Sect.~\ref{sec:discussion}.

\subsection{Mid-infrared emission and dust} \label{subsec:mir}
\begin{figure}
    \centering
    \includegraphics[width=0.99\linewidth]{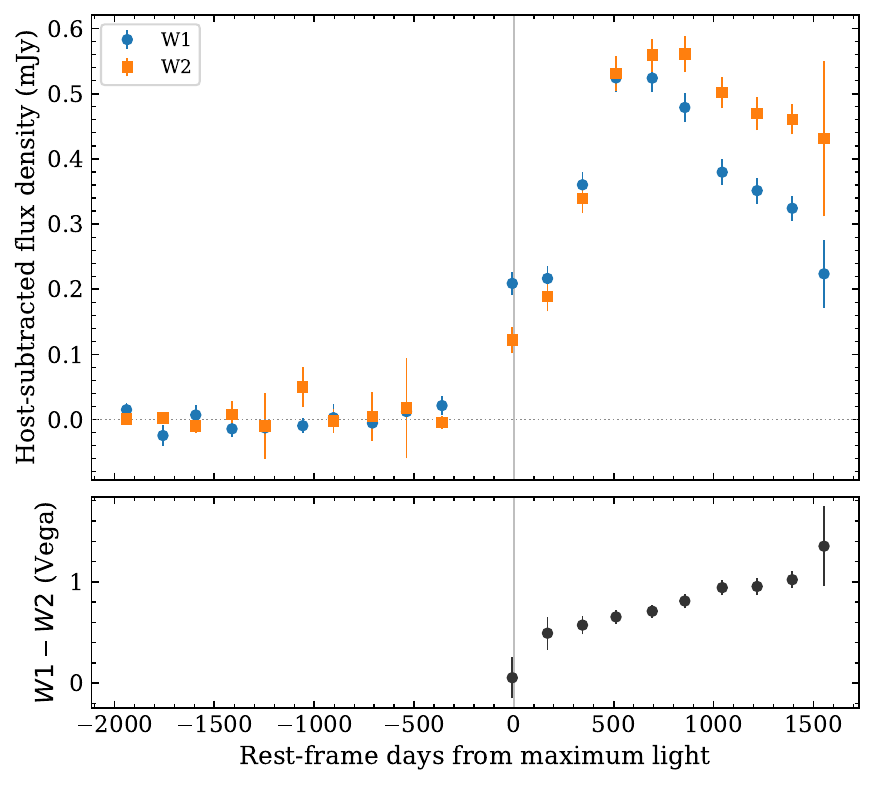}
    \caption{WISE light curve of SN 2019qem. The upper panel shows the host-subtracted $W1$ and $W2$ flux densities, including the pre-explosion visits (the dotted line marks zero excess), and the lower panel the $W1-W2$ colour of the excess (Vega). The vertical line marks maximum light.}
    \label{fig:wise}
\end{figure}

Figure \ref{fig:wise} shows the WISE light curve. We used the NEOWISE single-exposure photometry within $3''$ of the supernova, kept exposures with good frame quality and no contamination flags, and took the median flux of each visit. Before the explosion, the source was constant, and ten visits between 2014 and 2018 have a scatter of only 9\% in $W1$ and 14\% in $W2$, with no sign of pre-explosion variability. In particular, the visits 2.4 and 1.9\,yr before the explosion, which bracket the eruption epoch of our light-curve model (Sect.~\ref{sec:discussion}), show no excess down to $M_{W1}\approx-19.7$\,mag ($3\sigma$). This limit is too shallow to exclude a typical precursor outburst ($M\approx-13$ to $-15$\,mag; \citealt{2014ApJ...789..104O}). We subtracted this host level from all later visits.

The supernova is already detected at $-8$ d, when its mid-infrared colour is blue ($W1-W2 \approx 0.05$ mag). By +168 d the excess is much redder ($W1-W2 \approx 0.5$ mag), and it then brightens while the optical fades: between +168 and +693 d the $W1$ and $W2$ fluxes rise by factors of 2.4 and 3.0, whereas the $r$ band fades by 2.7 mag between +149 and +695 d. $W1$ levels off between +511 and +693 d and $W2$ peaks at +693 to +856 d, at $M_{W1} \approx -22.0$ and $M_{W2} \approx -22.7$ mag. The emission then declines slowly, by about 0.5 mag in $W1$ and 0.2 mag in $W2$ by +1394 d, and is still detected in both bands at +1553 d, the last NEOWISE visit. Throughout, the colour reddens steadily, from $W1-W2 \approx 0.5$ mag at +168 d to 1.0 mag at +1394 d.

\section{Discussion} \label{sec:discussion}
SN 2019qem is a long-lived Type IIn supernova. SN 2019qem exploded in the inner disc of a low-mass, star-forming galaxy with $\log(M_*/M_\odot) = 9.19$ and ${\rm SFR} = 0.19\,M_\odot$\,yr$^{-1}$ (Sect.~\ref{subsec:host}). The Type IIn hosts of the Palomar Transient Factory sample have median values of $\log(M_*/M_\odot) = 9.63$, $\log({\rm SFR}/M_\odot\,{\rm yr}^{-1}) = -0.09$ and $\log({\rm sSFR/yr^{-1}}) = -9.70$ \citep{2021ApJS..255...29S}. The host of SN 2019qem has about three times less stellar mass and four times less star formation than these medians, but a similar specific star formation rate, and it lies well within the distribution of Type IIn hosts. \citet{2021ApJS..255...29S} also showed that the masses of Type IIn hosts follow the distribution expected if these supernovae simply trace star formation, as for Type II hosts, so a low-mass host is not unusual for this class. Within their hosts, however, Type IIn supernovae follow the H$\alpha$ emission much less closely than Type Ic supernovae, which suggests that the class arises from a range of progenitors \citep{2014MNRAS.441.2230H}.

Narrow nebular lines at the position of the supernova show that star formation is ongoing there. About 60\% of Type IIn supernovae are associated with H$\alpha$ emission, consistent with massive progenitors such as luminous blue variables, while the other 40\% are not \citep{2022MNRAS.513.3564R}. The narrow host lines at the position of SN 2019qem are consistent with the first group, although we lack the H$\alpha$ imaging. The association is not universal for luminous-blue-variable-like events. SN 2009ip, also in our comparison sample, exploded in the remote outskirts of its host, with no sign of massive star formation within a few kpc, which \citet{2016MNRAS.463.2904S} interpreted as a progenitor rejuvenated by binary mass transfer or a merger.

Long-lasting Type IIn events similar to SN 1988Z and SN 2005ip tend to occur at lower metallicity than short-lived events similar to SN 1998S \citep{2015A&A...580A.131T}, and more luminous Type IIn supernovae tend to occur in environments with lower metallicity or younger stellar populations than fainter ones \citep{2023A&A...677A..20M}. In a volume-limited study of Type IIn environments with the \textit{Hubble Space Telescope}, the luminous events lie mainly in actively star-forming regions, whereas the faintest are associated with older populations and less massive progenitors \citep{2026ApJ..1001...44N}. SN 2019qem is both luminous and long-lived, and its star-forming site, with a sub-solar gas-phase metallicity (Sect.~\ref{subsec:host}), is consistent with both trends. The two closest analogues bracket it in metallicity. SN 2010jl exploded in a metal-poor dwarf galaxy with $12+\log({\rm O/H})=8.2\pm0.1$ \citep{2011ApJ...730...34S}, and SN 2015da in a large spiral with 8.48 near the explosion site \citep{2020A&A...635A..39T}, both on the \citet{2004MNRAS.348L..59P} scale. The host therefore supports a massive progenitor, although host properties alone cannot identify it.

The bolometric light curve peaks at $7.0\times10^{43}$\,erg\,s$^{-1}$ and radiates $6.9\times10^{50}$\,erg between $-29$ and $+545$\,d (Sect. \ref{subsec:lightcurve}), not counting the mid-infrared emission (Sect.~\ref{subsec:mir}). Powering the luminosity 151 rest-frame days after explosion ($+121$\,d) by $^{56}$Co decay alone would require about 4\,$M_\odot$ of $^{56}$Ni, an extreme amount, and the luminosity declines at about half the rate of $^{56}$Co decay. For such an unusual object these arguments alone are not conclusive, but the hydrogen-dominated spectra and the H$\alpha$ emission that lasts for two years show that SN~2019qem is powered by the interaction of the ejecta with a dense circumstellar medium \citep{1994ApJ...420..268C,2017hsn..book..403S}.

\begin{figure}
\centering
\includegraphics[width=\linewidth]{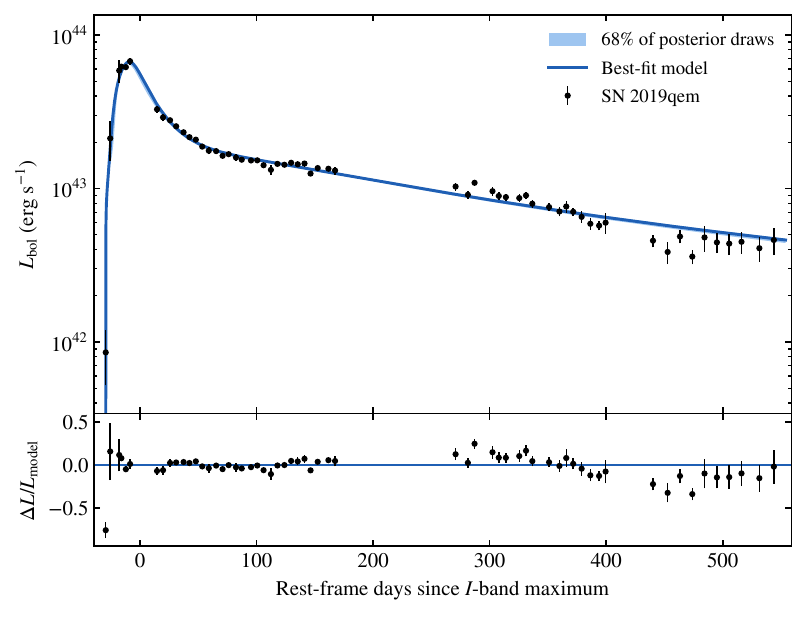}
\caption{Bolometric light curve of SN 2019qem (black) with the preferred CSM-interaction model (ejecta with a broken power-law density profile running into a steady wind plus a Gaussian eruption). The line is the maximum-likelihood model and the band encloses 68 per cent of 300 posterior draws. The lower panel shows the fractional residuals from the maximum-likelihood model.}
\label{fig:redback_fit}
\end{figure}

We fitted the bolometric light curve with the thin-shell interaction models of \texttt{REDBACK} \citep{2024MNRAS.531.1203S,2026MNRAS.552g1640S} to understand the circumstellar medium. In these models, the luminosity comes from kinetic energy that is converted into radiation at the forward and reverse shocks, and we required the velocity of the shocked shell to stay within limits set by the H$\alpha$ line widths. These limits do not fix the explosion energy directly but act through the shell velocity, which depends on the explosion energy, the ejecta mass and the wind density. The preferred model (\texttt{gausswind\_bpl}) has ejecta with a broken power-law density profile running into a steady wind, with an additional eruption in the mass-loss history described by a Gaussian in time. The fit used \texttt{bilby} \citep{2019ApJS..241...27A} with the nested sampler \texttt{PyMultiNest} \citep{2009MNRAS.398.1601F,2014A&A...564A.125B}, a Gaussian likelihood with an additional free scatter term, and a soft penalty that keeps the shell velocity between half the FWHM of the intermediate and of the broad H$\alpha$ components, with a 15 per cent systematic uncertainty on the widths. Table~\ref{tab:redback} lists the priors and fixed parameters. The preferred model is strongly favoured over a single power-law CSM, with a difference in Bayesian evidence $Z$ of $\Delta\ln Z = 30.5$, and the shell of the power-law model moves about twice as fast as the H$\alpha$ wings allow. The light curve alone also prefers the wind model, by $\Delta\ln\mathcal{L}\approx19$ in maximum likelihood $\mathcal{L}$. The fit is shown in Fig. \ref{fig:redback_fit} and the posterior in Fig. \ref{fig:redback_corner}. The preferred model gives an ejecta mass of $7.0^{+1.4}_{-0.9}\,M_\odot$, an explosion energy of $5.2^{+3.6}_{-1.4}\times10^{51}$\,erg and a conversion efficiency of $0.67^{+0.23}_{-0.27}$. For a wind speed of 100\,km\,s$^{-1}$, the progenitor lost $0.62^{+0.24}_{-0.15}\,M_\odot$\,yr$^{-1}$ and, in addition, ejected $9.3^{+2.0}_{-1.6}\,M_\odot$ in an eruption $2.2^{+0.4}_{-0.3}$\,yr before the explosion. By $+544$\,d the shock had reached $1.2\times10^{16}$\,cm and swept up about $32\,M_\odot$. The $^{56}$Ni mass is not constrained ($\lesssim0.07\,M_\odot$ at 95\% confidence, within a prior that extends to $0.1\,M_\odot$) and supplies less than 1\% of the radiated energy.

\begin{figure}
\centering
\includegraphics[width=\linewidth]{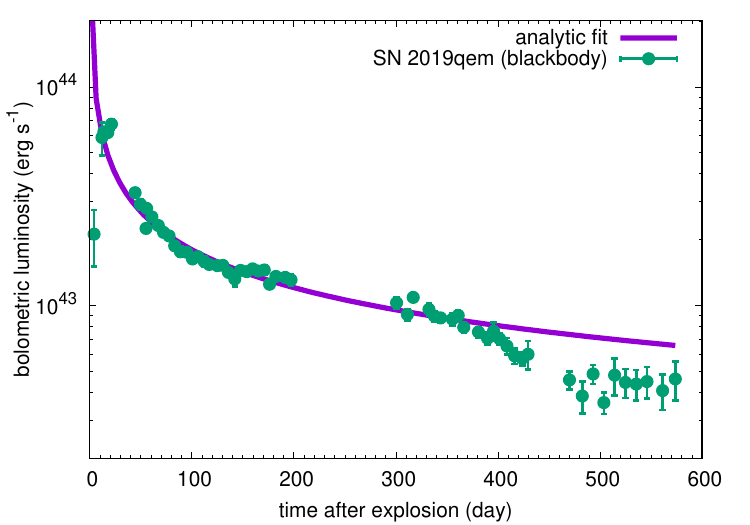}
\caption{Bolometric light curve of SN 2019qem (green) with the analytic interaction model of \citet{2013MNRAS.435.1520M} (purple). Time is measured from the explosion, assumed 30\,d before the $I$-band maximum.}
\label{fig:moriya_fit}
\end{figure}

These results can be compared directly with SN 2010jl and SN 2015da, the two closest analogues in our comparison sample (Sect.~\ref{subsec:comparison}). For SN 2010jl, \citet{2014ApJ...797..118F} found a peak bolometric luminosity of $\sim3\times10^{43}$\,erg\,s$^{-1}$, about half that of SN 2019qem, and a radiated energy of $\gtrsim6.5\times10^{50}$\,erg over $\sim$1100\,d, similar to the $6.9\times10^{50}$\,erg that SN 2019qem emitted within 575\,d. They modelled the light curve as a radiative shock in an $r^{-2}$ CSM with $\dot M\approx0.1\,M_\odot$\,yr$^{-1}$ for $v_w = 100$\,km\,s$^{-1}$ and $\gtrsim3\,M_\odot$ of CSM, measured narrow CSM lines at $\sim100$\,km\,s$^{-1}$, attributed the broad H$\alpha$ wings to electron scattering and the line blueshift to causes other than dust, and found a near-infrared excess from CSM dust and evidence for an asymmetric, luminous-blue-variable-like CSM. Each of these has a counterpart in SN 2019qem (Sects.~\ref{subsec:line_evo} and \ref{subsec:mir}). SN 2019qem therefore appears to be a more extreme version of SN 2010jl: it reached twice the peak luminosity and radiated the same energy in half the time, and our model requires a denser wind and more swept-up CSM. SN 2015da is closer still in power. Corrected for the missing ultraviolet, its bolometric light curve rose in about 30\,d to $6.2\times10^{43}$\,erg\,s$^{-1}$ \citep{2020A&A...635A..39T}, close to that of SN 2019qem, and it radiated at least $1.6\times10^{51}$\,erg over seven years \citep{2024MNRAS.530..405S}. After the peak its luminosity fell as $t^{-0.92}$ \citep{2020A&A...635A..39T}, faster than the $t^{-0.54}$ of SN 2010jl \citep{2014ApJ...797..118F} and the $t^{-0.58}$ of SN 2019qem (Fig.~\ref{fig:moriya_fit}). Its CSM expanded at 90-110\,km\,s$^{-1}$, and estimates of its mass range from 5.5--10 \citep{2020A&A...635A..39T} and 14 \citep{2026arXiv260708693C} to more than $20\,M_\odot$ \citep{2024MNRAS.530..405S}, for an explosion energy of (4-10)$\times10^{51}$\,erg. These values bracket our $\approx32\,M_\odot$ of swept-up CSM and $5\times10^{51}$\,erg. Late radio and X-ray limits show that the heavy mass loss of SN 2015da lasted at most a few centuries \citep{2026ApJ..1001..111H}, consistent with the $\lesssim40$\,yr over which the CSM swept up by SN 2019qem was lost.

The model reproduces the rise, the peak and the slow decline over almost 600\,d (Fig. \ref{fig:redback_fit}), with $\chi^2/N = 3.45$. The data lie about 10-15\% above the model between $+270$ and $+340$\,d and below it at $+440$-$475$\,d. The decline itself reflects the decelerating shell, because the shock luminosity scales as $\epsilon\,(\dot M/v_w)\,v_{\rm sh}^3$, and the interaction continues to the end of our coverage (Sect.~\ref{subsec:line_evo}). The dip at $+440$-$475$\,d is comparable to the systematic uncertainty of the late luminosities (Sect.~\ref{subsec:lightcurve}) and may also reflect density variations in the CSM. The explosion epoch sits at the latest date allowed by the pre-discovery OGLE photometry, the explosion energy and the efficiency are strongly correlated, and all mass-loss rates and times scale linearly with the assumed wind speed; only quantities that depend on $\dot{M}/v_w$, such as the swept-up mass, do not. Even so, the wind density, $\dot{M}/v_w \simeq 4\times10^{18}$\,g\,cm$^{-1}$, is far higher than the steady winds of massive stars can produce \citep{2014ARA&A..52..487S}, and the gas swept up by $+544$\,d was lost within the last $\sim$40\,yr before the explosion.

The fit also shares the limitations of thin-shell models \citep{2026MNRAS.552g1640S}. The light curve constrains mainly $\dot M/v_w$ and $E_{\rm K}\epsilon = (3.5\pm0.3)\times10^{51}$\,erg rather than the individual parameters; $\epsilon = 0.67$ is higher than the $\approx0.1$ of radiation-hydrodynamics calculations \citep{2014MNRAS.439.2917M}, and a lower $\epsilon$ would require a higher $E_{\rm K}$. The one-zone luminosity treatment does not reproduce the detailed shape of the rise, which likely explains the poor fit to the earliest point ($-29.4$\,d; Fig.~\ref{fig:redback_fit}). The eruption mass and epoch depend on the assumed constant CSM velocity and spherical geometry, so a disc-like or bipolar eruption would give different values, the CSM masses are spherical-equivalent, and the light curve omits the mid-infrared emission (Sect. \ref{subsec:mir}).

The H$\alpha$ profiles (Sect.~\ref{subsec:line_evo}) test this picture independently. The narrow line is unresolved (FWHM $\lesssim 200$\,km\,s$^{-1}$), so the gas ahead of the shock is slow, as assumed, although its speed remains unknown. In the model the eruption material lies at about $7\times10^{14}$\,cm and is swept up within a few weeks of the explosion, before maximum light. It powers the rise and the high peak luminosity, and it also accounts for the flash-ionised He\,{\sc ii} $\lambda$4686 emission at $-22$\,d (Sect.~\ref{subec:spectra}). The narrow H$\alpha$ line near and after maximum must therefore come from the steady wind further out. For a wind with density $\rho\propto r^{-2}$, the emission measure of the unshocked gas outside the shock falls as $1/R_{\rm sh}$, so the narrow line should weaken as the shock moves outwards, and faster still as the ionising luminosity declines. This agrees with its rapid fading after maximum.

The width of the broad H$\alpha$ emission agrees with the velocity of the shocked shell. In the model the shell slows from 3300\,km\,s$^{-1}$ at $+15$\,d to 2800\,km\,s$^{-1}$ at $+100$\,d and 1800\,km\,s$^{-1}$ at $+544$\,d. Around $+100$\,d the shell velocity lies between the half-widths of the intermediate Gaussian component (2500-2950\,km\,s$^{-1}$) and of the broad Lorentzian component (3100-3250\,km\,s$^{-1}$) at $+78$-$108$\,d. The observed wings also narrow with time, as expected for a decelerating shell, but faster than the model predicts: between $+91$ and $+487$\,d the half-width of the broad Lorentzian component falls by a factor of 2.4, whereas the model shell slows by a factor of only 1.5, so at $+445$-$487$\,d the shell is 25-40\% faster than this half-width. Near maximum, the opposite holds: at $+37$-$65$\,d the broad wings, with a half-width of about 5000\,km\,s$^{-1}$, are faster than the shell ($\approx$3000-3200\,km\,s$^{-1}$). Both differences suggest that the line wings are shaped by electron scattering as well as by the motion of the gas.

The model also predicts how opaque the unshocked wind is to electron scattering. If the wind is fully ionised, its optical depth outside the shock is $\tau_{\rm es} = \kappa_{\rm es}\dot{M}/(4\pi v_w R_{\rm sh})$, which depends on the wind density but not on the assumed wind speed. With $\kappa_{\rm es} = 0.34$\,cm$^2$\,g$^{-1}$ this gives $\tau_{\rm es}\approx 30$ at $+100$\,d, falling to $\approx 10$ at $+445$\,d and $\approx 9$ at $+544$\,d. Such optical depths naturally produce the Lorentzian profiles observed from $+78$ to $+445$\,d, whose wings extend beyond the velocity of the gas \citep{2001MNRAS.326.1448C,2009MNRAS.394...21D,2018MNRAS.475.1261H}, and a declining $\tau_{\rm es}$ helps to explain why the wings narrow faster than the shell slows down. After about $+490$\,d, however, the profiles become Gaussian-like again, although the model still gives $\tau_{\rm es}\approx 9$. This suggests that by then much of the gas ahead of the shock is no longer ionised, or that the circumstellar medium is less dense or more clumpy than the smooth wind of the model.

The line shift tells us about the geometry of the emitting region. H$\alpha$ is blueshifted at all epochs, with its centroid at $-60$ to $-500$\,km\,s$^{-1}$ near maximum and at $-420$ to $-690$\,km\,s$^{-1}$ from $+37$\,d onwards. At late times its red wing is also cut short: at $+545$\,d it reaches only $+1500$\,km\,s$^{-1}$, less than the model shell velocity of 1800\,km\,s$^{-1}$, whereas the blue wing reaches $-3300$\,km\,s$^{-1}$. Such blueshifts are often attributed to dust formation in the cold dense shell behind the shock, which absorbs light from the receding side; for SN 2010jl both this interpretation \citep{2012AJ....143...17S,2014Natur.511..326G} and bulk motion of the emitting gas \citep{2014ApJ...797..118F} have been proposed, and the same two views exist for SN 2015da \citep{2024MNRAS.530..405S,2020A&A...635A..39T}. In SN 2019qem, three results argue against newly formed dust as the main cause. The line is already blueshifted near maximum, when the shock, at about $10^{15}$\,cm in the model, lies far inside the radius of $4$-$7\times10^{16}$\,cm within which the peak luminosity destroys dust. In SN 2010jl and SN 2015da, by contrast, the blueshift appeared only about 50 and 110\,d after explosion \citep{2014ApJ...797..118F,2024MNRAS.530..405S}, whereas in SN 2019qem it is present about 30\,d after explosion. After the maximum, H$\beta$ is not more blueshifted than H$\alpha$, as dust extinction would require, although blends around H$\beta$ weaken this test. And the warm dust seen in the mid-infrared lies outside the shock, where it dims the light from the approaching and receding gas equally. Given the large electron-scattering optical depth in the model, we favour a geometric explanation where light from the receding side is partly hidden by the optically thick inner regions, and the circumstellar medium may not be spherical \citep{2021MNRAS.506.4715R}, as the continuum polarisation of 1.7--2.0 and about 3 per cent near maximum indicates for SN 2010jl and SN 2015da \citep{2011A&A...527L...6P,2024MNRAS.529.1104B}. A small contribution from new dust cannot be excluded.

Overall, the light-curve model and the H$\alpha$ profiles give a consistent picture. An energetic explosion ran first into about $9\,M_\odot$ of CSM ejected a few years earlier, and then into a dense, slow wind that it was still sweeping up two years later. The line widths support the moderate shock velocities of the model, and the line shapes point to an optically thick, electron-scattering and probably asymmetric interaction region. The remaining tensions, namely the correlated light-curve residuals, the fast narrowing of the line wings at late times and the Gaussian-like profiles after $+490$\,d, suggest that the real circumstellar medium is less smooth than a spherical wind with a single eruption.

We also fitted the light curve with the analytic interaction model of \citet{2013MNRAS.435.1520M,2014MNRAS.439.2917M}, shown in Fig.~\ref{fig:moriya_fit}.
The bolometric light curve from the peak luminosity to around 350~days can be fitted as $2.58\times 10^{44} (t/1~\mathrm{d})^{\alpha}~\mathrm{erg~s^{-1}}$ where $\alpha = -0.578$.

If we assume that the CSM density $\rho_\mathrm{CSM}$ follows $\rho_\mathrm{CSM}\propto r^{-2}$ (a steady wind) and the SN ejecta density structure $\rho_\mathrm{ej}$ follows the double power-law profile ($\rho_\mathrm{ej}\propto r^{-n}$ outside and $\rho_\mathrm{ej}\propto r^{-\delta}$ inside), $\alpha$ can be expressed as $\alpha=-3/(n-2)$. Given $\alpha = -0.578$, $n$ is estimated to be $n=7.2$, which is similar to the value estimated for SN~2010jl by \citet{2014ApJ...797..118F}. Assuming $\delta = 1$, $\epsilon = 0.3$, the ejecta mass of $5~\mathrm{M_\odot}$, and the explosion energy of $5\times 10^{51}~\mathrm{erg}$, we obtain $\dot{M}=0.02~\mathrm{M_\odot~yr^{-1}}$ with $v_w=100~\mathrm{km~s^{-1}}$ (see \citealt{2013MNRAS.435.1520M} for the detailed analytic formulation). 

This rate is 30 times lower than in the \texttt{REDBACK} fit. Because the shock luminosity scales as $\epsilon\,(\dot M/v_w)\,v_{\rm sh}^3$, the light curve alone cannot separate the wind density from the shell velocity \citep{2026MNRAS.552g1640S}. With its $\epsilon = 0.3$, compared with 0.67 in the \texttt{REDBACK} fit, the analytic model needs $v_{\rm sh}\approx9\times10^3$\,km\,s$^{-1}$ at $+100$\,d to produce the observed luminosity, more than three times the $\approx2800$\,km\,s$^{-1}$ allowed by the H$\alpha$ widths. Such a shell would produce H$\alpha$ wings extending to $\approx\pm9000$\,km\,s$^{-1}$, whereas the observed line falls to 10 per cent of its peak at $-7200$ and $+6200$\,km\,s$^{-1}$ at $+78$\,d (Sect.~\ref{subsec:line_evo}), even though electron scattering broadens it. The analytic solution is therefore inconsistent with the line widths of SN 2019qem, and we adopt the \texttt{REDBACK} solution. The comparison shows that, without a velocity constraint, mass-loss rates inferred from SN IIn light curves can differ by more than an order of magnitude, as also found for SN 2010jl \citep{2014MNRAS.439.2917M,2014ApJ...797..118F}.

The mid-infrared light curve shows the usual signature of warm dust (Sect.~\ref{subsec:mir}). Near maximum, the supernova is blue in the mid-infrared ($W1-W2\approx0.05$\,mag), as expected from its hot photosphere, but from $+168$\,d onwards the excess is much redder, keeps reddening, and brightens while the optical fades. Such a rising and reddening excess is thermal emission from dust that slowly cools, as seen in other long-lived Type IIn supernovae \citep{2011ApJ...741....7F,2019ApJS..241...38S}. This does not conflict with the argument above against dust as the cause of the H$\alpha$ blueshift. That argument concerns dust newly formed in the cold dense shell, whereas the mid-infrared excess can arise in pre-existing circumstellar dust outside the shock, which dims the light from the approaching and receding gas equally. At its peak the excess reaches $\nu L_\nu\approx2\times10^{42}$\,erg\,s$^{-1}$ in each of $W1$ and $W2$, comparable to the $\approx4$--$5\times10^{42}$\,erg\,s$^{-1}$ that the interaction still radiates in the optical at $+500$\,d (Fig.~\ref{fig:redback_fit}), so the dust must reprocess a substantial fraction of the interaction luminosity, but no additional energy source is needed. The dust needs a heat source that lasts for years: its emission peaks around $+700$\,d, when the $r$ band has faded by 2.7\,mag since $+149$\,d, and the dust emission is still detected at $+1553$\,d. The interaction is the natural source, because the H$\alpha$ profiles and the light-curve model show that the shock was still converting kinetic energy into radiation at these times. The infrared excesses of the two closest analogues have been interpreted in different ways. That of SN 2010jl appeared within 90\,d of explosion and has been attributed to an echo from pre-existing dust \citep{2011AJ....142...45A,2014ApJ...797..118F,2021ApJ...917...84D} and, after about day 380, to new dust formed behind the shock \citep{2014Natur.511..326G,2018ApJ...859...66S}, of which JWST found $\gtrsim0.1\,M_\odot$ after 13\,yr \citep{2026ApJ..1006..224S}. That of SN 2015da appeared only $\approx430$\,d after explosion and has been attributed to an echo \citep{2020A&A...635A..39T}, to new dust in the cold dense shell \citep{2024MNRAS.530..405S}, or to new dust in the unshocked ejecta \citep{2026arXiv260708693C}. In SN 2019qem the red excess was already present at $+168$\,d, the first WISE visit after maximum and about 200\,d after explosion, so its onset lies between those of the two analogues. Its luminosity of $\nu L_\nu\approx2\times10^{42}$\,erg\,s$^{-1}$ in each of $W1$ and $W2$ is of the same order as the dust luminosities of $\gtrsim5\times10^{42}$\,erg\,s$^{-1}$ at 500\,d in SN 2010jl \citep{2014ApJ...797..118F} and $1.6\times10^{42}$\,erg\,s$^{-1}$ at peak in SN 2015da \citep{2020A&A...635A..39T}, and in all three the emission lasts for years. With only two mid-infrared bands, our data cannot by themselves separate these origins. The lack of any link between the growing dust emission and the H$\alpha$ blueshift suggests, however, that any newly formed dust has little effect on the line profiles.

The \texttt{REDBACK} light-curve model implies that the $\approx7\,M_\odot$ of ejecta swept up about 32\,$M_\odot$ of circumstellar gas by $+544$\,d, that about 9\,$M_\odot$ of this was lost in a single episode about two years before core collapse, and that the progenitor lost about 0.6\,$M_\odot$\,yr$^{-1}$ in the preceding decades (for $v_w = 100$\,km\,s$^{-1}$). This wind is about six times denser than the $\approx0.1\,M_\odot$\,yr$^{-1}$ inferred for SN 2010jl, and the swept-up mass is also larger than the $\gtrsim3\,M_\odot$ of CSM found there \citep{2014ApJ...797..118F}. It is, however, similar to the $\approx0.6\,M_\odot$\,yr$^{-1}$ inferred for SN 2015da \citep{2020A&A...635A..39T}, whose CSM mass may exceed $20\,M_\odot$ \citep{2024MNRAS.530..405S}. Such mass loss is far beyond the steady winds of red supergiants and Wolf-Rayet stars and is comparable to the giant eruptions of luminous blue variables \citep{2014ARA&A..52..487S,2017hsn..book..403S}. Models of enhanced mass loss from red supergiants in their final centuries give $10^{-4}$-$10^{-2}\,M_\odot$\,yr$^{-1}$ \citep{2026ApJ...996...18S}, and Case~C mass transfer in wide binaries ejects only 0.01-0.2\,$M_\odot$ in the last $\sim10^3$\,yr \citep{2026ApJ..1005L..36T}. Even for a red-supergiant wind speed of 10-20\,km\,s$^{-1}$, our wind density implies 0.06-0.12\,$M_\odot$\,yr$^{-1}$, and the $\approx32\,M_\odot$ of CSM swept up by $+544$\,d, which does not depend on the wind speed, exceeds what either channel provides by an order of magnitude or more. Taken at face value, the model also constrains the progenitor mass. With $\approx32\,M_\odot$ of swept-up CSM, $\approx7\,M_\odot$ of ejecta and a compact remnant of at least $\sim1.5\,M_\odot$, the progenitor had an initial mass of at least $\approx40\,M_\odot$, without counting any mass lost before the dense CSM formed. A similar budget led \citet{2024MNRAS.530..405S} to a progenitor of more than $30\,M_\odot$ for SN 2015da. This estimate depends on the smooth, spherical CSM of the model. If similar mass budgets are found for other well-observed SNe IIn, some of the most massive stars missing from the red-supergiant progenitor population \citep{2015PASA...32...16S} may end their lives as strongly interacting supernovae.

The eruption itself may have been observable. If it radiated an energy comparable to its kinetic energy, $\frac{1}{2}M_{\rm er}v^2\approx10^{48}\,(v/100\,{\rm km\,s^{-1}})^2$\,erg, over $\sigma_{\rm er}\approx0.6\,(100\,{\rm km\,s^{-1}}/v)$\,yr, its luminosity was $\approx5\times10^{40}\,(v/100\,{\rm km\,s^{-1}})^3$\,erg\,s$^{-1}$, which corresponds to $M\approx-13$\,mag for an ejection speed of 100\,km\,s$^{-1}$ and $M\approx-16.6$\,mag for 300\,km\,s$^{-1}$. Gaia visited the position 53 times between November 2014 and September 2019, 13 times within the modelled eruption window, and never detected a source there \citep{2021A&A...652A..76H}. This limits any outburst at these epochs to $M_G\gtrsim-16$, and the WISE limit in the mid-infrared, $M_{W1}\approx-19.7$\,mag (Sect.~\ref{subsec:mir}), is shallower, so only the fastest and brightest eruptions are excluded. The VISTA survey of the Magellanic Clouds system \citep[VMC;][]{2011A&A...527A.116C} imaged the field in $Y$, $J$ and $K_{\rm s}$ on 19 nights between February 2017 and March 2018, during the modelled eruption, and could test it more deeply. An eruption with $M\approx-13$\,mag would have an apparent magnitude of about 23.5 at the distance of SN 2019qem, within reach of single visits of the Vera C. Rubin Observatory, which should therefore detect such precursors in future events of this kind. The star-forming, low-mass host is compatible with a very massive progenitor, but, as discussed above, host properties alone cannot establish it.

\section{Conclusion} \label{sec:conclusion}
We have presented optical photometry, 26 optical spectra and four years of mid-infrared photometry of the Type IIn SN 2019qem. Our main conclusions are as follows.
\begin{enumerate}

\item SN 2019qem rose in $\approx$30\,d to $M_I = -20.5$\,mag and $7.0\times10^{43}$\,erg\,s$^{-1}$, and radiated $6.9\times10^{50}$\,erg within 575\,d (from blackbody fits to the optical photometry, excluding the mid-infrared). Its light curve and spectra most closely resemble those of SN 2010jl and SN 2015da.

\item Radioactive decay would need $\approx4\,M_\odot$ of $^{56}$Ni to power the light curve. Instead, the H$\alpha$ emission that lasts two years, a \ion{He}{ii}/\ion{N}{iii} flash feature 8--15\,d after explosion and unresolved narrow Balmer lines (FWHM $\lesssim200$\,km\,s$^{-1}$) show interaction with dense, slow, ionised CSM.

\item Thin-shell models constrained by the H$\alpha$ line widths favour $7.0^{+1.4}_{-0.9}\,M_\odot$ of ejecta with $5.2^{+3.6}_{-1.4}\times10^{51}$\,erg, running first into $9.3^{+2.0}_{-1.6}\,M_\odot$ ejected $2.2^{+0.4}_{-0.3}$\,yr before explosion and then into a wind of $0.62^{+0.24}_{-0.15}\,M_\odot$\,yr$^{-1}$ (for $v_w = 100$\,km\,s$^{-1}$). By $+544$\,d the shock had swept up $\approx32\,M_\odot$ lost within the last $\sim$40\,yr. Only $\dot M/v_w$ and $E_{\rm K}\epsilon = (3.5\pm0.3)\times10^{51}$\,erg are well constrained. Taken at face value, the CSM and ejecta masses imply a progenitor of at least $\approx40\,M_\odot$.

\item The analytic model of \citet{2013MNRAS.435.1520M} reproduces the light curve to $\approx$350\,d with a 30 times lower mass-loss rate, but only with a shell more than three times faster than the H$\alpha$ widths allow. Because the interaction luminosity scales as $\epsilon\,(\dot M/v_w)\,v_{\rm sh}^3$, mass-loss rates of SNe IIn from light curves alone are uncertain by more than an order of magnitude, and spectroscopic shock velocities are needed to fix them.

\item The H$\alpha$ profiles are Lorentzian from $+78$ to $+445$\,d, their broad component narrows from about 6400 to 2700\,km\,s$^{-1}$, and from $+37$\,d onwards the line is blueshifted by 400--700\,km\,s$^{-1}$. This points to an optically thick, electron-scattering and probably asymmetric interaction region rather than mainly to newly formed dust.

\item The mid-infrared excess brightens for about two years while the optical fades and is still detected at $+1553$\,d. It is most likely pre-existing circumstellar dust heated by the ongoing interaction, which can supply its $\nu L_\nu\approx2\times10^{42}$\,erg\,s$^{-1}$ per band.

\item SN 2019qem exploded in the inner disc of a low-mass, star-forming galaxy, $\log(M_*/M_\odot) = 9.19$, at a site with a sub-solar gas-phase metallicity, $12+\log({\rm O/H})\approx8.3$-$8.4$.
\end{enumerate}

The progenitor thus lost mass in a dense wind for decades and in an eruption two years before core collapse, far beyond what red-supergiant winds or Case~C mass transfer in wide binaries are predicted to supply, and similar to the eruptions of luminous blue variables. The correlated light-curve residuals and the late-time H$\alpha$ profiles show that the real CSM is less smooth than our spherical wind-plus-eruption model, which multi-dimensional models and continued late-time monitoring can test.

\begin{acknowledgements}
BHTOM is based on the open-source TOM Toolkit by Las Cumbres Observatory and has been developed with funding from the OPTICON-RadioNet Pilot (ORP) of the European Union's Horizon 2020 research and innovation programme under grant agreement No 101004719 (2021-2025). This project has received funding from the European Union's Horizon Europe Research and Innovation programme ACME under grant agreement No 101131928 (2024-2028).
The OGLE project has received funding from the Polish National Science Centre grant OPUS-28 2024/55/B/ST9/00447 awarded to AU.
Based on observations collected at the European Organisation for Astronomical Research in the Southern Hemisphere, Chile, as part of ePESSTO+ (the advanced Public ESO Spectroscopic Survey for Transient Objects Survey – PI: Inserra). ePESSTO+ observations were obtained under ESO program ID 1103.D-0328 and 1106.D-0811.
This work makes use of observations from the Las Cumbres Observatory network. The Las Cumbres Observatory team is supported by NSF grants AST-2308113 and AST-1911151.
A part of this work is based on observations made with the Southern African Large Telescope (SALT) with the Large Science Programme on transients (2018-2-LSP-001; PI: DAHB). Polish participation in SALT is funded by grant No. MEiN 2021/WK/01 and DIR/WK/2016/07. DAHB acknowledges research support from the National Research Foundation.
The Legacy Surveys consist of three individual and complementary projects: the Dark Energy Camera Legacy Survey (DECaLS; Proposal ID \#2014B-0404; PIs: David Schlegel and Arjun Dey), the Beijing-Arizona Sky Survey (BASS; NOAO Prop. ID \#2015A-0801; PIs: Zhou Xu and Xiaohui Fan), and the Mayall z-band Legacy Survey (MzLS; Prop. ID \#2016A-0453; PI: Arjun Dey). DECaLS, BASS and MzLS together include data obtained, respectively, at the Blanco telescope, Cerro Tololo Inter-American Observatory, NSF's NOIRLab; the Bok telescope, Steward Observatory, University of Arizona; and the Mayall telescope, Kitt Peak National Observatory, NOIRLab. Pipeline processing and analyses of the data were supported by NOIRLab and the Lawrence Berkeley National Laboratory (LBNL). The Legacy Surveys project is honored to be permitted to conduct astronomical research on I'oligam Du'ag (Kitt Peak), a mountain with particular significance to the Tohono O'odham Nation.

NOIRLab is operated by the Association of Universities for Research in Astronomy (AURA) under a cooperative agreement with the National Science Foundation. LBNL is managed by the Regents of the University of California under contract to the U.S. Department of Energy.

This project used data obtained with the Dark Energy Camera (DECam), which was constructed by the Dark Energy Survey (DES) collaboration. Funding for the DES Projects has been provided by the U.S. Department of Energy, the U.S. National Science Foundation, the Ministry of Science and Education of Spain, the Science and Technology Facilities Council of the United Kingdom, the Higher Education Funding Council for England, the National Center for Supercomputing Applications at the University of Illinois at Urbana-Champaign, the Kavli Institute of Cosmological Physics at the University of Chicago, Center for Cosmology and Astro-Particle Physics at the Ohio State University, the Mitchell Institute for Fundamental Physics and Astronomy at Texas A\&M University, Financiadora de Estudos e Projetos, Fundacao Carlos Chagas Filho de Amparo, Financiadora de Estudos e Projetos, Fundacao Carlos Chagas Filho de Amparo a Pesquisa do Estado do Rio de Janeiro, Conselho Nacional de Desenvolvimento Cientifico e Tecnologico and the Ministerio da Ciencia, Tecnologia e Inovacao, the Deutsche Forschungsgemeinschaft and the Collaborating Institutions in the Dark Energy Survey. The Collaborating Institutions are Argonne National Laboratory, the University of California at Santa Cruz, the University of Cambridge, Centro de Investigaciones Energeticas, Medioambientales y Tecnologicas-Madrid, the University of Chicago, University College London, the DES-Brazil Consortium, the University of Edinburgh, the Eidgenossische Technische Hochschule (ETH) Zurich, Fermi National Accelerator Laboratory, the University of Illinois at Urbana-Champaign, the Institut de Ciencies de l'Espai (IEEC/CSIC), the Institut de Fisica d'Altes Energies, Lawrence Berkeley National Laboratory, the Ludwig Maximilians Universitat Munchen and the associated Excellence Cluster Universe, the University of Michigan, NSF's NOIRLab, the University of Nottingham, the Ohio State University, the University of Pennsylvania, the University of Portsmouth, SLAC National Accelerator Laboratory, Stanford University, the University of Sussex, and Texas A\&M University.

BASS is a key project of the Telescope Access Program (TAP), which has been funded by the National Astronomical Observatories of China, the Chinese Academy of Sciences (the Strategic Priority Research Program ``The Emergence of Cosmological Structures'' Grant \# XDB09000000), and the Special Fund for Astronomy from the Ministry of Finance. The BASS is also supported by the External Cooperation Program of Chinese Academy of Sciences (Grant \# 114A11KYSB20160057), and Chinese National Natural Science Foundation (Grant \# 12120101003, \# 11433005).

The Legacy Survey team makes use of data products from the Near-Earth Object Wide-field Infrared Survey Explorer (NEOWISE), which is a project of the Jet Propulsion Laboratory/California Institute of Technology. NEOWISE is funded by the National Aeronautics and Space Administration.

The Legacy Surveys imaging of the DESI footprint is supported by the Director, Office of Science, Office of High Energy Physics of the U.S. Department of Energy under Contract No. DE-AC02-05CH11231, by the National Energy Research Scientific Computing Center, a DOE Office of Science User Facility under the same contract; and by the U.S. National Science Foundation, Division of Astronomical Sciences under Contract No. AST-0950945 to NOAO.

T.E.M.B. is funded by Horizon Europe ERC grant no. 101125877. 

P.C. acknowledges financial support from the Secretary of Universities and Research (Government of Catalonia) and by the Horizon 2020 Research and Innovation Programme of the European Union under the Marie Sk\l{}odowska-Curie and the Beatriu de Pin\'os 2024 BP 00125 programme, and from the Centro Superior de Investigaciones Cient\'ificas (CSIC) under the Spanish program Unidad de Excelencia María de Maeztu CEX2020-001058-M, financed by MCIN/AEI/10.13039/501100011033, and by the MaX-CSIC Excellence Award MaX4-SOMMA-ICE. L.G. acknowledges financial support from MCIN and AEI 10.13039/501100011033 under projects PID2023-151307NB-I00, CEX2020-001058-M, and by the MaX-CSIC Excellence Award MaX4-SOMMA-ICE. C.P.G. acknowledges financial support from grant RYC2024-050959-I, funded by MICIU/AEI/10.13039/501100011033 and FSE+, and from grant PID2025-172154NA-I00 funded by MICIU/AEI/10.13039/501100011033, FSE+, ERDF, and the EU. Additional support was provided by CSIC through PIE 20215AT016, the María de Maeztu Unit of Excellence programme (CEX2020-001058-M), and the MaX-CSIC Excellence Award (MaX4-SOMMA-ICE). J.H.T. is funded by Horizon Europe ERC grant no. 101125877. T.-W.C. acknowledges financial support from the Yushan Fellow Program by the Ministry of Education, Taiwan (MOE-111-YSFMS-0008-001-P1), and the National Science and Technology Council, Taiwan (NSTC grant 114-2112-M-008-021-MY3). H.K. was funded by the Research Council of Finland projects 358691 and 358692.  

D.H. is supported by STScI grants HST-GO17770.002, JWST-GO-12468.001, and JWST-GO09964.001.

\end{acknowledgements}

\bibliographystyle{aa}
\bibliography{ref}

@ARTICLE{1990MNRAS.244..269S,
       author = {{Schlegel}, Eric M.},
        title = "{A new subclass of type II supernovae ?}",
      journal = {\mnras},
         year = 1990,
        month = may,
       volume = {244},
        pages = {269-271},
       adsurl = {https://ui.adsabs.harvard.edu/abs/1990MNRAS.244..269S}
}

@ARTICLE{1997ARA&A..35..309F,
       author = {{Filippenko}, Alexei V.},
        title = "{Optical Spectra of Supernovae}",
      journal = {\araa},
         year = 1997,
        month = jan,
       volume = {35},
        pages = {309-355},
          doi = {10.1146/annurev.astro.35.1.309},
       adsurl = {https://ui.adsabs.harvard.edu/abs/1997ARA&A..35..309F}
}

@ARTICLE{1994MNRAS.268..173C,
       author = {{Chugai}, N.~N. and {Danziger}, I.~J.},
        title = "{SN 1988Z: low-mass ejecta colliding with the clumpy wind?}",
      journal = {\mnras},
         year = 1994,
        month = may,
       volume = {268},
        pages = {173-180},
          doi = {10.1093/mnras/268.1.173},
       adsurl = {https://ui.adsabs.harvard.edu/abs/1994MNRAS.268..173C}
}

@INCOLLECTION{2017hsn..book..403S,
       author = {{Smith}, Nathan},
        title = "{Interacting Supernovae: Types IIn and Ibn}",
    booktitle = {Handbook of Supernovae},
         year = 2017,
       editor = {{Alsabti}, Athem W. and {Murdin}, Paul},
       publisher="Springer International Publishing",
        pages = {403},
          doi = {10.1007/978-3-319-21846-5_38},
       adsurl = {https://ui.adsabs.harvard.edu/abs/2017hsn..book..403S}
}

@ARTICLE{1982ApJ...258..790C,
       author = {{Chevalier}, R.~A.},
        title = "{Self-similar solutions for the interaction of stellar ejecta with an external medium.}",
      journal = {\apj},
         year = 1982,
        month = jul,
       volume = {258},
        pages = {790-797},
          doi = {10.1086/160126},
       adsurl = {https://ui.adsabs.harvard.edu/abs/1982ApJ...258..790C}
}

@INCOLLECTION{2017hsn..book..875C,
       author = {{Chevalier}, Roger A. and {Fransson}, Claes},
        title = "{Thermal and Non-thermal Emission from Circumstellar Interaction}",
    booktitle = {Handbook of Supernovae},
         year = 2017,
       editor = {{Alsabti}, Athem W. and {Murdin}, Paul},
        pages = {875},
        publisher="Springer International Publishing",
          doi = {10.1007/978-3-319-21846-5_34},
       adsurl = {https://ui.adsabs.harvard.edu/abs/2017hsn..book..875C}
}

@ARTICLE{2020A&A...637A..73N,
       author = {{Nyholm}, A. and {Sollerman}, J. and {Tartaglia}, L. and {Taddia}, F. and {Fremling}, C. and {Blagorodnova}, N. and {Filippenko}, A.~V. and {Gal-Yam}, A. and {Howell}, D.~A. and {Karamehmetoglu}, E. and {Kulkarni}, S.~R. and {Laher}, R. and {Leloudas}, G. and {Masci}, F. and {Kasliwal}, M.~M. and {Mor{\r{a}}}, K. and {Moriya}, T.~J. and {Ofek}, E.~O. and {Papadogiannakis}, S. and {Quimby}, R. and {Rebbapragada}, U. and {Schulze}, S.},
        title = "{Type IIn supernova light-curve properties measured from an untargeted survey sample}",
      journal = {\aap},
         year = 2020,
        month = may,
       volume = {637},
          eid = {A73},
        pages = {A73},
          doi = {10.1051/0004-6361/201936097},
archivePrefix = {arXiv},
       eprint = {1906.05812},
 primaryClass = {astro-ph.SR},
       adsurl = {https://ui.adsabs.harvard.edu/abs/2020A&A...637A..73N}
}

@ARTICLE{2014Natur.509..471G,
       author = {{Gal-Yam}, Avishay and {Arcavi}, I. and {Ofek}, E.~O. and {Ben-Ami}, S. and {Cenko}, S.~B. and {Kasliwal}, M.~M. and {Cao}, Y. and {Yaron}, O. and {Tal}, D. and {Silverman}, J.~M. and {Horesh}, A. and {De Cia}, A. and {Taddia}, F. and {Sollerman}, J. and {Perley}, D. and {Vreeswijk}, P.~M. and {Kulkarni}, S.~R. and {Nugent}, P.~E. and {Filippenko}, A.~V. and {Wheeler}, J.~C.},
        title = "{A Wolf-Rayet-like progenitor of SN 2013cu from spectral observations of a stellar wind}",
      journal = {\nat},
         year = 2014,
        month = may,
       volume = {509},
       number = {7501},
        pages = {471-474},
          doi = {10.1038/nature13304},
archivePrefix = {arXiv},
       eprint = {1406.7640},
 primaryClass = {astro-ph.HE},
       adsurl = {https://ui.adsabs.harvard.edu/abs/2014Natur.509..471G}
}

@ARTICLE{2016ApJ...818....3K,
       author = {{Khazov}, D. and {Yaron}, O. and {Gal-Yam}, A. and {Manulis}, I. and {Rubin}, A. and {Kulkarni}, S.~R. and {Arcavi}, I. and {Kasliwal}, M.~M. and {Ofek}, E.~O. and {Cao}, Y. and {Perley}, D. and {Sollerman}, J. and {Horesh}, A. and {Sullivan}, M. and {Filippenko}, A.~V. and {Nugent}, P.~E. and {Howell}, D.~A. and {Cenko}, S.~B. and {Silverman}, J.~M. and {Ebeling}, H. and {Taddia}, F. and {Johansson}, J. and {Laher}, R.~R. and {Surace}, J. and {Rebbapragada}, U.~D. and {Wozniak}, P.~R. and {Matheson}, T.},
        title = "{Flash Spectroscopy: Emission Lines from the Ionized Circumstellar Material around <10-day-old Type II Supernovae}",
      journal = {\apj},
         year = 2016,
        month = feb,
       volume = {818},
       number = {1},
          eid = {3},
        pages = {3},
          doi = {10.3847/0004-637X/818/1/3},
archivePrefix = {arXiv},
       eprint = {1512.00846},
 primaryClass = {astro-ph.HE},
       adsurl = {https://ui.adsabs.harvard.edu/abs/2016ApJ...818....3K}
}

@ARTICLE{2014ARA&A..52..487S,
       author = {{Smith}, Nathan},
        title = "{Mass Loss: Its Effect on the Evolution and Fate of High-Mass Stars}",
      journal = {\araa},
         year = 2014,
        month = aug,
       volume = {52},
        pages = {487-528},
          doi = {10.1146/annurev-astro-081913-040025},
archivePrefix = {arXiv},
       eprint = {1402.1237},
 primaryClass = {astro-ph.SR},
       adsurl = {https://ui.adsabs.harvard.edu/abs/2014ARA&A..52..487S}
}

@ARTICLE{2014ApJ...789..104O,
       author = {{Ofek}, Eran O. and {Sullivan}, Mark and {Shaviv}, Nir J. and {Steinbok}, Aviram and {Arcavi}, Iair and {Gal-Yam}, Avishay and {Tal}, David and {Kulkarni}, Shrinivas R. and {Nugent}, Peter E. and {Ben-Ami}, Sagi and {Kasliwal}, Mansi M. and {Cenko}, S. Bradley and {Laher}, Russ and {Surace}, Jason and {Bloom}, Joshua S. and {Filippenko}, Alexei V. and {Silverman}, Jeffrey M. and {Yaron}, Ofer},
        title = "{Precursors Prior to Type IIn Supernova Explosions are Common: Precursor Rates, Properties, and Correlations}",
      journal = {\apj},
         year = 2014,
        month = jul,
       volume = {789},
       number = {2},
          eid = {104},
        pages = {104},
          doi = {10.1088/0004-637X/789/2/104},
archivePrefix = {arXiv},
       eprint = {1401.5468},
 primaryClass = {astro-ph.HE},
       adsurl = {https://ui.adsabs.harvard.edu/abs/2014ApJ...789..104O}
}

@ARTICLE{2011ApJ...732...32F,
       author = {{Foley}, Ryan J. and {Berger}, Edo and {Fox}, Ori and {Levesque}, Emily M. and {Challis}, Peter J. and {Ivans}, Inese I. and {Rhoads}, James E. and {Soderberg}, Alicia M.},
        title = "{The Diversity of Massive Star Outbursts. I. Observations of SN2009ip, UGC 2773 OT2009-1, and Their Progenitors}",
      journal = {\apj},
         year = 2011,
        month = may,
       volume = {732},
       number = {1},
          eid = {32},
        pages = {32},
          doi = {10.1088/0004-637X/732/1/32},
archivePrefix = {arXiv},
       eprint = {1002.0635},
 primaryClass = {astro-ph.SR},
       adsurl = {https://ui.adsabs.harvard.edu/abs/2011ApJ...732...32F}
}

@ARTICLE{2013MNRAS.430.1801M,
       author = {{Mauerhan}, Jon C. and {Smith}, Nathan and {Filippenko}, Alexei V. and {Blanchard}, Kyle B. and {Blanchard}, Peter K. and {Casper}, Chadwick F.~E. and {Cenko}, S. Bradley and {Clubb}, Kelsey I. and {Cohen}, Daniel P. and {Fuller}, Kiera L. and {Li}, Gary Z. and {Silverman}, Jeffrey M.},
        title = "{The unprecedented 2012 outburst of SN 2009ip: a luminous blue variable star becomes a true supernova}",
      journal = {\mnras},
         year = 2013,
        month = apr,
       volume = {430},
       number = {3},
        pages = {1801-1810},
          doi = {10.1093/mnras/stt009},
archivePrefix = {arXiv},
       eprint = {1209.6320},
 primaryClass = {astro-ph.SR},
       adsurl = {https://ui.adsabs.harvard.edu/abs/2013MNRAS.430.1801M}
}

@ARTICLE{2009ApJ...695.1334S,
       author = {{Smith}, Nathan and {Silverman}, Jeffrey M. and {Chornock}, Ryan and {Filippenko}, Alexei V. and {Wang}, Xiaofeng and {Li}, Weidong and {Ganeshalingam}, Mohan and {Foley}, Ryan J. and {Rex}, Jacob and {Steele}, Thea N.},
        title = "{Coronal Lines and Dust Formation in SN 2005ip: Not the Brightest, but the Hottest Type IIn Supernova}",
      journal = {\apj},
         year = 2009,
        month = apr,
       volume = {695},
       number = {2},
        pages = {1334-1350},
          doi = {10.1088/0004-637X/695/2/1334},
archivePrefix = {arXiv},
       eprint = {0809.5079},
 primaryClass = {astro-ph},
       adsurl = {https://ui.adsabs.harvard.edu/abs/2009ApJ...695.1334S}
}

@ARTICLE{2012ApJ...756..173S,
       author = {{Stritzinger}, Maximilian and {Taddia}, Francesco and {Fransson}, Claes and {Fox}, Ori D. and {Morrell}, Nidia and {Phillips}, M.~M. and {Sollerman}, Jesper and {Anderson}, J.~P. and {Boldt}, Luis and {Brown}, Peter J. and {Campillay}, Abdo and {Castellon}, Sergio and {Contreras}, Carlos and {Folatelli}, Gast{\'o}n and {Habergham}, S.~M. and {Hamuy}, Mario and {Hjorth}, Jens and {James}, Phil A. and {Krzeminski}, Wojtek and {Mattila}, Seppo and {Persson}, Sven E. and {Roth}, Miguel},
        title = "{Multi-wavelength Observations of the Enduring Type IIn Supernovae 2005ip and 2006jd}",
      journal = {\apj},
         year = 2012,
        month = sep,
       volume = {756},
       number = {2},
          eid = {173},
        pages = {173},
          doi = {10.1088/0004-637X/756/2/173},
archivePrefix = {arXiv},
       eprint = {1206.5575},
 primaryClass = {astro-ph.CO},
       adsurl = {https://ui.adsabs.harvard.edu/abs/2012ApJ...756..173S}
}

@ARTICLE{2007ApJ...666.1116S,
       author = {{Smith}, Nathan and {Li}, Weidong and {Foley}, Ryan J. and {Wheeler}, J. Craig and {Pooley}, David and {Chornock}, Ryan and {Filippenko}, Alexei V. and {Silverman}, Jeffrey M. and {Quimby}, Robert and {Bloom}, Joshua S. and {Hansen}, Charles},
        title = "{SN 2006gy: Discovery of the Most Luminous Supernova Ever Recorded, Powered by the Death of an Extremely Massive Star like {\ensuremath{\eta}} Carinae}",
      journal = {\apj},
         year = 2007,
        month = sep,
       volume = {666},
       number = {2},
        pages = {1116-1128},
          doi = {10.1086/519949},
archivePrefix = {arXiv},
       eprint = {astro-ph/0612617},
 primaryClass = {astro-ph},
       adsurl = {https://ui.adsabs.harvard.edu/abs/2007ApJ...666.1116S}
}

@ARTICLE{2012Sci...337..927G,
       author = {{Gal-Yam}, Avishay},
        title = "{Luminous Supernovae}",
      journal = {Science},
         year = 2012,
        month = aug,
       volume = {337},
       number = {6097},
        pages = {927},
          doi = {10.1126/science.1203601},
archivePrefix = {arXiv},
       eprint = {1208.3217},
 primaryClass = {astro-ph.CO},
       adsurl = {https://ui.adsabs.harvard.edu/abs/2012Sci...337..927G}
}

@ARTICLE{2017A&A...605A...6N,
       author = {{Nyholm}, A. and {Sollerman}, J. and {Taddia}, F. and {Fremling}, C. and {Moriya}, T.~J. and {Ofek}, E.~O. and {Gal-Yam}, A. and {De Cia}, A. and {Roy}, R. and {Kasliwal}, M.~M. and {Cao}, Y. and {Nugent}, P.~E. and {Masci}, F.~J.},
        title = "{The bumpy light curve of Type IIn supernova iPTF13z over 3 years}",
      journal = {\aap},
         year = 2017,
        month = aug,
       volume = {605},
          eid = {A6},
        pages = {A6},
          doi = {10.1051/0004-6361/201629906},
archivePrefix = {arXiv},
       eprint = {1703.09679},
 primaryClass = {astro-ph.SR},
       adsurl = {https://ui.adsabs.harvard.edu/abs/2017A&A...605A...6N}
}

@ARTICLE{2020A&A...638A..92T,
       author = {{Taddia}, F. and {Stritzinger}, M.~D. and {Fransson}, C. and {Brown}, P.~J. and {Contreras}, C. and {Holmbo}, S. and {Moriya}, T.~J. and {Phillips}, M.~M. and {Sollerman}, J. and {Suntzeff}, N.~B. and {Ashall}, C. and {Burns}, C.~R. and {Busta}, L. and {Campillay}, A. and {Castell{\'o}n}, S. and {Corco}, C. and {Di Mille}, F. and {Gall}, C. and {Gonz{\'a}lez}, C. and {Hsiao}, E.~Y. and {Morrell}, N. and {Nyholm}, A. and {Simon}, J.~D. and {Ser{\'o}n}, J.},
        title = "{The Carnegie Supernova Project II. The shock wave revealed through the fog: The strongly interacting Type IIn SN 2013L}",
      journal = {\aap},
         year = 2020,
        month = jun,
       volume = {638},
          eid = {A92},
        pages = {A92},
          doi = {10.1051/0004-6361/201936654},
archivePrefix = {arXiv},
       eprint = {2003.09709},
 primaryClass = {astro-ph.SR},
       adsurl = {https://ui.adsabs.harvard.edu/abs/2020A&A...638A..92T}
}

@ARTICLE{2026ApJ...999..176H,
       author = {{Hu}, Yang and {Lunnan}, Ragnhild and {Pessi}, Priscila J. and {Saldana-Lopez}, Alberto and {Jerkstrand}, Anders and {Sollerman}, Jesper and {Schulze}, Steve and {Anderson}, Joseph P. and {Brennan}, Se{\'a}n J. and {Cosentino}, Stefano P. and {Gangopadhyay}, Anjasha and {Gkini}, Anamaria and {Gromadzki}, Mariusz and {Hayes}, Matthew J. and {Inserra}, Cosimo and {M{\"u}ller-Bravo}, Tom{\'a}s E. and {Nicholl}, Matt and {Pignata}, Giuliano and {Singh}, Avinash and {Wise}, Jacob L. and {Yan}, Lin and {Adler}, Judy and {Chen}, Ting-Wan and {Chen}, Tracy X. and {Kasliwal}, Mansi M. and {Pessi}, Thallis and {Salmaso}, Irene and {Young}, David R.},
        title = "{SN 2021aaev: A Hydrogen-rich Superluminous Supernova with Early Flash and Long-lived Circumstellar Interaction in an Unusual Host Environment}",
      journal = {\apj},
         year = 2026,
        month = mar,
       volume = {999},
       number = {2},
          eid = {176},
        pages = {176},
          doi = {10.3847/1538-4357/ae374c},
archivePrefix = {arXiv},
       eprint = {2508.11559},
 primaryClass = {astro-ph.HE},
       adsurl = {https://ui.adsabs.harvard.edu/abs/2026ApJ...999..176H}
}

@ARTICLE{2012AJ....143...17S,
       author = {{Smith}, Nathan and {Silverman}, Jeffrey M. and {Filippenko}, Alexei V. and {Cooper}, Michael C. and {Matheson}, Thomas and {Bian}, Fuyan and {Weiner}, Benjamin J. and {Comerford}, Julia M.},
        title = "{Systematic Blueshift of Line Profiles in the Type IIn Supernova 2010jl: Evidence for Post-shock Dust Formation?}",
      journal = {\aj},
         year = 2012,
        month = jan,
       volume = {143},
       number = {1},
          eid = {17},
        pages = {17},
          doi = {10.1088/0004-6256/143/1/17},
archivePrefix = {arXiv},
       eprint = {1108.2869},
 primaryClass = {astro-ph.HE},
       adsurl = {https://ui.adsabs.harvard.edu/abs/2012AJ....143...17S}
}

@ARTICLE{2011A&ARv..19...43G,
       author = {{Gall}, C. and {Hjorth}, J. and {Andersen}, A.~C.},
        title = "{Production of dust by massive stars at high redshift}",
      journal = {\aapr},
         year = 2011,
        month = sep,
       volume = {19},
          eid = {43},
        pages = {43},
          doi = {10.1007/s00159-011-0043-7},
archivePrefix = {arXiv},
       eprint = {1108.0403},
 primaryClass = {astro-ph.CO},
       adsurl = {https://ui.adsabs.harvard.edu/abs/2011A&ARv..19...43G}
}

@ARTICLE{2018A&A...611A..67B,
       author = {{Bak Nielsen}, Ann-Sofie and {Hjorth}, Jens and {Gall}, Christa},
        title = "{Early gray dust formation in the type IIn SN 2005ip}",
      journal = {\aap},
         year = 2018,
        month = mar,
       volume = {611},
          eid = {A67},
        pages = {A67},
          doi = {10.1051/0004-6361/201629904},
archivePrefix = {arXiv},
       eprint = {1801.04217},
 primaryClass = {astro-ph.SR},
       adsurl = {https://ui.adsabs.harvard.edu/abs/2018A&A...611A..67B}
}

@ARTICLE{2022MNRAS.511.2021L,
       author = {{Li}, Jun and {Gao}, Jian and {Jiang}, Biwei and {Lin}, Zesen},
        title = "{Dust models for the extinction of Type IIn supernova SN 2010jl}",
      journal = {\mnras},
         year = 2022,
        month = apr,
       volume = {511},
       number = {2},
        pages = {2021-2032},
          doi = {10.1093/mnras/stac220},
archivePrefix = {arXiv},
       eprint = {2201.09298},
 primaryClass = {astro-ph.SR},
       adsurl = {https://ui.adsabs.harvard.edu/abs/2022MNRAS.511.2021L}
}

@ARTICLE{2013AJ....146....2F,
       author = {{Fox}, Ori D. and {Filippenko}, Alexei V. and {Skrutskie}, Michael F. and {Silverman}, Jeffrey M. and {Ganeshalingam}, Mohan and {Cenko}, S. Bradley and {Clubb}, Kelsey I.},
        title = "{Late-time Circumstellar Interaction in a Spitzer Selected Sample of Type IIn Supernovae}",
      journal = {\aj},
         year = 2013,
        month = jul,
       volume = {146},
       number = {1},
          eid = {2},
        pages = {2},
          doi = {10.1088/0004-6256/146/1/2},
archivePrefix = {arXiv},
       eprint = {1304.0248},
 primaryClass = {astro-ph.SR},
       adsurl = {https://ui.adsabs.harvard.edu/abs/2013AJ....146....2F}
}

@ARTICLE{2014AcA....64..197W,
       author = {{Wyrzykowski}, {\L}. and {Kostrzewa-Rutkowska}, Z. and {Koz{\l}owski}, S. and {Udalski}, A. and {Poleski}, R. and {Skowron}, J. and {Blagorodnova}, N. and {Kubiak}, M. and {Szyma{\'n}ski}, M.~K. and {Pietrzy{\'n}ski}, G. and {Soszy{\'n}ski}, I. and {Ulaczyk}, K. and {Pietrukowicz}, P. and {Mr{\'o}z}, P.},
        title = "{OGLE-IV Real-Time Transient Search}",
      journal = {\actaa},
         year = 2014,
        month = sep,
       volume = {64},
       number = {3},
        pages = {197-232},
          doi = {10.48550/arXiv.1409.1095},
archivePrefix = {arXiv},
       eprint = {1409.1095},
 primaryClass = {astro-ph.GA},
       adsurl = {https://ui.adsabs.harvard.edu/abs/2014AcA....64..197W}
}

@ARTICLE{2019TNSTR1836....1G,
       author = {{Gromadzki}, M. and {Wyrzykowski}, {\L}.},
        title = "{OGLE Transient Discovery Report for 2019-09-16}",
      journal = {Transient Name Server Discovery Report},
         year = 2019,
        month = sep,
       volume = {2019-1836},
        pages = {1},
       adsurl = {https://ui.adsabs.harvard.edu/abs/2019TNSTR1836....1G}
}

@ARTICLE{2019TNSCR2193....1G,
       author = {{Gromadzki}, M.},
        title = "{Transient Classification Report for 2019-10-26}",
      journal = {Transient Name Server Classification Report},
         year = 2019,
        month = oct,
       volume = {2019-2193},
        pages = {1},
       adsurl = {https://ui.adsabs.harvard.edu/abs/2019TNSCR2193....1G}
}

@ARTICLE{2020A&A...641A...6P,
       author = {{Planck Collaboration} and {Aghanim}, N. and {Akrami}, Y. and {Ashdown}, M. and {Aumont}, J. and {Baccigalupi}, C. and {Ballardini}, M. and {Banday}, A.~J. and {Barreiro}, R.~B. and {Bartolo}, N. and {Basak}, S. and {Battye}, R. and {Benabed}, K. and {Bernard}, J.-P. and {Bersanelli}, M. and {Bielewicz}, P. and {Bock}, J.~J. and {Bond}, J.~R. and {Borrill}, J. and {Bouchet}, F.~R. and {Boulanger}, F. and {Bucher}, M. and {Burigana}, C. and {Butler}, R.~C. and {Calabrese}, E. and {Cardoso}, J.-F. and {Carron}, J. and {Challinor}, A. and {Chiang}, H.~C. and {Chluba}, J. and {Colombo}, L.~P.~L. and {Combet}, C. and {Contreras}, D. and {Crill}, B.~P. and {Cuttaia}, F. and {de Bernardis}, P. and {de Zotti}, G. and {Delabrouille}, J. and {Delouis}, J.-M. and {Di Valentino}, E. and {Diego}, J.~M. and {Dor{\'e}}, O. and {Douspis}, M. and {Ducout}, A. and {Dupac}, X. and {Dusini}, S. and {Efstathiou}, G. and {Elsner}, F. and {En{\ss}lin}, T.~A. and {Eriksen}, H.~K. and {Fantaye}, Y. and {Farhang}, M. and {Fergusson}, J. and {Fernandez-Cobos}, R. and {Finelli}, F. and {Forastieri}, F. and {Frailis}, M. and {Fraisse}, A.~A. and {Franceschi}, E. and {Frolov}, A. and {Galeotta}, S. and {Galli}, S. and {Ganga}, K. and {G{\'e}nova-Santos}, R.~T. and {Gerbino}, M. and {Ghosh}, T. and {Gonz{\'a}lez-Nuevo}, J. and {G{\'o}rski}, K.~M. and {Gratton}, S. and {Gruppuso}, A. and {Gudmundsson}, J.~E. and {Hamann}, J. and {Handley}, W. and {Hansen}, F.~K. and {Herranz}, D. and {Hildebrandt}, S.~R. and {Hivon}, E. and {Huang}, Z. and {Jaffe}, A.~H. and {Jones}, W.~C. and {Karakci}, A. and {Keih{\"a}nen}, E. and {Keskitalo}, R. and {Kiiveri}, K. and {Kim}, J. and {Kisner}, T.~S. and {Knox}, L. and {Krachmalnicoff}, N. and {Kunz}, M. and {Kurki-Suonio}, H. and {Lagache}, G. and {Lamarre}, J.-M. and {Lasenby}, A. and {Lattanzi}, M. and {Lawrence}, C.~R. and {Le Jeune}, M. and {Lemos}, P. and {Lesgourgues}, J. and {Levrier}, F. and {Lewis}, A. and {Liguori}, M. and {Lilje}, P.~B. and {Lilley}, M. and {Lindholm}, V. and {L{\'o}pez-Caniego}, M. and {Lubin}, P.~M. and {Ma}, Y.-Z. and {Mac{\'\i}as-P{\'e}rez}, J.~F. and {Maggio}, G. and {Maino}, D. and {Mandolesi}, N. and {Mangilli}, A. and {Marcos-Caballero}, A. and {Maris}, M. and {Martin}, P.~G. and {Martinelli}, M. and {Mart{\'\i}nez-Gonz{\'a}lez}, E. and {Matarrese}, S. and {Mauri}, N. and {McEwen}, J.~D. and {Meinhold}, P.~R. and {Melchiorri}, A. and {Mennella}, A. and {Migliaccio}, M. and {Millea}, M. and {Mitra}, S. and {Miville-Desch{\^e}nes}, M.-A. and {Molinari}, D. and {Montier}, L. and {Morgante}, G. and {Moss}, A. and {Natoli}, P. and {N{\o}rgaard-Nielsen}, H.~U. and {Pagano}, L. and {Paoletti}, D. and {Partridge}, B. and {Patanchon}, G. and {Peiris}, H.~V. and {Perrotta}, F. and {Pettorino}, V. and {Piacentini}, F. and {Polastri}, L. and {Polenta}, G. and {Puget}, J.-L. and {Rachen}, J.~P. and {Reinecke}, M. and {Remazeilles}, M. and {Renzi}, A. and {Rocha}, G. and {Rosset}, C. and {Roudier}, G. and {Rubi{\~n}o-Mart{\'\i}n}, J.~A. and {Ruiz-Granados}, B. and {Salvati}, L. and {Sandri}, M. and {Savelainen}, M. and {Scott}, D. and {Shellard}, E.~P.~S. and {Sirignano}, C. and {Sirri}, G. and {Spencer}, L.~D. and {Sunyaev}, R. and {Suur-Uski}, A.-S. and {Tauber}, J.~A. and {Tavagnacco}, D. and {Tenti}, M. and {Toffolatti}, L. and {Tomasi}, M. and {Trombetti}, T. and {Valenziano}, L. and {Valiviita}, J. and {Van Tent}, B. and {Vibert}, L. and {Vielva}, P. and {Villa}, F. and {Vittorio}, N. and {Wandelt}, B.~D. and {Wehus}, I.~K. and {White}, M. and {White}, S.~D.~M. and {Zacchei}, A. and {Zonca}, A.},
        title = "{Planck 2018 results. VI. Cosmological parameters}",
      journal = {\aap},
         year = 2020,
        month = sep,
       volume = {641},
          eid = {A6},
        pages = {A6},
          doi = {10.1051/0004-6361/201833910},
archivePrefix = {arXiv},
       eprint = {1807.06209},
 primaryClass = {astro-ph.CO},
       adsurl = {https://ui.adsabs.harvard.edu/abs/2020A&A...641A...6P}
}

@INPROCEEDINGS{2024lsst.confE...4W,
       author = {{Wyrzykowski}, Lukasz},
        title = "{Power of many - BHTOM telescope network for time-domain astronomy}",
    booktitle = {What Was That? - Planning ESO Follow up for Transients, Variables, and Solar System Objects in the Era of LSST},
         year = 2024,
        month = jan,
          eid = {4},
        pages = {4},
          doi = {10.5281/zenodo.10571539},
       adsurl = {https://ui.adsabs.harvard.edu/abs/2024lsst.confE...4W}
}

@ARTICLE{2013PASP..125.1031B,
       author = {{Brown}, T.~M. and {Baliber}, N. and {Bianco}, F.~B. and {Bowman}, M. and {Burleson}, B. and {Conway}, P. and {Crellin}, M. and {Depagne}, {\'E}. and {De Vera}, J. and {Dilday}, B. and {Dragomir}, D. and {Dubberley}, M. and {Eastman}, J.~D. and {Elphick}, M. and {Falarski}, M. and {Foale}, S. and {Ford}, M. and {Fulton}, B.~J. and {Garza}, J. and {Gomez}, E.~L. and {Graham}, M. and {Greene}, R. and {Haldeman}, B. and {Hawkins}, E. and {Haworth}, B. and {Haynes}, R. and {Hidas}, M. and {Hjelstrom}, A.~E. and {Howell}, D.~A. and {Hygelund}, J. and {Lister}, T.~A. and {Lobdill}, R. and {Martinez}, J. and {Mullins}, D.~S. and {Norbury}, M. and {Parrent}, J. and {Paulson}, R. and {Petry}, D.~L. and {Pickles}, A. and {Posner}, V. and {Rosing}, W.~E. and {Ross}, R. and {Sand}, D.~J. and {Saunders}, E.~S. and {Shobbrook}, J. and {Shporer}, A. and {Street}, R.~A. and {Thomas}, D. and {Tsapras}, Y. and {Tufts}, J.~R. and {Valenti}, S. and {Vander Horst}, K. and {Walker}, Z. and {White}, G. and {Willis}, M.},
        title = "{Las Cumbres Observatory Global Telescope Network}",
      journal = {\pasp},
         year = 2013,
        month = sep,
       volume = {125},
       number = {931},
        pages = {1031},
          doi = {10.1086/673168},
archivePrefix = {arXiv},
       eprint = {1305.2437},
 primaryClass = {astro-ph.IM},
       adsurl = {https://ui.adsabs.harvard.edu/abs/2013PASP..125.1031B}
}

@ARTICLE{2015A&A...579A..40S,
       author = {{Smartt}, S.~J. and {Valenti}, S. and {Fraser}, M. and {Inserra}, C. and {Young}, D.~R. and {Sullivan}, M. and {Pastorello}, A. and {Benetti}, S. and {Gal-Yam}, A. and {Knapic}, C. and {Molinaro}, M. and {Smareglia}, R. and {Smith}, K.~W. and {Taubenberger}, S. and {Yaron}, O. and {Anderson}, J.~P. and {Ashall}, C. and {Balland}, C. and {Baltay}, C. and {Barbarino}, C. and {Bauer}, F.~E. and {Baumont}, S. and {Bersier}, D. and {Blagorodnova}, N. and {Bongard}, S. and {Botticella}, M.~T. and {Bufano}, F. and {Bulla}, M. and {Cappellaro}, E. and {Campbell}, H. and {Cellier-Holzem}, F. and {Chen}, T.-W. and {Childress}, M.~J. and {Clocchiatti}, A. and {Contreras}, C. and {Dall'Ora}, M. and {Danziger}, J. and {de Jaeger}, T. and {De Cia}, A. and {Della Valle}, M. and {Dennefeld}, M. and {Elias-Rosa}, N. and {Elman}, N. and {Feindt}, U. and {Fleury}, M. and {Gall}, E. and {Gonzalez-Gaitan}, S. and {Galbany}, L. and {Morales Garoffolo}, A. and {Greggio}, L. and {Guillou}, L.~L. and {Hachinger}, S. and {Hadjiyska}, E. and {Hage}, P.~E. and {Hillebrandt}, W. and {Hodgkin}, S. and {Hsiao}, E.~Y. and {James}, P.~A. and {Jerkstrand}, A. and {Kangas}, T. and {Kankare}, E. and {Kotak}, R. and {Kromer}, M. and {Kuncarayakti}, H. and {Leloudas}, G. and {Lundqvist}, P. and {Lyman}, J.~D. and {Hook}, I.~M. and {Maguire}, K. and {Manulis}, I. and {Margheim}, S.~J. and {Mattila}, S. and {Maund}, J.~R. and {Mazzali}, P.~A. and {McCrum}, M. and {McKinnon}, R. and {Moreno-Raya}, M.~E. and {Nicholl}, M. and {Nugent}, P. and {Pain}, R. and {Pignata}, G. and {Phillips}, M.~M. and {Polshaw}, J. and {Pumo}, M.~L. and {Rabinowitz}, D. and {Reilly}, E. and {Romero-Ca{\~n}izales}, C. and {Scalzo}, R. and {Schmidt}, B. and {Schulze}, S. and {Sim}, S. and {Sollerman}, J. and {Taddia}, F. and {Tartaglia}, L. and {Terreran}, G. and {Tomasella}, L. and {Turatto}, M. and {Walker}, E. and {Walton}, N.~A. and {Wyrzykowski}, L. and {Yuan}, F. and {Zampieri}, L.},
        title = "{PESSTO: survey description and products from the first data release by the Public ESO Spectroscopic Survey of Transient Objects}",
      journal = {\aap},
         year = 2015,
        month = jul,
       volume = {579},
          eid = {A40},
        pages = {A40},
          doi = {10.1051/0004-6361/201425237},
archivePrefix = {arXiv},
       eprint = {1411.0299},
 primaryClass = {astro-ph.SR},
       adsurl = {https://ui.adsabs.harvard.edu/abs/2015A&A...579A..40S}
}

@ARTICLE{2019CoSka..49..125Z,
       author = {{Zieli{\'n}ski}, P. and {Wyrzykowski}, {\L}. and {Rybicki}, K. and {Ko{\l}aczkowski}, Z. and {Bru{\'s}}, P. and {Miko{\l}ajczyk}, P.},
        title = "{CPCS 2.0 {\textemdash} new automatic tool for time-domain astronomy}",
      journal = {Contributions of the Astronomical Observatory Skalnate Pleso},
         year = 2019,
        month = may,
       volume = {49},
       number = {2},
        pages = {125-131},
       adsurl = {https://ui.adsabs.harvard.edu/abs/2019CoSka..49..125Z}
}

@INPROCEEDINGS{2020past.conf..190Z,
       author = {{Zieli{\'n}ski}, Pawe{\l} and {Wyrzykowski}, {\l}ukasz and {Miko{\l}ajczyk}, Przemys{\l}aw and {Rybicki}, Krzysztof and {Ko{\l}aczkowski}, Zbigniew},
        title = "{Towards an automatic processing of CCD images with CPCS 2.0}",
    booktitle = {XXXIX Polish Astronomical Society Meeting},
         year = 2020,
       editor = {{Ma{\l}ek}, Katarzyna and {Poli{\'n}ska}, Magdalena and {Majczyna}, Agnieszka and {Stachowski}, Grzegorz and {Poleski}, Rados{\l}aw and {Wyrzykowski}, {\L}ukasz and {R{\'o}{\.z}a{\'n}ska}, Agata},
       volume = {10},
        month = oct,
        pages = {190-193},
          doi = {10.48550/arXiv.2006.05160},
archivePrefix = {arXiv},
       eprint = {2006.05160},
 primaryClass = {astro-ph.IM},
       adsurl = {https://ui.adsabs.harvard.edu/abs/2020past.conf..190Z}
}

@ARTICLE{2023A&A...674A..33G,
       author = {{Gaia Collaboration} and {Montegriffo}, P. and {Bellazzini}, M. and {De Angeli}, F. and {Andrae}, R. and {Barstow}, M.~A. and {Bossini}, D. and {Bragaglia}, A. and {Burgess}, P.~W. and {Cacciari}, C. and {Carrasco}, J.~M. and {Chornay}, N. and {Delchambre}, L. and {Evans}, D.~W. and {Fouesneau}, M. and {Fr{\'e}mat}, Y. and {Garabato}, D. and {Jordi}, C. and {Manteiga}, M. and {Massari}, D. and {Palaversa}, L. and {Pancino}, E. and {Riello}, M. and {Ruz Mieres}, D. and {Sanna}, N. and {Santove{\~n}a}, R. and {Sordo}, R. and {Vallenari}, A. and {Walton}, N.~A. and {Brown}, A.~G.~A. and {Prusti}, T. and {de Bruijne}, J.~H.~J. and {Arenou}, F. and {Babusiaux}, C. and {Biermann}, M. and {Creevey}, O.~L. and {Ducourant}, C. and {Eyer}, L. and {Guerra}, R. and {Hutton}, A. and {Klioner}, S.~A. and {Lammers}, U.~L. and {Lindegren}, L. and {Luri}, X. and {Mignard}, F. and {Panem}, C. and {Pourbaix}, D. and {Randich}, S. and {Sartoretti}, P. and {Soubiran}, C. and {Tanga}, P. and {Bailer-Jones}, C.~A.~L. and {Bastian}, U. and {Drimmel}, R. and {Jansen}, F. and {Katz}, D. and {Lattanzi}, M.~G. and {van Leeuwen}, F. and {Bakker}, J. and {Casta{\~n}eda}, J. and {Fabricius}, C. and {Galluccio}, L. and {Guerrier}, A. and {Heiter}, U. and {Masana}, E. and {Messineo}, R. and {Mowlavi}, N. and {Nicolas}, C. and {Nienartowicz}, K. and {Pailler}, F. and {Panuzzo}, P. and {Riclet}, F. and {Roux}, W. and {Seabroke}, G.~M. and {Th{\'e}venin}, F. and {Gracia-Abril}, G. and {Portell}, J. and {Teyssier}, D. and {Altmann}, M. and {Audard}, M. and {Bellas-Velidis}, I. and {Benson}, K. and {Berthier}, J. and {Blomme}, R. and {Busonero}, D. and {Busso}, G. and {C{\'a}novas}, H. and {Carry}, B. and {Cellino}, A. and {Cheek}, N. and {Clementini}, G. and {Damerdji}, Y. and {Davidson}, M. and {de Teodoro}, P. and {Nu{\~n}ez Campos}, M. and {Dell'Oro}, A. and {Esquej}, P. and {Fern{\'a}ndez-Hern{\'a}ndez}, J. and {Fraile}, E. and {Garc{\'\i}a-Lario}, P. and {Gosset}, E. and {Haigron}, R. and {Halbwachs}, J.-L. and {Hambly}, N.~C. and {Harrison}, D.~L. and {Hern{\'a}ndez}, J. and {Hestroffer}, D. and {Hodgkin}, S.~T. and {Holl}, B. and {Jan{\ss}en}, K. and {Jevardat de Fombelle}, G. and {Jordan}, S. and {Krone-Martins}, A. and {Lanzafame}, A.~C. and {L{\"o}ffler}, W. and {Marchal}, O. and {Marrese}, P.~M. and {Moitinho}, A. and {Muinonen}, K. and {Osborne}, P. and {Pauwels}, T. and {Recio-Blanco}, A. and {Reyl{\'e}}, C. and {Rimoldini}, L. and {Roegiers}, T. and {Rybizki}, J. and {Sarro}, L.~M. and {Siopis}, C. and {Smith}, M. and {Sozzetti}, A. and {Utrilla}, E. and {van Leeuwen}, M. and {Abbas}, U. and {{\'A}brah{\'a}m}, P. and {Abreu Aramburu}, A. and {Aerts}, C. and {Aguado}, J.~J. and {Ajaj}, M. and {Aldea-Montero}, F. and {Altavilla}, G. and {{\'A}lvarez}, M.~A. and {Alves}, J. and {Anderson}, R.~I. and {Anglada Varela}, E. and {Antoja}, T. and {Baines}, D. and {Baker}, S.~G. and {Balaguer-N{\'u}{\~n}ez}, L. and {Balbinot}, E. and {Balog}, Z. and {Barache}, C. and {Barbato}, D. and {Barros}, M. and {Bartolom{\'e}}, S. and {Bassilana}, J.-L. and {Bauchet}, N. and {Becciani}, U. and {Berihuete}, A. and {Bernet}, M. and {Bertone}, S. and {Bianchi}, L. and {Binnenfeld}, A. and {Blanco-Cuaresma}, S. and {Boch}, T. and {Bombrun}, A. and {Bouquillon}, S. and {Bramante}, L. and {Breedt}, E. and {Bressan}, A. and {Brouillet}, N. and {Brugaletta}, E. and {Bucciarelli}, B. and {Burlacu}, A. and {Butkevich}, A.~G. and {Buzzi}, R. and {Caffau}, E. and {Cancelliere}, R. and {Cantat-Gaudin}, T. and {Carballo}, R. and {Carlucci}, T. and {Carnerero}, M.~I. and {Casamiquela}, L. and {Castellani}, M. and {Castro-Ginard}, A. and {Chaoul}, L. and {Charlot}, P. and {Chemin}, L. and {Chiaramida}, V. and {Chiavassa}, A. and {Comoretto}, G. and {Contursi}, G. and {Cooper}, W.~J. and {Cornez}, T. and {Cowell}, S. and {Crifo}, F. and {Cropper}, M. and {Crosta}, M. and {Crowley}, C. and {Dafonte}, C. and {Dapergolas}, A.},
        title = "{Gaia Data Release 3. The Galaxy in your preferred colours: Synthetic photometry from Gaia low-resolution spectra}",
      journal = {\aap},
         year = 2023,
        month = jun,
       volume = {674},
          eid = {A33},
        pages = {A33},
          doi = {10.1051/0004-6361/202243709},
archivePrefix = {arXiv},
       eprint = {2206.06215},
 primaryClass = {astro-ph.SR},
       adsurl = {https://ui.adsabs.harvard.edu/abs/2023A&A...674A..33G}
}

@ARTICLE{2008AN....329..295C,
       author = {{Cavanagh}, B. and {Jenness}, T. and {Economou}, F. and {Currie}, M.~J.},
        title = "{The ORAC-DR data reduction pipeline}",
      journal = {Astronomische Nachrichten},
         year = 2008,
        month = mar,
       volume = {329},
       number = {3},
        pages = {295-297},
          doi = {10.1002/asna.200710944},
       adsurl = {https://ui.adsabs.harvard.edu/abs/2008AN....329..295C}
}

@software{2015ascl.soft04004B,
       author = {{Becker}, Andrew},
        title = "{HOTPANTS: High Order Transform of PSF ANd Template Subtraction}",
 howpublished = {Astrophysics Source Code Library, record ascl:1504.004},
         year = 2015,
        month = apr,
          eid = {ascl:1504.004},
archivePrefix = {ascl},
       eprint = {1504.004},
       adsurl = {https://ui.adsabs.harvard.edu/abs/2015ascl.soft04004B}
}

@ARTICLE{2014ApJ...797..118F,
       author = {{Fransson}, Claes and {Ergon}, Mattias and {Challis}, Peter J. and {Chevalier}, Roger A. and {France}, Kevin and {Kirshner}, Robert P. and {Marion}, G.~H. and {Milisavljevic}, Dan and {Smith}, Nathan and {Bufano}, Filomena and {Friedman}, Andrew S. and {Kangas}, Tuomas and {Larsson}, Josefin and {Mattila}, Seppo and {Benetti}, Stefano and {Chornock}, Ryan and {Czekala}, Ian and {Soderberg}, Alicia and {Sollerman}, Jesper},
        title = "{High-density Circumstellar Interaction in the Luminous Type IIn SN 2010jl: The First 1100 Days}",
      journal = {\apj},
         year = 2014,
        month = dec,
       volume = {797},
       number = {2},
          eid = {118},
        pages = {118},
          doi = {10.1088/0004-637X/797/2/118},
archivePrefix = {arXiv},
       eprint = {1312.6617},
 primaryClass = {astro-ph.HE},
       adsurl = {https://ui.adsabs.harvard.edu/abs/2014ApJ...797..118F}
}

@ARTICLE{1999MNRAS.309..343A,
       author = {{Aretxaga}, Itziar and {Benetti}, S. and {Terlevich}, R.~J. and {Fabian}, A.~C. and {Cappellaro}, E. and {Turatto}, M. and {della Valle}, M.},
        title = "{SN 1988Z: spectro-photometric catalogue and energy estimates$^{*}$}",
      journal = {\mnras},
         year = 1999,
        month = oct,
       volume = {309},
       number = {2},
        pages = {343-354},
          doi = {10.1046/j.1365-8711.1999.02830.x},
archivePrefix = {arXiv},
       eprint = {astro-ph/9905313},
 primaryClass = {astro-ph},
       adsurl = {https://ui.adsabs.harvard.edu/abs/1999MNRAS.309..343A}
}

@ARTICLE{2001MNRAS.326.1448C,
       author = {{Chugai}, N.~N.},
        title = "{Broad emission lines from the opaque electron-scattering environment of SN 1998S}",
      journal = {\mnras},
         year = 2001,
        month = oct,
       volume = {326},
       number = {4},
        pages = {1448-1454},
          doi = {10.1111/j.1365-2966.2001.04717.x},
archivePrefix = {arXiv},
       eprint = {astro-ph/0106234},
 primaryClass = {astro-ph},
       adsurl = {https://ui.adsabs.harvard.edu/abs/2001MNRAS.326.1448C}
}

@ARTICLE{1994ApJ...420..268C,
       author = {{Chevalier}, Roger A. and {Fransson}, Claes},
        title = "{Emission from Circumstellar Interaction in Normal Type II Supernovae}",
      journal = {\apj},
         year = 1994,
        month = jan,
       volume = {420},
        pages = {268},
          doi = {10.1086/173557},
       adsurl = {https://ui.adsabs.harvard.edu/abs/1994ApJ...420..268C}
}

@ARTICLE{2014Natur.511..326G,
       author = {{Gall}, Christa and {Hjorth}, Jens and {Watson}, Darach and {Dwek}, Eli and {Maund}, Justyn R. and {Fox}, Ori and {Leloudas}, Giorgos and {Malesani}, Daniele and {Day-Jones}, Avril C.},
        title = "{Rapid formation of large dust grains in the luminous supernova 2010jl}",
      journal = {\nat},
         year = 2014,
        month = jul,
       volume = {511},
       number = {7509},
        pages = {326-329},
          doi = {10.1038/nature13558},
archivePrefix = {arXiv},
       eprint = {1407.4447},
 primaryClass = {astro-ph.SR},
       adsurl = {https://ui.adsabs.harvard.edu/abs/2014Natur.511..326G}
}

@ARTICLE{2012PASP..124..668Y,
       author = {{Yaron}, Ofer and {Gal-Yam}, Avishay},
        title = "{WISeREP{\textemdash}An Interactive Supernova Data Repository}",
      journal = {\pasp},
         year = 2012,
        month = jul,
       volume = {124},
       number = {917},
        pages = {668},
          doi = {10.1086/666656},
archivePrefix = {arXiv},
       eprint = {1204.1891},
 primaryClass = {astro-ph.IM},
       adsurl = {https://ui.adsabs.harvard.edu/abs/2012PASP..124..668Y}
}

@ARTICLE{2011ApJ...737..103S,
       author = {{Schlafly}, Edward F. and {Finkbeiner}, Douglas P.},
        title = "{Measuring Reddening with Sloan Digital Sky Survey Stellar Spectra and Recalibrating SFD}",
      journal = {\apj},
         year = 2011,
        month = aug,
       volume = {737},
       number = {2},
          eid = {103},
        pages = {103},
          doi = {10.1088/0004-637X/737/2/103},
archivePrefix = {arXiv},
       eprint = {1012.4804},
 primaryClass = {astro-ph.GA},
       adsurl = {https://ui.adsabs.harvard.edu/abs/2011ApJ...737..103S}
}

@ARTICLE{1999PASP..111...63F,
       author = {{Fitzpatrick}, Edward L.},
        title = "{Correcting for the Effects of Interstellar Extinction}",
      journal = {\pasp},
         year = 1999,
        month = jan,
       volume = {111},
       number = {755},
        pages = {63-75},
          doi = {10.1086/316293},
archivePrefix = {arXiv},
       eprint = {astro-ph/9809387},
 primaryClass = {astro-ph},
       adsurl = {https://ui.adsabs.harvard.edu/abs/1999PASP..111...63F}
}

@ARTICLE{2014MNRAS.439.2917M,
       author = {{Moriya}, Takashi J. and {Maeda}, Keiichi and {Taddia}, Francesco and {Sollerman}, Jesper and {Blinnikov}, Sergei I. and {Sorokina}, Elena I.},
        title = "{Mass-loss histories of Type IIn supernova progenitors within decades before their explosion}",
      journal = {\mnras},
         year = 2014,
        month = apr,
       volume = {439},
       number = {3},
        pages = {2917-2926},
          doi = {10.1093/mnras/stu163},
archivePrefix = {arXiv},
       eprint = {1401.4893},
 primaryClass = {astro-ph.SR},
       adsurl = {https://ui.adsabs.harvard.edu/abs/2014MNRAS.439.2917M}
}

@ARTICLE{2010ApJ...709..856S,
       author = {{Smith}, Nathan and {Chornock}, Ryan and {Silverman}, Jeffrey M. and {Filippenko}, Alexei V. and {Foley}, Ryan J.},
        title = "{Spectral Evolution of the Extraordinary Type IIn Supernova 2006gy}",
      journal = {\apj},
         year = 2010,
        month = feb,
       volume = {709},
       number = {2},
        pages = {856-883},
          doi = {10.1088/0004-637X/709/2/856},
archivePrefix = {arXiv},
       eprint = {0906.2200},
 primaryClass = {astro-ph.HE},
       adsurl = {https://ui.adsabs.harvard.edu/abs/2010ApJ...709..856S}
}

@ARTICLE{2019ApJS..241...38S,
       author = {{Szalai}, Tam{\'a}s and {Zs{\'\i}ros}, Szanna and {Fox}, Ori D. and {Pejcha}, Ond{\v{r}}ej and {M{\"u}ller}, Tom{\'a}s},
        title = "{A Comprehensive Analysis of Spitzer Supernovae}",
      journal = {\apjs},
         year = 2019,
        month = apr,
       volume = {241},
       number = {2},
          eid = {38},
        pages = {38},
          doi = {10.3847/1538-4365/ab10df},
archivePrefix = {arXiv},
       eprint = {1803.02571},
 primaryClass = {astro-ph.HE},
       adsurl = {https://ui.adsabs.harvard.edu/abs/2019ApJS..241...38S}
}

@ARTICLE{2011ApJ...741....7F,
       author = {{Fox}, Ori D. and {Chevalier}, Roger A. and {Skrutskie}, Michael F. and {Soderberg}, Alicia M. and {Filippenko}, Alexei V. and {Ganeshalingam}, Mohan and {Silverman}, Jeffrey M. and {Smith}, Nathan and {Steele}, Thea N.},
        title = "{A Spitzer Survey for Dust in Type IIn Supernovae}",
      journal = {\apj},
         year = 2011,
        month = nov,
       volume = {741},
       number = {1},
          eid = {7},
        pages = {7},
          doi = {10.1088/0004-637X/741/1/7},
archivePrefix = {arXiv},
       eprint = {1104.5012},
 primaryClass = {astro-ph.SR},
       adsurl = {https://ui.adsabs.harvard.edu/abs/2011ApJ...741....7F}
}

@ARTICLE{2021AJ....161..192N,
       author = {{Nidever}, David L. and {Dey}, Arjun and {Fasbender}, Katie and {Juneau}, St{\'e}phanie and {Meisner}, Aaron M. and {Wishart}, Joseph and {Scott}, Adam and {Matt}, Kyle and {Nikutta}, Robert and {Pucha}, Ragadeepika},
        title = "{Second Data Release of the All-sky NOIRLab Source Catalog}",
      journal = {\aj},
         year = 2021,
        month = apr,
       volume = {161},
       number = {4},
          eid = {192},
        pages = {192},
          doi = {10.3847/1538-3881/abd6e1},
archivePrefix = {arXiv},
       eprint = {2011.08868},
 primaryClass = {astro-ph.GA},
       adsurl = {https://ui.adsabs.harvard.edu/abs/2021AJ....161..192N}
}

@ARTICLE{2002AJ....124..266P,
       author = {{Peng}, Chien Y. and {Ho}, Luis C. and {Impey}, Chris D. and {Rix}, Hans-Walter},
        title = "{Detailed Structural Decomposition of Galaxy Images}",
      journal = {\aj},
         year = 2002,
        month = jul,
       volume = {124},
       number = {1},
        pages = {266-293},
          doi = {10.1086/340952},
archivePrefix = {arXiv},
       eprint = {astro-ph/0204182},
 primaryClass = {astro-ph},
       adsurl = {https://ui.adsabs.harvard.edu/abs/2002AJ....124..266P}
}

@ARTICLE{2010AJ....139.2097P,
       author = {{Peng}, Chien Y. and {Ho}, Luis C. and {Impey}, Chris D. and {Rix}, Hans-Walter},
        title = "{Detailed Decomposition of Galaxy Images. II. Beyond Axisymmetric Models}",
      journal = {\aj},
         year = 2010,
        month = jun,
       volume = {139},
       number = {6},
        pages = {2097-2129},
          doi = {10.1088/0004-6256/139/6/2097},
archivePrefix = {arXiv},
       eprint = {0912.0731},
 primaryClass = {astro-ph.CO},
       adsurl = {https://ui.adsabs.harvard.edu/abs/2010AJ....139.2097P}
}

@ARTICLE{2015AcA....65....1U,
       author = {{Udalski}, A. and {Szyma{\'n}ski}, M.~K. and {Szyma{\'n}ski}, G.},
        title = "{OGLE-IV: Fourth Phase of the Optical Gravitational Lensing Experiment}",
      journal = {\actaa},
         year = 2015,
        month = mar,
       volume = {65},
       number = {1},
        pages = {1-38},
          doi = {10.48550/arXiv.1504.05966},
archivePrefix = {arXiv},
       eprint = {1504.05966},
 primaryClass = {astro-ph.SR},
       adsurl = {https://ui.adsabs.harvard.edu/abs/2015AcA....65....1U}
}

@ARTICLE{2014ApJS..214...15S,
       author = {{Speagle}, J.~S. and {Steinhardt}, C.~L. and {Capak}, P.~L. and {Silverman}, J.~D.},
        title = "{A Highly Consistent Framework for the Evolution of the Star-Forming ``Main Sequence'' from z \raisebox{-0.5ex}\textasciitilde 0-6}",
      journal = {\apjs},
         year = 2014,
        month = oct,
       volume = {214},
       number = {2},
          eid = {15},
        pages = {15},
          doi = {10.1088/0067-0049/214/2/15},
archivePrefix = {arXiv},
       eprint = {1405.2041},
 primaryClass = {astro-ph.GA},
       adsurl = {https://ui.adsabs.harvard.edu/abs/2014ApJS..214...15S}
}

@ARTICLE{2024MNRAS.531.1203S,
       author = {{Sarin}, Nikhil and {H{\"u}bner}, Moritz and {Omand}, Conor M.~B. and {Setzer}, Christian N. and {Schulze}, Steve and {Adhikari}, Naresh and {Sagu{\'e}s-Carracedo}, Ana and {Galaudage}, Shanika and {Wallace}, Wendy F. and {Lamb}, Gavin P. and {Lin}, En-Tzu},
        title = "{REDBACK: a Bayesian inference software package for electromagnetic transients}",
      journal = {\mnras},
         year = 2024,
        month = jun,
       volume = {531},
       number = {1},
        pages = {1203-1227},
          doi = {10.1093/mnras/stae1238},
archivePrefix = {arXiv},
       eprint = {2308.12806},
 primaryClass = {astro-ph.HE},
       adsurl = {https://ui.adsabs.harvard.edu/abs/2024MNRAS.531.1203S}
}

@ARTICLE{2021ApJS..255...29S,
       author = {{Schulze}, Steve and {Yaron}, Ofer and {Sollerman}, Jesper and {Leloudas}, Giorgos and {Gal}, Amit and {Wright}, Angus H. and {Lunnan}, Ragnhild and {Gal-Yam}, Avishay and {Ofek}, Eran O. and {Perley}, Daniel A. and {Filippenko}, Alexei V. and {Kasliwal}, Mansi M. and {Kulkarni}, Shrinivas R. and {Neill}, James D. and {Nugent}, Peter E. and {Quimby}, Robert M. and {Sullivan}, Mark and {Strotjohann}, Nora Linn and {Arcavi}, Iair and {Ben-Ami}, Sagi and {Bianco}, Federica and {Bloom}, Joshua S. and {De}, Kishalay and {Fraser}, Morgan and {Fremling}, Christoffer U. and {Horesh}, Assaf and {Johansson}, Joel and {Kelly}, Patrick L. and {Kne{\v{z}}evi{\'c}}, Nikola and {Kne{\v{z}}evi{\'c}}, Sladjana and {Maguire}, Kate and {Nyholm}, Anders and {Papadogiannakis}, Sem{\'e}li and {Petrushevska}, Tanja and {Rubin}, Adam and {Yan}, Lin and {Yang}, Yi and {Adams}, Scott M. and {Bufano}, Filomena and {Clubb}, Kelsey I. and {Foley}, Ryan J. and {Green}, Yoav and {Harmanen}, Jussi and {Ho}, Anna Y.~Q. and {Hook}, Isobel M. and {Hosseinzadeh}, Griffin and {Howell}, D. Andrew and {Kong}, Albert K.~H. and {Kotak}, Rubina and {Matheson}, Thomas and {McCully}, Curtis and {Milisavljevic}, Dan and {Pan}, Yen-Chen and {Poznanski}, Dovi and {Shivvers}, Isaac and {van Velzen}, Sjoert and {Verbeek}, Kars K.},
        title = "{The Palomar Transient Factory Core-collapse Supernova Host-galaxy Sample. I. Host-galaxy Distribution Functions and Environment Dependence of Core-collapse Supernovae}",
      journal = {\apjs},
         year = 2021,
        month = aug,
       volume = {255},
       number = {2},
          eid = {29},
        pages = {29},
          doi = {10.3847/1538-4365/abff5e},
archivePrefix = {arXiv},
       eprint = {2008.05988},
 primaryClass = {astro-ph.GA},
       adsurl = {https://ui.adsabs.harvard.edu/abs/2021ApJS..255...29S}
}

@ARTICLE{2022MNRAS.513.3564R,
       author = {{Ransome}, C.~L. and {Habergham-Mawson}, S.~M. and {Darnley}, M.~J. and {James}, P.~A. and {Percival}, S.~M.},
        title = "{An H{\ensuremath{\alpha}} survey of the host environments of 77 type IIn supernovae within z < 0.02}",
      journal = {\mnras},
         year = 2022,
        month = jul,
       volume = {513},
       number = {3},
        pages = {3564-3576},
          doi = {10.1093/mnras/stac1093},
archivePrefix = {arXiv},
       eprint = {2204.09706},
 primaryClass = {astro-ph.HE},
       adsurl = {https://ui.adsabs.harvard.edu/abs/2022MNRAS.513.3564R}
}

@ARTICLE{2015A&A...580A.131T,
       author = {{Taddia}, F. and {Sollerman}, J. and {Fremling}, C. and {Pastorello}, A. and {Leloudas}, G. and {Fransson}, C. and {Nyholm}, A. and {Stritzinger}, M.~D. and {Ergon}, M. and {Roy}, R. and {Migotto}, K.},
        title = "{Metallicity at the explosion sites of interacting transients}",
      journal = {\aap},
         year = 2015,
        month = aug,
       volume = {580},
          eid = {A131},
        pages = {A131},
          doi = {10.1051/0004-6361/201525989},
archivePrefix = {arXiv},
       eprint = {1505.04719},
 primaryClass = {astro-ph.HE},
       adsurl = {https://ui.adsabs.harvard.edu/abs/2015A&A...580A.131T}
}

@ARTICLE{2023A&A...677A..20M,
       author = {{Moriya}, Takashi J. and {Galbany}, Llu{\'\i}s and {Jim{\'e}nez-Palau}, Cristina and {Anderson}, Joseph P. and {Kuncarayakti}, Hanindyo and {S{\'a}nchez}, Sebasti{\'a}n F. and {Lyman}, Joseph D. and {Pessi}, Thallis and {Prieto}, Jose L. and {Kochanek}, Christopher S. and {Dong}, Subo and {Chen}, Ping},
        title = "{Environmental dependence of Type IIn supernova properties}",
      journal = {\aap},
         year = 2023,
        month = sep,
       volume = {677},
          eid = {A20},
        pages = {A20},
          doi = {10.1051/0004-6361/202346703},
archivePrefix = {arXiv},
       eprint = {2306.09647},
 primaryClass = {astro-ph.HE},
       adsurl = {https://ui.adsabs.harvard.edu/abs/2023A&A...677A..20M}
}

@ARTICLE{2021MNRAS.506.4715R,
       author = {{Ransome}, C.~L. and {Habergham-Mawson}, S.~M. and {Darnley}, M.~J. and {James}, P.~A. and {Filippenko}, A.~V. and {Schlegel}, E.~M.},
        title = "{A systematic reclassification of Type IIn supernovae}",
      journal = {\mnras},
         year = 2021,
        month = oct,
       volume = {506},
       number = {4},
        pages = {4715-4734},
          doi = {10.1093/mnras/stab1938},
archivePrefix = {arXiv},
       eprint = {2107.02179},
 primaryClass = {astro-ph.SR},
       adsurl = {https://ui.adsabs.harvard.edu/abs/2021MNRAS.506.4715R}
}

@ARTICLE{2026MNRAS.552g1640S,
       author = {{Sarin}, Nikhil and {Hirai}, Ryosuke},
        title = "{From mass-loss histories to light curves: a generalized framework for interaction-powered transients}",
      journal = {\mnras},
         year = 2026,
        month = oct,
       volume = {552},
       number = {1},
          eid = {stag1640},
        pages = {stag1640},
          doi = {10.1093/mnras/stag1640},
archivePrefix = {arXiv},
       eprint = {2605.19571},
 primaryClass = {astro-ph.HE},
       adsurl = {https://ui.adsabs.harvard.edu/abs/2026MNRAS.552g1640S}
}

@ARTICLE{2009MNRAS.394...21D,
       author = {{Dessart}, Luc and {Hillier}, D. John and {Gezari}, Suvi and {Basa}, St{\'e}phane and {Matheson}, Tom},
        title = "{SN 1994W: an interacting supernova or two interacting shells?}",
      journal = {\mnras},
         year = 2009,
        month = mar,
       volume = {394},
       number = {1},
        pages = {21-37},
          doi = {10.1111/j.1365-2966.2008.14042.x},
archivePrefix = {arXiv},
       eprint = {0804.0428},
 primaryClass = {astro-ph},
       adsurl = {https://ui.adsabs.harvard.edu/abs/2009MNRAS.394...21D}
}

@ARTICLE{2018MNRAS.475.1261H,
       author = {{Huang}, Chenliang and {Chevalier}, Roger A.},
        title = "{Electron scattering wings on lines in interacting supernovae}",
      journal = {\mnras},
         year = 2018,
        month = mar,
       volume = {475},
       number = {1},
        pages = {1261-1273},
          doi = {10.1093/mnras/stx3163},
archivePrefix = {arXiv},
       eprint = {1712.01237},
 primaryClass = {astro-ph.HE},
       adsurl = {https://ui.adsabs.harvard.edu/abs/2018MNRAS.475.1261H}
}

@ARTICLE{2013MNRAS.435.1520M,
       author = {{Moriya}, Takashi J. and {Maeda}, Keiichi and {Taddia}, Francesco and {Sollerman}, Jesper and {Blinnikov}, Sergei I. and {Sorokina}, Elena I.},
        title = "{An analytic bolometric light curve model of interaction-powered supernovae and its application to Type IIn supernovae}",
      journal = {\mnras},
         year = 2013,
        month = oct,
       volume = {435},
       number = {2},
        pages = {1520-1535},
          doi = {10.1093/mnras/stt1392},
archivePrefix = {arXiv},
       eprint = {1307.2644},
 primaryClass = {astro-ph.HE},
       adsurl = {https://ui.adsabs.harvard.edu/abs/2013MNRAS.435.1520M}
}

@ARTICLE{2019A&A...622A.103B,
       author = {{Boquien}, M. and {Burgarella}, D. and {Roehlly}, Y. and {Buat}, V. and {Ciesla}, L. and {Corre}, D. and {Inoue}, A.~K. and {Salas}, H.},
        title = "{CIGALE: a python Code Investigating GALaxy Emission}",
      journal = {\aap},
         year = 2019,
        month = feb,
       volume = {622},
          eid = {A103},
        pages = {A103},
          doi = {10.1051/0004-6361/201834156},
archivePrefix = {arXiv},
       eprint = {1811.03094},
 primaryClass = {astro-ph.GA},
       adsurl = {https://ui.adsabs.harvard.edu/abs/2019A&A...622A.103B}
}

@ARTICLE{2026ApJ..1005L..36T,
       author = {{Tsai}, Sung-Han and {Chen}, Ke-Jung and {Maeda}, Keiichi and {Ou}, Po-Sheng and {R{\"o}pke}, Friedrich K.},
        title = "{Interacting Binary Stars as Progenitors for Interacting Supernovae}",
      journal = {\apjl},
         year = 2026,
        month = jul,
       volume = {1005},
       number = {2},
          eid = {L36},
        pages = {L36},
          doi = {10.3847/2041-8213/ae7e84},
archivePrefix = {arXiv},
       eprint = {2605.11635},
 primaryClass = {astro-ph.HE},
       adsurl = {https://ui.adsabs.harvard.edu/abs/2026ApJ..1005L..36T}
}

@ARTICLE{2026ApJ...996...18S,
       author = {{Sengupta}, Sutirtha and {Sujit}, Das and {Sarangi}, Arkaprabha},
        title = "{Dance to Demise{\textemdash}How Massive Stars May Form Dense Circumstellar Shells before Explosion}",
      journal = {\apj},
         year = 2026,
        month = jan,
       volume = {996},
       number = {1},
          eid = {18},
        pages = {18},
          doi = {10.3847/1538-4357/ae129c},
archivePrefix = {arXiv},
       eprint = {2508.04497},
 primaryClass = {astro-ph.SR},
       adsurl = {https://ui.adsabs.harvard.edu/abs/2026ApJ...996...18S}
}

@ARTICLE{2026ApJ..1001...44N,
       author = {{Niu}, Zexi and {Sun}, Ning-Chen and {Zapartas}, Emmanouil and {Ransome}, Conor L. and {Maund}, Justyn R. and {Rojas-Bravo}, Cesar and {Liu}, Jifeng},
        title = "{Revealing the Diversity of Type IIn Supernova Progenitors through Their Environments}",
      journal = {\apj},
         year = 2026,
        month = apr,
       volume = {1001},
       number = {1},
          eid = {44},
        pages = {44},
          doi = {10.3847/1538-4357/ae4bd6},
archivePrefix = {arXiv},
       eprint = {2601.07542},
 primaryClass = {astro-ph.HE},
       adsurl = {https://ui.adsabs.harvard.edu/abs/2026ApJ..1001...44N}
}

@ARTICLE{2016MNRAS.456.2622J,
       author = {{Jencson}, J.~E. and {Prieto}, J.~L. and {Kochanek}, C.~S. and {Shappee}, B.~J. and {Stanek}, K.~Z. and {Pogge}, R.~W.},
        title = "{Optical observations of the luminous Type IIn Supernova 2010jl for over 900 d}",
      journal = {\mnras},
         year = 2016,
        month = mar,
       volume = {456},
       number = {3},
        pages = {2622-2635},
          doi = {10.1093/mnras/stv2795},
archivePrefix = {arXiv},
       eprint = {1505.01186},
 primaryClass = {astro-ph.HE},
       adsurl = {https://ui.adsabs.harvard.edu/abs/2016MNRAS.456.2622J}
}

@ARTICLE{2009ApJ...697..747K,
       author = {{Kawabata}, Koji S. and {Tanaka}, Masaomi and {Maeda}, Keiichi and {Hattori}, Takashi and {Nomoto}, Ken'ichi and {Tominaga}, Nozomu and {Yamanaka}, Masayuki},
        title = "{Extremely Luminous Supernova 2006gy at Late Phase: Detection of Optical Emission from Supernova}",
      journal = {\apj},
         year = 2009,
        month = may,
       volume = {697},
       number = {1},
        pages = {747-757},
          doi = {10.1088/0004-637X/697/1/747},
archivePrefix = {arXiv},
       eprint = {0902.1440},
 primaryClass = {astro-ph.SR},
       adsurl = {https://ui.adsabs.harvard.edu/abs/2009ApJ...697..747K}
}

@ARTICLE{2011JMLR...12.2825P,
       author = {{Pedregosa}, Fabian and {Varoquaux}, Ga{\"e}l and {Gramfort}, Alexandre and {Michel}, Vincent and {Thirion}, Bertrand and {Grisel}, Olivier and {Blondel}, Mathieu and {M{\"u}ller}, Andreas and {Nothman}, Joel and {Louppe}, Gilles and {Prettenhofer}, Peter and {Weiss}, Ron and {Dubourg}, Vincent and {Vanderplas}, Jake and {Passos}, Alexandre and {Cournapeau}, David and {Brucher}, Matthieu and {Perrot}, Matthieu and {Duchesnay}, {\'E}douard},
        title = "{Scikit-learn: Machine Learning in Python}",
      journal = {Journal of Machine Learning Research},
         year = 2011,
        month = oct,
       volume = {12},
        pages = {2825-2830},
          doi = {10.48550/arXiv.1201.0490},
archivePrefix = {arXiv},
       eprint = {1201.0490},
 primaryClass = {cs.LG},
       adsurl = {https://ui.adsabs.harvard.edu/abs/2011JMLR...12.2825P}
}

@ARTICLE{2008ApJ...686..467S,
       author = {{Smith}, Nathan and {Chornock}, Ryan and {Li}, Weidong and {Ganeshalingam}, Mohan and {Silverman}, Jeffrey M. and {Foley}, Ryan J. and {Filippenko}, Alexei V. and {Barth}, Aaron J.},
        title = "{SN 2006tf: Precursor Eruptions and the Optically Thick Regime of Extremely Luminous Type IIn Supernovae}",
      journal = {\apj},
         year = 2008,
        month = oct,
       volume = {686},
       number = {1},
        pages = {467-484},
          doi = {10.1086/591021},
archivePrefix = {arXiv},
       eprint = {0804.0042},
 primaryClass = {astro-ph},
       adsurl = {https://ui.adsabs.harvard.edu/abs/2008ApJ...686..467S}
}

@ARTICLE{2026ApJ..1005...82H,
       author = {{Hiramatsu}, Daichi and {Berger}, Edo and {Gomez}, Sebastian and {Blanchard}, Peter K. and {Kumar}, Harsh and {Athukoralalage}, Wasundara},
        title = "{Type IIn Supernovae. I. Uniform Light-curve Characterization and a Bimodality in the Radiated Energy Distribution}",
      journal = {\apj},
         year = 2026,
        month = jul,
       volume = {1005},
       number = {1},
          eid = {82},
        pages = {82},
          doi = {10.3847/1538-4357/ae743c},
archivePrefix = {arXiv},
       eprint = {2411.07287},
 primaryClass = {astro-ph.HE},
       adsurl = {https://ui.adsabs.harvard.edu/abs/2026ApJ..1005...82H}
}

@ARTICLE{2021ApJ...907...99S,
       author = {{Strotjohann}, Nora L. and {Ofek}, Eran O. and {Gal-Yam}, Avishay and {Bruch}, Rachel and {Schulze}, Steve and {Shaviv}, Nir and {Sollerman}, Jesper and {Filippenko}, Alexei V. and {Yaron}, Ofer and {Fremling}, Christoffer and {Nordin}, Jakob and {Kool}, Erik C. and {Perley}, Dan A. and {Ho}, Anna Y.~Q. and {Yang}, Yi and {Yao}, Yuhan and {Soumagnac}, Maayane T. and {Graham}, Melissa L. and {Barbarino}, Cristina and {Tartaglia}, Leonardo and {De}, Kishalay and {Goldstein}, Daniel A. and {Cook}, David O. and {Brink}, Thomas G. and {Taggart}, Kirsty and {Yan}, Lin and {Lunnan}, Ragnhild and {Kasliwal}, Mansi and {Kulkarni}, Shri R. and {Nugent}, Peter E. and {Masci}, Frank J. and {Rosnet}, Philippe and {Adams}, Scott M. and {Andreoni}, Igor and {Bagdasaryan}, Ashot and {Bellm}, Eric C. and {Burdge}, Kevin and {Duev}, Dmitry A. and {Dugas}, Alison and {Frederick}, Sara and {Goldwasser}, Samantha and {Hankins}, Matthew and {Irani}, Ido and {Karambelkar}, Viraj and {Kupfer}, Thomas and {Liang}, Jingyi and {Neill}, James D. and {Porter}, Michael and {Riddle}, Reed L. and {Sharma}, Yashvi and {Short}, Phil and {Taddia}, Francesco and {Tzanidakis}, Anastasios and {van Roestel}, Jan and {Walters}, Richard and {Zhuang}, Zhuyun},
        title = "{Bright, Months-long Stellar Outbursts Announce the Explosion of Interaction-powered Supernovae}",
      journal = {\apj},
         year = 2021,
        month = feb,
       volume = {907},
       number = {2},
          eid = {99},
        pages = {99},
          doi = {10.3847/1538-4357/abd032},
archivePrefix = {arXiv},
       eprint = {2010.11196},
 primaryClass = {astro-ph.HE},
       adsurl = {https://ui.adsabs.harvard.edu/abs/2021ApJ...907...99S}
}

@ARTICLE{2009Natur.458..865G,
       author = {{Gal-Yam}, A. and {Leonard}, D.~C.},
        title = "{A massive hypergiant star as the progenitor of the supernova SN 2005gl}",
      journal = {\nat},
         year = 2009,
        month = apr,
       volume = {458},
       number = {7240},
        pages = {865-867},
          doi = {10.1038/nature07934},
       adsurl = {https://ui.adsabs.harvard.edu/abs/2009Natur.458..865G}
}

@ARTICLE{2007ApJ...656..372G,
       author = {{Gal-Yam}, Avishay and {Leonard}, D.~C. and {Fox}, D.~B. and {Cenko}, S.~B. and {Soderberg}, A.~M. and {Moon}, D.-S. and {Sand}, D.~J. and {Caltech Core Collapse Program} and {Li}, Weidong and {Filippenko}, Alexei V. and {Aldering}, G. and {Copin}, Y.},
        title = "{On the Progenitor of SN 2005gl and the Nature of Type IIn Supernovae}",
      journal = {\apj},
         year = 2007,
        month = feb,
       volume = {656},
       number = {1},
        pages = {372-381},
          doi = {10.1086/510523},
archivePrefix = {arXiv},
       eprint = {astro-ph/0608029},
 primaryClass = {astro-ph},
       adsurl = {https://ui.adsabs.harvard.edu/abs/2007ApJ...656..372G}
}

@ARTICLE{2024ApJ...970..189J,
       author = {{Jacobson-Gal{\'a}n}, W.~V. and {Dessart}, L. and {Davis}, K.~W. and {Kilpatrick}, C.~D. and {Margutti}, R. and {Foley}, R.~J. and {Chornock}, R. and {Terreran}, G. and {Hiramatsu}, D. and {Newsome}, M. and {Padilla Gonzalez}, E. and {Pellegrino}, C. and {Howell}, D.~A. and {Filippenko}, A.~V. and {Anderson}, J.~P. and {Angus}, C.~R. and {Auchettl}, K. and {Bostroem}, K.~A. and {Brink}, T.~G. and {Cartier}, R. and {Coulter}, D.~A. and {de Boer}, T. and {Drout}, M.~R. and {Earl}, N. and {Ertini}, K. and {Farah}, J.~R. and {Farias}, D. and {Gall}, C. and {Gao}, H. and {Gerlach}, M.~A. and {Guo}, F. and {Haynie}, A. and {Hosseinzadeh}, G. and {Ibik}, A.~L. and {Jha}, S.~W. and {Jones}, D.~O. and {Langeroodi}, D. and {LeBaron}, N. and {Magnier}, E.~A. and {Piro}, A.~L. and {Raimundo}, S.~I. and {Rest}, A. and {Rest}, S. and {Rich}, R. Michael and {Rojas-Bravo}, C. and {Sears}, H. and {Taggart}, K. and {Villar}, V.~A. and {Wainscoat}, R.~J. and {Wang}, X.-F. and {Wasserman}, A.~R. and {Yan}, S. and {Yang}, Y. and {Zhang}, J. and {Zheng}, W.},
        title = "{Final Moments. II. Observational Properties and Physical Modeling of Circumstellar-material-interacting Type II Supernovae}",
      journal = {\apj},
         year = 2024,
        month = aug,
       volume = {970},
       number = {2},
          eid = {189},
        pages = {189},
          doi = {10.3847/1538-4357/ad4a2a},
archivePrefix = {arXiv},
       eprint = {2403.02382},
 primaryClass = {astro-ph.HE},
       adsurl = {https://ui.adsabs.harvard.edu/abs/2024ApJ...970..189J}
}

@ARTICLE{2023ApJ...952..119B,
       author = {{Bruch}, Rachel J. and {Gal-Yam}, Avishay and {Yaron}, Ofer and {Chen}, Ping and {Strotjohann}, Nora L. and {Irani}, Ido and {Zimmerman}, Erez and {Schulze}, Steve and {Yang}, Yi and {Kim}, Young-Lo and {Bulla}, Mattia and {Sollerman}, Jesper and {Rigault}, Mickael and {Ofek}, Eran and {Soumagnac}, Maayane and {Masci}, Frank J. and {Fremling}, Christoffer and {Perley}, Daniel and {Nordin}, Jakob and {Cenko}, S. Bradley and {Ho}, Anna Y.~Q. and {Adams}, S. and {Adreoni}, Igor and {Bellm}, Eric C. and {Blagorodnova}, Nadia and {Burdge}, Kevin and {De}, Kishalay and {Dekany}, Richard G. and {Dhawan}, Suhail and {Drake}, Andrew J. and {Duev}, Dmitry A. and {Graham}, Matthew and {Graham}, Melissa L. and {Jencson}, Jacob and {Karamehmetoglu}, Emir and {Kasliwal}, Mansi M. and {Kulkarni}, Shrinivas and {Miller}, A.~A. and {Neill}, James D. and {Prince}, Thomas A. and {Riddle}, Reed and {Rusholme}, Benjamin and {Sharma}, Y. and {Smith}, Roger and {Sravan}, Niharika and {Taggart}, Kirsty and {Walters}, Richard and {Yan}, Lin},
        title = "{The Prevalence and Influence of Circumstellar Material around Hydrogen-rich Supernova Progenitors}",
      journal = {\apj},
         year = 2023,
        month = aug,
       volume = {952},
       number = {2},
          eid = {119},
        pages = {119},
          doi = {10.3847/1538-4357/acd8be},
archivePrefix = {arXiv},
       eprint = {2212.03313},
 primaryClass = {astro-ph.HE},
       adsurl = {https://ui.adsabs.harvard.edu/abs/2023ApJ...952..119B}
}

@ARTICLE{2021ApJ...912...46B,
       author = {{Bruch}, Rachel J. and {Gal-Yam}, Avishay and {Schulze}, Steve and {Yaron}, Ofer and {Yang}, Yi and {Soumagnac}, Maayane and {Rigault}, Mickael and {Strotjohann}, Nora L. and {Ofek}, Eran and {Sollerman}, Jesper and {Masci}, Frank J. and {Barbarino}, Cristina and {Ho}, Anna Y.~Q. and {Fremling}, Christoffer and {Perley}, Daniel and {Nordin}, Jakob and {Cenko}, S. Bradley and {Adams}, S. and {Adreoni}, Igor and {Bellm}, Eric C. and {Blagorodnova}, Nadia and {Bulla}, Mattia and {Burdge}, Kevin and {De}, Kishalay and {Dhawan}, Suhail and {Drake}, Andrew J. and {Duev}, Dmitry A. and {Dugas}, Alison and {Graham}, Matthew and {Graham}, Melissa L. and {Irani}, Ido and {Jencson}, Jacob and {Karamehmetoglu}, Emir and {Kasliwal}, Mansi and {Kim}, Young-Lo and {Kulkarni}, Shrinivas and {Kupfer}, Thomas and {Liang}, Jingyi and {Mahabal}, Ashish and {Miller}, A.~A. and {Prince}, Thomas A. and {Riddle}, Reed and {Sharma}, Y. and {Smith}, Roger and {Taddia}, Francesco and {Taggart}, Kirsty and {Walters}, Richard and {Yan}, Lin},
        title = "{A Large Fraction of Hydrogen-rich Supernova Progenitors Experience Elevated Mass Loss Shortly Prior to Explosion}",
      journal = {\apj},
         year = 2021,
        month = may,
       volume = {912},
       number = {1},
          eid = {46},
        pages = {46},
          doi = {10.3847/1538-4357/abef05},
archivePrefix = {arXiv},
       eprint = {2008.09986},
 primaryClass = {astro-ph.HE},
       adsurl = {https://ui.adsabs.harvard.edu/abs/2021ApJ...912...46B}
}

@ARTICLE{2016ApJ...823..127N,
       author = {{Nakar}, Ehud and {Poznanski}, Dovi and {Katz}, Boaz},
        title = "{The Importance of $^{56}$Ni in Shaping the Light Curves of Type II Supernovae}",
      journal = {\apj},
         year = 2016,
        month = jun,
       volume = {823},
       number = {2},
          eid = {127},
        pages = {127},
          doi = {10.3847/0004-637X/823/2/127},
archivePrefix = {arXiv},
       eprint = {1506.07185},
 primaryClass = {astro-ph.HE},
       adsurl = {https://ui.adsabs.harvard.edu/abs/2016ApJ...823..127N}
}

@INCOLLECTION{2017hsn..book..195G,
       author = {{Gal-Yam}, Avishay},
        title = "{Observational and Physical Classification of Supernovae}",
    booktitle = {Handbook of Supernovae},
         year = 2017,
       editor = {{Alsabti}, Athem W. and {Murdin}, Paul},
        pages = {195},
        publisher="Springer International Publishing",
          doi = {10.1007/978-3-319-21846-5_35},
       adsurl = {https://ui.adsabs.harvard.edu/abs/2017hsn..book..195G}
}

@INPROCEEDINGS{2024eas..conf.2399W,
       author = {{Wyrzykowski}, Lukasz and {Zielinski}, Pawel and {Mikolajczyk}, Przemyslaw and {Kotysz}, Krzysztof},
        title = "{BHTOM global network of telescopes - a research infrastructure open for telescopes and scientists}",
    booktitle = {EAS2024, European Astronomical Society Annual Meeting},
         year = 2024,
        month = jul,
          eid = {2399},
        pages = {2399},
       adsurl = {https://ui.adsabs.harvard.edu/abs/2024eas..conf.2399W}
}

@ARTICLE{2007Natur.447..829P,
       author = {{Pastorello}, A. and {Smartt}, S.~J. and {Mattila}, S. and {Eldridge}, J.~J. and {Young}, D. and {Itagaki}, K. and {Yamaoka}, H. and {Navasardyan}, H. and {Valenti}, S. and {Patat}, F. and et al.},
        title = "{A giant outburst two years before the core-collapse of a massive star}",
      journal = {\nat},
         year = 2007,
        month = jun,
       volume = {447},
       number = {7146},
        pages = {829-832},
          doi = {10.1038/nature05825},
archivePrefix = {arXiv},
       eprint = {astro-ph/0703663},
 primaryClass = {astro-ph},
       adsurl = {https://ui.adsabs.harvard.edu/abs/2007Natur.447..829P}
}

@ARTICLE{1941PASP...53..224M,
       author = {{Minkowski}, R.},
        title = "{Spectra of Supernovae}",
      journal = {\pasp},
         year = 1941,
        month = aug,
       volume = {53},
       number = {314},
        pages = {224},
          doi = {10.1086/125315},
       adsurl = {https://ui.adsabs.harvard.edu/abs/1941PASP...53..224M}
}

@ARTICLE{2026A&A...706A.169E,
       author = {{Ercolino}, Andrea and {Jin}, Harim and {Langer}, Norbert and {Gal-Yam}, Avishay and {Schootemeijer}, Abel and {Mannes}, Caroline},
        title = "{The demographics of core-collapse supernovae: The role of binary evolution and interaction with the circumstellar medium}",
      journal = {\aap},
         year = 2026,
        month = feb,
       volume = {706},
          eid = {A169},
        pages = {A169},
          doi = {10.1051/0004-6361/202557572},
archivePrefix = {arXiv},
       eprint = {2510.04872},
 primaryClass = {astro-ph.SR},
       adsurl = {https://ui.adsabs.harvard.edu/abs/2026A&A...706A.169E}
}

@ARTICLE{2021A&A...652A..76H,
       author = {{Hodgkin}, S.~T. and {Harrison}, D.~L. and {Breedt}, E. and {Wevers}, T. and {Rixon}, G. and {Delgado}, A. and {Yoldas}, A. and {Kostrzewa-Rutkowska}, Z. and {Wyrzykowski}, {\L}. and {van Leeuwen}, M. and {Blagorodnova}, N. and {Campbell}, H. and {Eappachen}, D. and {Fraser}, M. and {Ihanec}, N. and {Koposov}, S.~E. and {Kruszy{\'n}ska}, K. and {Marton}, G. and {Rybicki}, K.~A. and {Brown}, A.~G.~A. and {Burgess}, P.~W. and {Busso}, G. and {Cowell}, S. and {De Angeli}, F. and {Diener}, C. and {Evans}, D.~W. and {Gilmore}, G. and {Holland}, G. and {Jonker}, P.~G. and {van Leeuwen}, F. and {Mignard}, F. and {Osborne}, P.~J. and {Portell}, J. and {Prusti}, T. and {Richards}, P.~J. and {Riello}, M. and {Seabroke}, G.~M. and {Walton}, N.~A. and {{\'A}brah{\'a}m}, P. and {Altavilla}, G. and {Baker}, S.~G. and {Bastian}, U. and {O'Brien}, P. and {de Bruijne}, J. and {Butterley}, T. and {Carrasco}, J.~M. and {Casta{\~n}eda}, J. and {Clark}, J.~S. and {Clementini}, G. and {Copperwheat}, C.~M. and {Cropper}, M. and {Damljanovic}, G. and {Davidson}, M. and {Davis}, C.~J. and {Dennefeld}, M. and {Dhillon}, V.~S. and {Dolding}, C. and {Dominik}, M. and {Esquej}, P. and {Eyer}, L. and {Fabricius}, C. and {Fridman}, M. and {Froebrich}, D. and {Garralda}, N. and {Gomboc}, A. and {Gonz{\'a}lez-Vidal}, J.~J. and {Guerra}, R. and {Hambly}, N.~C. and {Hardy}, L.~K. and {Holl}, B. and {Hourihane}, A. and {Japelj}, J. and {Kann}, D.~A. and {Kiss}, C. and {Knigge}, C. and {Kolb}, U. and {Komossa}, S. and {K{\'o}sp{\'a}l}, {\'A}. and {Kov{\'a}cs}, G. and {Kun}, M. and {Leto}, G. and {Lewis}, F. and {Littlefair}, S.~P. and {Mahabal}, A.~A. and {Mundell}, C.~G. and {Nagy}, Z. and {Padeletti}, D. and {Palaversa}, L. and {Pigulski}, A. and {Pretorius}, M.~L. and {van Reeven}, W. and {Ribeiro}, V.~A.~R.~M. and {Roelens}, M. and {Rowell}, N. and {Schartel}, N. and {Scholz}, A. and {Schwope}, A. and {Sip{\H{o}}cz}, B.~M. and {Smartt}, S.~J. and {Smith}, M.~D. and {Serraller}, I. and {Steeghs}, D. and {Sullivan}, M. and {Szabados}, L. and {Szegedi-Elek}, E. and {Tisserand}, P. and {Tomasella}, L. and {van Velzen}, S. and {Whitelock}, P.~A. and {Wilson}, R.~W. and {Young}, D.~R.},
        title = "{Gaia Early Data Release 3. Gaia photometric science alerts}",
      journal = {\aap},
         year = 2021,
        month = aug,
       volume = {652},
          eid = {A76},
        pages = {A76},
          doi = {10.1051/0004-6361/202140735},
archivePrefix = {arXiv},
       eprint = {2106.01394},
 primaryClass = {astro-ph.IM},
       adsurl = {https://ui.adsabs.harvard.edu/abs/2021A&A...652A..76H}
}

@ARTICLE{2014ApJ...792...30M,
       author = {{Mainzer}, A. and {Bauer}, J. and {Cutri}, R.~M. and {Grav}, T. and {Masiero}, J. and {Beck}, R. and {Clarkson}, P. and {Conrow}, T. and {Dailey}, J. and {Eisenhardt}, P. and {Fabinsky}, B. and {Fajardo-Acosta}, S. and {Fowler}, J. and {Gelino}, C. and {Grillmair}, C. and {Heinrichsen}, I. and {Kendall}, M. and {Kirkpatrick}, J. Davy and {Liu}, F. and {Masci}, F. and {McCallon}, H. and {Nugent}, C.~R. and {Papin}, M. and {Rice}, E. and {Royer}, D. and {Ryan}, T. and {Sevilla}, P. and {Sonnett}, S. and {Stevenson}, R. and {Thompson}, D.~B. and {Wheelock}, S. and {Wiemer}, D. and {Wittman}, M. and {Wright}, E. and {Yan}, L.},
        title = "{Initial Performance of the NEOWISE Reactivation Mission}",
      journal = {\apj},
         year = 2014,
        month = sep,
       volume = {792},
       number = {1},
          eid = {30},
        pages = {30},
          doi = {10.1088/0004-637X/792/1/30},
archivePrefix = {arXiv},
       eprint = {1406.6025},
 primaryClass = {astro-ph.EP},
       adsurl = {https://ui.adsabs.harvard.edu/abs/2014ApJ...792...30M}
}

@ARTICLE{2005SSRv..120...95R,
       author = {{Roming}, Peter W.~A. and {Kennedy}, Thomas E. and {Mason}, Keith O. and {Nousek}, John A. and {Ahr}, Lindy and {Bingham}, Richard E. and {Broos}, Patrick S. and {Carter}, Mary J. and {Hancock}, Barry K. and {Huckle}, Howard E. and {Hunsberger}, S.~D. and {Kawakami}, Hajime and {Killough}, Ronnie and {Koch}, T. Scott and {McLelland}, Michael K. and {Smith}, Kelly and {Smith}, Philip J. and {Soto}, Juan Carlos and {Boyd}, Patricia T. and {Breeveld}, Alice A. and {Holland}, Stephen T. and {Ivanushkina}, Mariya and {Pryzby}, Michael S. and {Still}, Martin D. and {Stock}, Joseph},
        title = "{The Swift Ultra-Violet/Optical Telescope}",
      journal = {\ssr},
         year = 2005,
        month = oct,
       volume = {120},
       number = {3-4},
        pages = {95-142},
          doi = {10.1007/s11214-005-5095-4},
archivePrefix = {arXiv},
       eprint = {astro-ph/0507413},
 primaryClass = {astro-ph},
       adsurl = {https://ui.adsabs.harvard.edu/abs/2005SSRv..120...95R}
}

@ARTICLE{1998Msngr..91....9M,
       author = {{Moorwood}, A. and {Cuby}, J.-G. and {Lidman}, C.},
        title = "{SOFI sees first light at the NTT.}",
      journal = {The Messenger},
         year = 1998,
        month = mar,
       volume = {91},
        pages = {9-13},
       adsurl = {https://ui.adsabs.harvard.edu/abs/1998Msngr..91....9M}
}

@ARTICLE{2022ApJ...941L..32K,
       author = {{Kuncarayakti}, H. and {Maeda}, K. and {Dessart}, L. and {Nagao}, T. and {Fulton}, M. and {Guti{\'e}rrez}, C.~P. and {Huber}, M.~E. and {Young}, D.~R. and {Kotak}, R. and {Mattila}, S. and {Anderson}, J.~P. and {Ferrari}, L. and {Folatelli}, G. and {Gao}, H. and {Magnier}, E. and {Smith}, K.~W. and {Srivastav}, S.},
        title = "{Late-time H/He-poor Circumstellar Interaction in the Type Ic Supernova SN 2021ocs: An Exposed Oxygen-Magnesium Layer and Extreme Stripping of the Progenitor}",
      journal = {\apjl},
         year = 2022,
        month = dec,
       volume = {941},
       number = {2},
          eid = {L32},
        pages = {L32},
          doi = {10.3847/2041-8213/aca672},
archivePrefix = {arXiv},
       eprint = {2210.01755},
 primaryClass = {astro-ph.SR},
       adsurl = {https://ui.adsabs.harvard.edu/abs/2022ApJ...941L..32K}
}

@software{2023zndo...7872772H,
       author = {{Hosseinzadeh}, Griffin and {Bostroem}, K. Azalee and {Gomez}, Sebastian},
        title = "{Light Curve Fitting}",
         year = 2023,
        month = apr,
          eid = {10.5281/zenodo.7872772},
          doi = {10.5281/zenodo.7872772},
      version = {v0.8.0},
    publisher = {Zenodo},
       adsurl = {https://ui.adsabs.harvard.edu/abs/2023zndo...7872772H}
}

@ARTICLE{2014ApJ...780..172D,
       author = {{Draine}, B.~T. and {Aniano}, G. and {Krause}, Oliver and {Groves}, Brent and {Sandstrom}, Karin and {Braun}, Robert and {Leroy}, Adam and {Klaas}, Ulrich and {Linz}, Hendrik and {Rix}, Hans-Walter and et al.},
        title = "{Andromeda's Dust}",
      journal = {\apj},
         year = 2014,
        month = jan,
       volume = {780},
       number = {2},
          eid = {172},
        pages = {172},
          doi = {10.1088/0004-637X/780/2/172},
archivePrefix = {arXiv},
       eprint = {1306.2304},
 primaryClass = {astro-ph.CO},
       adsurl = {https://ui.adsabs.harvard.edu/abs/2014ApJ...780..172D}
}

@ARTICLE{2000ApJ...533..682C,
       author = {{Calzetti}, Daniela and {Armus}, Lee and {Bohlin}, Ralph C. and {Kinney}, Anne L. and {Koornneef}, Jan and {Storchi-Bergmann}, Thaisa},
        title = "{The Dust Content and Opacity of Actively Star-forming Galaxies}",
      journal = {\apj},
         year = 2000,
        month = apr,
       volume = {533},
       number = {2},
        pages = {682-695},
          doi = {10.1086/308692},
archivePrefix = {arXiv},
       eprint = {astro-ph/9911459},
 primaryClass = {astro-ph},
       adsurl = {https://ui.adsabs.harvard.edu/abs/2000ApJ...533..682C}
}

@ARTICLE{2003PASP..115..763C,
       author = {{Chabrier}, Gilles},
        title = "{Galactic Stellar and Substellar Initial Mass Function}",
      journal = {\pasp},
         year = 2003,
        month = jul,
       volume = {115},
       number = {809},
        pages = {763-795},
          doi = {10.1086/376392},
archivePrefix = {arXiv},
       eprint = {astro-ph/0304382},
 primaryClass = {astro-ph},
       adsurl = {https://ui.adsabs.harvard.edu/abs/2003PASP..115..763C}
}

@ARTICLE{2003MNRAS.344.1000B,
       author = {{Bruzual}, G. and {Charlot}, S.},
        title = "{Stellar population synthesis at the resolution of 2003}",
      journal = {\mnras},
         year = 2003,
        month = oct,
       volume = {344},
       number = {4},
        pages = {1000-1028},
          doi = {10.1046/j.1365-8711.2003.06897.x},
archivePrefix = {arXiv},
       eprint = {astro-ph/0309134},
 primaryClass = {astro-ph},
       adsurl = {https://ui.adsabs.harvard.edu/abs/2003MNRAS.344.1000B}
}

@ARTICLE{2004MNRAS.348L..59P,
       author = {{Pettini}, Max and {Pagel}, Bernard E.~J.},
        title = "{[OIII]/[NII] as an abundance indicator at high redshift}",
      journal = {\mnras},
         year = 2004,
        month = mar,
       volume = {348},
       number = {3},
        pages = {L59-L63},
          doi = {10.1111/j.1365-2966.2004.07591.x},
archivePrefix = {arXiv},
       eprint = {astro-ph/0401128},
 primaryClass = {astro-ph},
       adsurl = {https://ui.adsabs.harvard.edu/abs/2004MNRAS.348L..59P}
}

@ARTICLE{2016Ap&SS.361...61D,
       author = {{Dopita}, Michael A. and {Kewley}, Lisa J. and {Sutherland}, Ralph S. and {Nicholls}, David C.},
        title = "{Chemical abundances in high-redshift galaxies: a powerful new emission line diagnostic}",
      journal = {\apss},
         year = 2016,
        month = feb,
       volume = {361},
          eid = {61},
        pages = {61},
          doi = {10.1007/s10509-016-2657-8},
archivePrefix = {arXiv},
       eprint = {1601.01337},
 primaryClass = {astro-ph.GA},
       adsurl = {https://ui.adsabs.harvard.edu/abs/2016Ap&SS.361...61D}
}

@ARTICLE{2009ARA&A..47..481A,
       author = {{Asplund}, Martin and {Grevesse}, Nicolas and {Sauval}, A. Jacques and {Scott}, Pat},
        title = "{The Chemical Composition of the Sun}",
      journal = {\araa},
         year = 2009,
        month = sep,
       volume = {47},
       number = {1},
        pages = {481-522},
          doi = {10.1146/annurev.astro.46.060407.145222},
archivePrefix = {arXiv},
       eprint = {0909.0948},
 primaryClass = {astro-ph.SR},
       adsurl = {https://ui.adsabs.harvard.edu/abs/2009ARA&A..47..481A}
}

@ARTICLE{2026arXiv260630390C,
       author = {{Cai}, Y.-Z. and {Pastorello}, A. and {Moriya}, T.~J. and {Wang}, X.-F. and {Reguitti}, A. and {Filippenko}, A.~V. and {Tomasella}, L. and {{\v{C}}otar}, K. and {Siviero}, A. and {Elias-Rosa}, N. and {Brink}, T.~G. and {Valerin}, G. and {Benetti}, S. and {Zhao}, J.-W. and {Peng}, Z.-H. and {Wang}, Z.-Y. and {Altunin}, I. and {Baer-Way}, R. and {Baron}, E. and {Chander}, V. and {Chu}, M. and {deGraw}, A. and {DerKacy}, J.~M. and {Isern}, J. and {Jennings}, C. and {Kotak}, R. and {Li}, L.-P. and {Marziani}, P. and {May}, M. and {Mazzali}, P.~A. and {Morales-Garoffolo}, A. and {Moran}, S. and {Ochner}, P. and {Salmaso}, I. and {Tartaglia}, L. and {Turatto}, M. and {Wu}, H.-Y. and {Xiang}, D.-F. and {Yan}, S.-Y. and {Zhang}, J.-J. and {Zheng}, W.},
        title = "{SN 2019vxm: A luminous and long-lived Type IIn supernova with early flash-ionisation features}",
      journal = {arXiv e-prints},
         year = 2026,
        month = jun,
          eid = {arXiv:2606.30390},
        pages = {arXiv:2606.30390},
          doi = {10.48550/arXiv.2606.30390},
archivePrefix = {arXiv},
       eprint = {2606.30390},
 primaryClass = {astro-ph.SR},
       adsurl = {https://ui.adsabs.harvard.edu/abs/2026arXiv260630390C}
}

@ARTICLE{2020A&A...635A..39T,
       author = {{Tartaglia}, L. and {Pastorello}, A. and {Sollerman}, J. and {Fransson}, C. and {Mattila}, S. and {Fraser}, M. and {Taddia}, F. and {Tomasella}, L. and {Turatto}, M. and {Morales-Garoffolo}, A. and {Elias-Rosa}, N. and {Lundqvist}, P. and {Harmanen}, J. and {Reynolds}, T. and {Cappellaro}, E. and {Barbarino}, C. and {Nyholm}, A. and {Kool}, E. and {Ofek}, E. and {Gao}, X. and {Jin}, Z. and {Tan}, H. and {Sand}, D.~J. and {Ciabattari}, F. and {Wang}, X. and {Zhang}, J. and {Huang}, F. and {Li}, W. and {Mo}, J. and {Rui}, L. and {Xiang}, D. and {Zhang}, T. and {Hosseinzadeh}, G. and {Howell}, D.~A. and {McCully}, C. and {Valenti}, S. and {Benetti}, S. and {Callis}, E. and {Carracedo}, A.~S. and {Fremling}, C. and {Kangas}, T. and {Rubin}, A. and {Somero}, A. and {Terreran}, G.},
        title = "{The long-lived Type IIn SN 2015da: Infrared echoes and strong interaction within an extended massive shell}",
      journal = {\aap},
         year = 2020,
        month = mar,
       volume = {635},
          eid = {A39},
        pages = {A39},
          doi = {10.1051/0004-6361/201936553},
archivePrefix = {arXiv},
       eprint = {1908.08580},
 primaryClass = {astro-ph.HE},
       adsurl = {https://ui.adsabs.harvard.edu/abs/2020A&A...635A..39T}
}

@ARTICLE{2014MNRAS.441.2230H,
       author = {{Habergham}, S.~M. and {Anderson}, J.~P. and {James}, P.~A. and {Lyman}, J.~D.},
        title = "{Environments of interacting transients: impostors and Type IIn supernovae}",
      journal = {\mnras},
         year = 2014,
        month = jul,
       volume = {441},
       number = {3},
        pages = {2230-2252},
          doi = {10.1093/mnras/stu684},
archivePrefix = {arXiv},
       eprint = {1404.2282},
 primaryClass = {astro-ph.HE},
       adsurl = {https://ui.adsabs.harvard.edu/abs/2014MNRAS.441.2230H}
}

@ARTICLE{2016MNRAS.463.2904S,
       author = {{Smith}, Nathan and {Andrews}, Jennifer E. and {Mauerhan}, Jon C.},
        title = "{Massive stars dying alone: the extremely remote environment of SN 2009ip}",
      journal = {\mnras},
         year = 2016,
        month = dec,
       volume = {463},
       number = {3},
        pages = {2904-2911},
          doi = {10.1093/mnras/stw2190},
archivePrefix = {arXiv},
       eprint = {1607.01056},
 primaryClass = {astro-ph.SR},
       adsurl = {https://ui.adsabs.harvard.edu/abs/2016MNRAS.463.2904S}
}

@ARTICLE{2011ApJ...730...34S,
       author = {{Stoll}, R. and {Prieto}, J.~L. and {Stanek}, K.~Z. and {Pogge}, R.~W. and {Szczygie{\l}}, D.~M. and {Pojma{\'n}ski}, G. and {Antognini}, J. and {Yan}, H.},
        title = "{SN 2010jl in UGC 5189: Yet Another Luminous Type IIn Supernova in a Metal-poor Galaxy}",
      journal = {\apj},
         year = 2011,
        month = mar,
       volume = {730},
       number = {1},
          eid = {34},
        pages = {34},
          doi = {10.1088/0004-637X/730/1/34},
archivePrefix = {arXiv},
       eprint = {1012.3461},
 primaryClass = {astro-ph.CO},
       adsurl = {https://ui.adsabs.harvard.edu/abs/2011ApJ...730...34S}
}

@ARTICLE{2014A&A...564A.125B,
       author = {{Buchner}, J. and {Georgakakis}, A. and {Nandra}, K. and {Hsu}, L. and {Rangel}, C. and {Brightman}, M. and {Merloni}, A. and {Salvato}, M. and {Donley}, J. and {Kocevski}, D.},
        title = "{X-ray spectral modelling of the AGN obscuring region in the CDFS: Bayesian model selection and catalogue}",
      journal = {\aap},
         year = 2014,
        month = apr,
       volume = {564},
          eid = {A125},
        pages = {A125},
          doi = {10.1051/0004-6361/201322971},
archivePrefix = {arXiv},
       eprint = {1402.0004},
 primaryClass = {astro-ph.HE},
       adsurl = {https://ui.adsabs.harvard.edu/abs/2014A&A...564A.125B}
}

@ARTICLE{2009MNRAS.398.1601F,
       author = {{Feroz}, F. and {Hobson}, M.~P. and {Bridges}, M.},
        title = "{MULTINEST: an efficient and robust Bayesian inference tool for cosmology and particle physics}",
      journal = {\mnras},
         year = 2009,
        month = oct,
       volume = {398},
       number = {4},
        pages = {1601-1614},
          doi = {10.1111/j.1365-2966.2009.14548.x},
archivePrefix = {arXiv},
       eprint = {0809.3437},
 primaryClass = {astro-ph},
       adsurl = {https://ui.adsabs.harvard.edu/abs/2009MNRAS.398.1601F}
}

@ARTICLE{2019ApJS..241...27A,
       author = {{Ashton}, Gregory and {H{\"u}bner}, Moritz and {Lasky}, Paul D. and {Talbot}, Colm and {Ackley}, Kendall and {Biscoveanu}, Sylvia and {Chu}, Qi and {Divakarla}, Atul and {Easter}, Paul J. and {Goncharov}, Boris and {Hernandez Vivanco}, Francisco and {Harms}, Jan and {Lower}, Marcus E. and {Meadors}, Grant D. and {Melchor}, Denyz and {Payne}, Ethan and {Pitkin}, Matthew D. and {Powell}, Jade and {Sarin}, Nikhil and {Smith}, Rory J.~E. and {Thrane}, Eric},
        title = "{BILBY: A User-friendly Bayesian Inference Library for Gravitational-wave Astronomy}",
      journal = {\apjs},
         year = 2019,
        month = apr,
       volume = {241},
       number = {2},
          eid = {27},
        pages = {27},
          doi = {10.3847/1538-4365/ab06fc},
archivePrefix = {arXiv},
       eprint = {1811.02042},
 primaryClass = {astro-ph.IM},
       adsurl = {https://ui.adsabs.harvard.edu/abs/2019ApJS..241...27A}
}

@ARTICLE{2026ApJ..1001..111H,
       author = {{Hillenkamp}, Elizabeth and {Baer-Way}, Raphael and {Chandra}, Poonam and {Sarangi}, Arkaprabha and {Chevalier}, Roger and {A.~J.}, Nayana and {Deutsch}, Annika and {Maeda}, Keiichi and {Smith}, Nathan},
        title = "{Fading Echoes of Interaction: Probing Centuries of Preexplosion Mass Loss in Four Type IIn Supernovae}",
      journal = {\apj},
         year = 2026,
        month = apr,
       volume = {1001},
       number = {1},
          eid = {111},
        pages = {111},
          doi = {10.3847/1538-4357/ae4e22},
archivePrefix = {arXiv},
       eprint = {2601.19891},
 primaryClass = {astro-ph.HE},
       adsurl = {https://ui.adsabs.harvard.edu/abs/2026ApJ..1001..111H}
}

@ARTICLE{2026arXiv260708693C,
       author = {{Chugai}, Nikolai},
        title = "{H{\ensuremath{\alpha}} clarifies picture of IIn supernova SN 2015da}",
      journal = {arXiv e-prints},
         year = 2026,
        month = jul,
          eid = {arXiv:2607.08693},
        pages = {arXiv:2607.08693},
          doi = {10.48550/arXiv.2607.08693},
archivePrefix = {arXiv},
       eprint = {2607.08693},
 primaryClass = {astro-ph.HE},
       adsurl = {https://ui.adsabs.harvard.edu/abs/2026arXiv260708693C}
}

@ARTICLE{2024MNRAS.530..405S,
       author = {{Smith}, Nathan and {Andrews}, Jennifer E. and {Milne}, Peter and {Filippenko}, Alexei V. and {Brink}, Thomas G. and {Kelly}, Patrick L. and {Yuk}, Heechan and {Jencson}, Jacob E.},
        title = "{SN 2015da: late-time observations of a persistent superluminous Type IIn supernova with post-shock dust formation}",
      journal = {\mnras},
         year = 2024,
        month = may,
       volume = {530},
       number = {1},
        pages = {405-423},
          doi = {10.1093/mnras/stae726},
archivePrefix = {arXiv},
       eprint = {2312.00253},
 primaryClass = {astro-ph.HE},
       adsurl = {https://ui.adsabs.harvard.edu/abs/2024MNRAS.530..405S}
}

@ARTICLE{2024MNRAS.529.1104B,
       author = {{Bilinski}, Christopher and {Smith}, Nathan and {Williams}, G. Grant and {Smith}, Paul S. and {Leonard}, Douglas C. and {Hoffman}, Jennifer L. and {Andrews}, Jennifer E. and {Milne}, Peter},
        title = "{Multi-epoch spectropolarimetry for a sample of Type IIn Supernovae: persistent asymmetry in dusty circumstellar material}",
      journal = {\mnras},
         year = 2024,
        month = apr,
       volume = {529},
       number = {2},
        pages = {1104-1129},
          doi = {10.1093/mnras/stae380},
archivePrefix = {arXiv},
       eprint = {2304.13034},
 primaryClass = {astro-ph.HE},
       adsurl = {https://ui.adsabs.harvard.edu/abs/2024MNRAS.529.1104B}
}

@ARTICLE{2011A&A...527L...6P,
       author = {{Patat}, F. and {Taubenberger}, S. and {Benetti}, S. and {Pastorello}, A. and {Harutyunyan}, A.},
        title = "{Asymmetries in the type IIn SN 2010jl}",
      journal = {\aap},
         year = 2011,
        month = mar,
       volume = {527},
          eid = {L6},
        pages = {L6},
          doi = {10.1051/0004-6361/201016217},
archivePrefix = {arXiv},
       eprint = {1011.5926},
 primaryClass = {astro-ph.CO},
       adsurl = {https://ui.adsabs.harvard.edu/abs/2011A&A...527L...6P}
}

@ARTICLE{2011A&A...527A.116C,
       author = {{Cioni}, M.-R.~L. and {Clementini}, G. and {Girardi}, L. and {Guandalini}, R. and {Gullieuszik}, M. and {Miszalski}, B. and {Moretti}, M.-I. and {Ripepi}, V. and {Rubele}, S. and {Bagheri}, G. and {Bekki}, K. and {Cross}, N. and {de Blok}, W.~J.~G. and {de Grijs}, R. and {Emerson}, J.~P. and {Evans}, C.~J. and {Gibson}, B. and {Gonzales-Solares}, E. and {Groenewegen}, M.~A.~T. and {Irwin}, M. and {Ivanov}, V.~D. and {Lewis}, J. and {Marconi}, M. and {Marquette}, J.-B. and {Mastropietro}, C. and {Moore}, B. and {Napiwotzki}, R. and {Naylor}, T. and {Oliveira}, J.~M. and {Read}, M. and {Sutorius}, E. and {van Loon}, J. Th. and {Wilkinson}, M.~I. and {Wood}, P.~R.},
        title = "{The VMC survey. I. Strategy and first data}",
      journal = {\aap},
         year = 2011,
        month = mar,
       volume = {527},
          eid = {A116},
        pages = {A116},
          doi = {10.1051/0004-6361/201016137},
archivePrefix = {arXiv},
       eprint = {1012.5193},
 primaryClass = {astro-ph.CO},
       adsurl = {https://ui.adsabs.harvard.edu/abs/2011A&A...527A.116C}
}

@ARTICLE{2015PASA...32...16S,
       author = {{Smartt}, S.~J.},
        title = "{Observational Constraints on the Progenitors of Core-Collapse Supernovae: The Case for Missing High-Mass Stars}",
      journal = {\pasa},
         year = 2015,
        month = apr,
       volume = {32},
          eid = {e016},
        pages = {e016},
          doi = {10.1017/pasa.2015.17},
archivePrefix = {arXiv},
       eprint = {1504.02635},
 primaryClass = {astro-ph.SR},
       adsurl = {https://ui.adsabs.harvard.edu/abs/2015PASA...32...16S}
}

@ARTICLE{2026ApJ..1006..224S,
       author = {{Smith}, Nathan and {Shahbandeh}, Melissa and {Fox}, Ori D. and {Moore}, Thomas and {Andrews}, Jennifer E. and {Brink}, Thomas G. and {Dwek}, Eli and {Engesser}, Michael and {Filippenko}, Alexei V. and {Nickson}, Bryony and {Temim}, Tea and {Yang}, Yi and {Zheng}, WeiKang and {Ashall}, Chris and {Baer-Way}, Raphael and {Clayton}, Geoffrey C. and {Jencson}, Jacob E. and {Johansson}, Joel and {Lane}, Zachary G. and {Milisavljevic}, Dan and {Rest}, Armin and {Sarangi}, Arkaprabha and {Szalai}, Tam{\'a}s and {Van Dyk}, Schuyler D. and {Williams}, Brian J.},
        title = "{JWST Spectra Indicate a Large Mass of Postshock Dust Formed by SN 2010jl}",
      journal = {\apj},
         year = 2026,
        month = aug,
       volume = {1006},
       number = {2},
          eid = {224},
        pages = {224},
          doi = {10.3847/1538-4357/ae74c5},
archivePrefix = {arXiv},
       eprint = {2607.27468},
 primaryClass = {astro-ph.HE},
       adsurl = {https://ui.adsabs.harvard.edu/abs/2026ApJ..1006..224S}
}

@ARTICLE{2018ApJ...859...66S,
       author = {{Sarangi}, Arkaprabha and {Dwek}, Eli and {Arendt}, Richard G.},
        title = "{Delayed Shock-induced Dust Formation in the Dense Circumstellar Shell Surrounding the Type IIn Supernova SN 2010jl}",
      journal = {\apj},
         year = 2018,
        month = may,
       volume = {859},
       number = {1},
          eid = {66},
        pages = {66},
          doi = {10.3847/1538-4357/aabfc3},
archivePrefix = {arXiv},
       eprint = {1804.06878},
 primaryClass = {astro-ph.SR},
       adsurl = {https://ui.adsabs.harvard.edu/abs/2018ApJ...859...66S}
}

@ARTICLE{2021ApJ...917...84D,
       author = {{Dwek}, Eli and {Sarangi}, Arkaprabha and {Arendt}, Richard G. and {Kallman}, Timothy and {Kazanas}, Demos and {Fox}, Ori D.},
        title = "{The Infrared Echo of SN2010jl and Its Implications for Shock Breakout Characteristics}",
      journal = {\apj},
         year = 2021,
        month = aug,
       volume = {917},
       number = {2},
          eid = {84},
        pages = {84},
          doi = {10.3847/1538-4357/ac09ea},
archivePrefix = {arXiv},
       eprint = {2106.06531},
 primaryClass = {astro-ph.HE},
       adsurl = {https://ui.adsabs.harvard.edu/abs/2021ApJ...917...84D}
}

@ARTICLE{2011AJ....142...45A,
       author = {{Andrews}, J.~E. and {Clayton}, Geoffrey C. and {Wesson}, R. and {Sugerman}, B.~E.~K. and {Barlow}, M.~J. and {Clem}, J. and {Ercolano}, B. and {Fabbri}, J. and {Gallagher}, J.~S. and {Landolt}, A. and {Meixner}, M. and {Otsuka}, M. and {Riebel}, D. and {Welch}, D.~L.},
        title = "{Evidence for Pre-existing Dust in the Bright Type IIn SN 2010jl}",
      journal = {\aj},
         year = 2011,
        month = aug,
       volume = {142},
       number = {2},
          eid = {45},
        pages = {45},
          doi = {10.1088/0004-6256/142/2/45},
archivePrefix = {arXiv},
       eprint = {1106.0537},
 primaryClass = {astro-ph.CO},
       adsurl = {https://ui.adsabs.harvard.edu/abs/2011AJ....142...45A}
}

@ARTICLE{2019AJ....157..168D,
       author = {{Dey}, Arjun and {Schlegel}, David J. and {Lang}, Dustin and {Blum}, Robert and {Burleigh}, Kaylan and {Fan}, Xiaohui and {Findlay}, Joseph R. and {Finkbeiner}, Doug and {Herrera}, David and {Juneau}, St{\'e}phanie and {Landriau}, Martin and {Levi}, Michael and {McGreer}, Ian and {Meisner}, Aaron and {Myers}, Adam D. and {Moustakas}, John and {Nugent}, Peter and {Patej}, Anna and {Schlafly}, Edward F. and {Walker}, Alistair R. and {Valdes}, Francisco and {Weaver}, Benjamin A. and {Y{\`e}che}, Christophe and {Zou}, Hu and {Zhou}, Xu and {Abareshi}, Behzad and {Abbott}, T.~M.~C. and {Abolfathi}, Bela and {Aguilera}, C. and {Alam}, Shadab and {Allen}, Lori and {Alvarez}, A. and {Annis}, James and {Ansarinejad}, Behzad and {Aubert}, Marie and {Beechert}, Jacqueline and {Bell}, Eric F. and {BenZvi}, Segev Y. and {Beutler}, Florian and {Bielby}, Richard M. and {Bolton}, Adam S. and {Brice{\~n}o}, C{\'e}sar and {Buckley-Geer}, Elizabeth J. and {Butler}, Karen and {Calamida}, Annalisa and {Carlberg}, Raymond G. and {Carter}, Paul and {Casas}, Ricard and {Castander}, Francisco J. and {Choi}, Yumi and {Comparat}, Johan and {Cukanovaite}, Elena and {Delubac}, Timoth{\'e}e and {DeVries}, Kaitlin and {Dey}, Sharmila and {Dhungana}, Govinda and {Dickinson}, Mark and {Ding}, Zhejie and {Donaldson}, John B. and {Duan}, Yutong and {Duckworth}, Christopher J. and {Eftekharzadeh}, Sarah and {Eisenstein}, Daniel J. and {Etourneau}, Thomas and {Fagrelius}, Parker A. and {Farihi}, Jay and {Fitzpatrick}, Mike and {Font-Ribera}, Andreu and {Fulmer}, Leah and {G{\"a}nsicke}, Boris T. and {Gaztanaga}, Enrique and {George}, Koshy and {Gerdes}, David W. and {Gontcho}, Satya Gontcho A. and {Gorgoni}, Claudio and {Green}, Gregory and {Guy}, Julien and {Harmer}, Diane and {Hernandez}, M. and {Honscheid}, Klaus and {Huang}, Lijuan Wendy and {James}, David J. and {Jannuzi}, Buell T. and {Jiang}, Linhua and {Joyce}, Richard and {Karcher}, Armin and {Karkar}, Sonia and {Kehoe}, Robert and {Kneib}, Jean-Paul and {Kueter-Young}, Andrea and {Lan}, Ting-Wen and {Lauer}, Tod R. and {Le Guillou}, Laurent and {Le Van Suu}, Auguste and {Lee}, Jae Hyeon and {Lesser}, Michael and {Perreault Levasseur}, Laurence and {Li}, Ting S. and {Mann}, Justin L. and {Marshall}, Robert and {Mart{\'\i}nez-V{\'a}zquez}, C.~E. and {Martini}, Paul and {du Mas des Bourboux}, H{\'e}lion and {McManus}, Sean and {Meier}, Tobias Gabriel and {M{\'e}nard}, Brice and {Metcalfe}, Nigel and {Mu{\~n}oz-Guti{\'e}rrez}, Andrea and {Najita}, Joan and {Napier}, Kevin and {Narayan}, Gautham and {Newman}, Jeffrey A. and {Nie}, Jundan and {Nord}, Brian and {Norman}, Dara J. and {Olsen}, Knut A.~G. and {Paat}, Anthony and {Palanque-Delabrouille}, Nathalie and {Peng}, Xiyan and {Poppett}, Claire L. and {Poremba}, Megan R. and {Prakash}, Abhishek and {Rabinowitz}, David and {Raichoor}, Anand and {Rezaie}, Mehdi and {Robertson}, A.~N. and {Roe}, Natalie A. and {Ross}, Ashley J. and {Ross}, Nicholas P. and {Rudnick}, Gregory and {Safonova}, Sasha and {Saha}, Abhijit and {S{\'a}nchez}, F. Javier and {Savary}, Elodie and {Schweiker}, Heidi and {Scott}, Adam and {Seo}, Hee-Jong and {Shan}, Huanyuan and {Silva}, David R. and {Slepian}, Zachary and {Soto}, Christian and {Sprayberry}, David and {Staten}, Ryan and {Stillman}, Coley M. and {Stupak}, Robert J. and {Summers}, David L. and {Sien Tie}, Suk and {Tirado}, H. and {Vargas-Maga{\~n}a}, Mariana and {Vivas}, A. Katherina and {Wechsler}, Risa H. and {Williams}, Doug and {Yang}, Jinyi and {Yang}, Qian and {Yapici}, Tolga and {Zaritsky}, Dennis and {Zenteno}, A. and {Zhang}, Kai and {Zhang}, Tianmeng and {Zhou}, Rongpu and {Zhou}, Zhimin},
        title = "{Overview of the DESI Legacy Imaging Surveys}",
      journal = {\aj},
         year = 2019,
        month = may,
       volume = {157},
       number = {5},
          eid = {168},
        pages = {168},
          doi = {10.3847/1538-3881/ab089d},
archivePrefix = {arXiv},
       eprint = {1804.08657},
 primaryClass = {astro-ph.IM},
       adsurl = {https://ui.adsabs.harvard.edu/abs/2019AJ....157..168D}
}

@INPROCEEDINGS{Burgh2003,
       author = {{Burgh}, Eric B. and {Nordsieck}, Kenneth H. and
{Kobulnicky}, Henry A. and
         {Williams}, Ted B. and {O'Donoghue}, Darragh and {Smith},
Michael P. and
         {Percival}, Jeffrey W.},
        title = "{Prime Focus Imaging Spectrograph for the Southern
African Large Telescope: optical design}",
    booktitle = {Instrument Design and Performance for
Optical/Infrared Ground-based Telescopes},
         year = 2003,
       editor = {{Iye}, Masanori and {Moorwood}, Alan F.~M.},
       series = {Society of Photo-Optical Instrumentation Engineers
(SPIE) Conference Series},
       volume = {4841},
        month = mar,
        pages = {1463-1471},
          doi = {10.1117/12.460312},
       adsurl = {https://ui.adsabs.harvard.edu/abs/2003SPIE.4841.1463B}
}

@INPROCEEDINGS{Buckley2006,
       author = {{Buckley}, David A.~H. and {Swart}, Gerhard P. and
{Meiring}, Jacobus G.},
        title = "{Completion and commissioning of the Southern African
Large Telescope}",
    booktitle = {Society of Photo-Optical Instrumentation Engineers
(SPIE) Conference Series},
         year = 2006,
       series = {Society of Photo-Optical Instrumentation Engineers
(SPIE) Conference Series},
       volume = {6267},
        month = jun,
          eid = {62670Z},
        pages = {62670Z},
          doi = {10.1117/12.673750},
       adsurl = {https://ui.adsabs.harvard.edu/abs/2006SPIE.6267E..0ZB}
}

@INPROCEEDINGS{Crawford2010,
   author = {{Crawford}, S.~M. and {Still}, M. and {Schellart}, P. and
{Balona}, L. and
    {Buckley}, D.~A.~H. and {Dugmore}, G. and {Gulbis}, A.~A.~S. and
    {Kniazev}, A. and {Kotze}, M. and {Loaring}, N. and {Nordsieck}, K.~H. and
    {Pickering}, T.~E. and {Potter}, S. and {Romero Colmenero}, E. and
    {Vaisanen}, P. and {Williams}, T. and {Zietsman}, E.},
    title = "{PySALT: the SALT science pipeline}",
booktitle = {Observatory Operations: Strategies, Processes, and Systems III},
     year = 2010,
   series = {\procspie},
   volume = 7737,
    month = jul,
      eid = {773725},
    pages = {773725},
      doi = {10.1117/12.857000},
   adsurl = {http://adsabs.harvard.edu/abs/2010SPIE.7737E..25C}
}

\begin{appendix}
\renewcommand{\topfraction}{0.95}\renewcommand{\bottomfraction}{0.9}\renewcommand{\textfraction}{0.05}
\renewcommand{\floatpagefraction}{0.8}\renewcommand{\dbltopfraction}{0.95}\renewcommand{\dblfloatpagefraction}{0.7}
\setcounter{topnumber}{3}\setcounter{dbltopnumber}{3}\setcounter{totalnumber}{5}

\section{Photometry, explosion epoch and bolometric light curve} \label{app:phot}

\begin{figure}[!h]
\centering
\includegraphics[width=\linewidth]{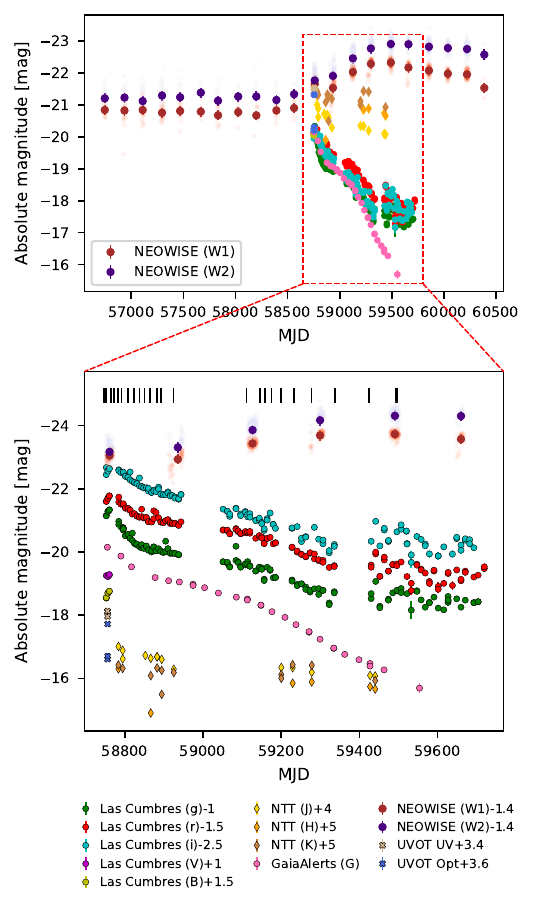}  
\caption{The upper panel shows the absolute magnitude light curve of SN 2019qem along with very long WISE data. The second panel shows the zoomed optical light curve of SN 2019qem. Black vertical lines represent the spectroscopic observations. For clarity, each band is shifted by the constant given in the legend. The magnitudes are corrected for Galactic extinction only. The photometry in this figure is not host-subtracted, so the late-time optical points are dominated by the host galaxy (see Fig. \ref{fig:diff_lc} for the host-subtracted light curves). The mid-infrared excess is analysed in Sect.~\ref{subsec:mir}.}
\label{fig:lightcurve}
\end{figure}

\begin{figure}
    \centering
    \includegraphics[width=0.99\linewidth]{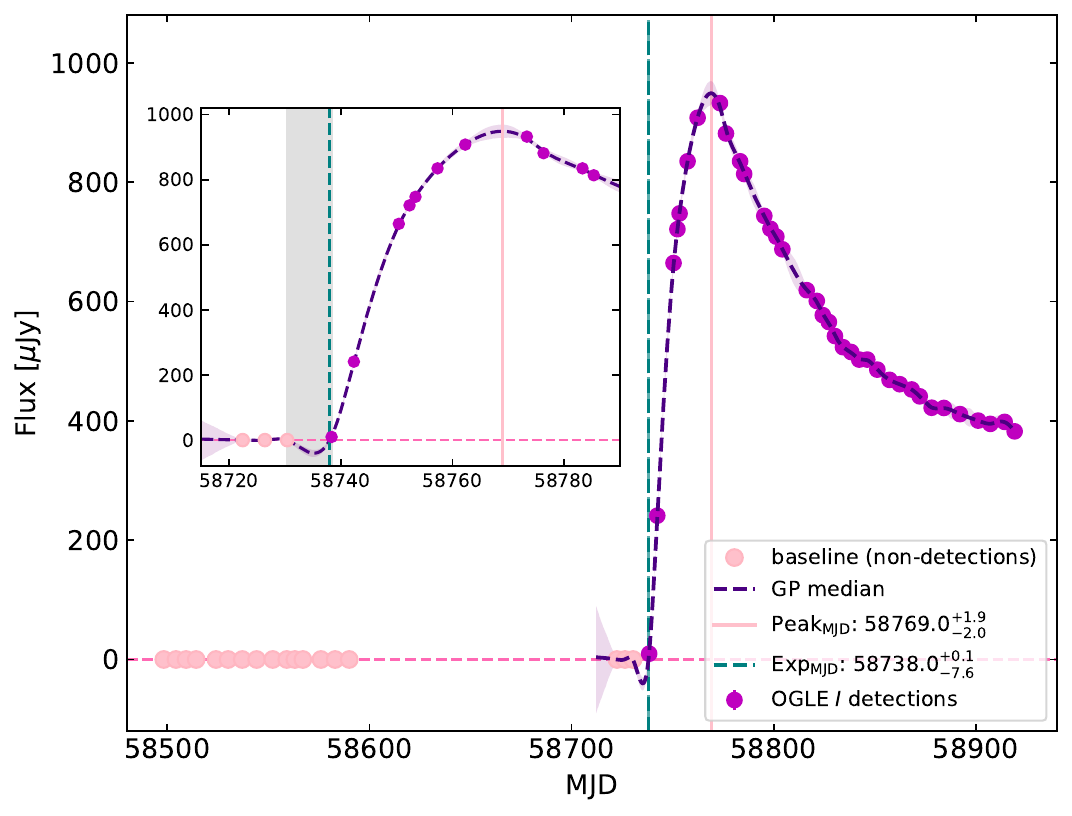}
    \caption{Gaussian-process fit to the OGLE $I$-band flux light curve. Pink points are the pre-explosion non-detections used as the baseline, and purple points the detections. The dashed line is the GP median with its uncertainty band; the vertical lines mark the peak (MJD $58\,769.0^{+1.9}_{-2.0}$) and the explosion epoch (MJD $58\,738.0^{+0.1}_{-7.6}$). The inset zooms on the rise. Before the explosion the band extends below zero flux, reflecting the uncertainty of the baseline.}
    \label{fig:ogle_gp_fit}
\end{figure}

\begin{figure}
\centering
\includegraphics[width=0.99\linewidth]{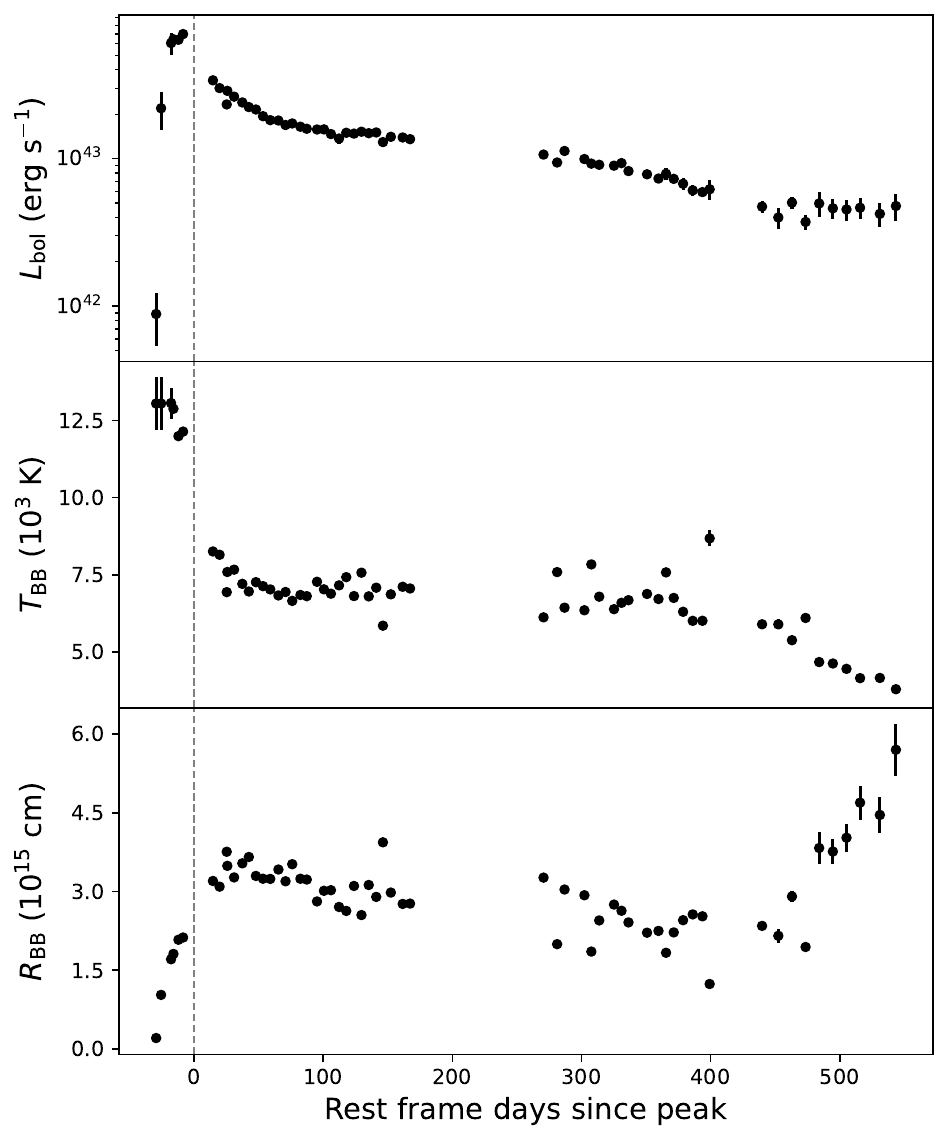} 
\caption{Blackbody bolometric light curve of SN 2019qem from the host-subtracted $g$, $r$, $i$ and OGLE $I$ photometry, with the luminosity (top), blackbody temperature (middle) and radius (bottom). The three points before $-16$\,d are OGLE-only epochs with the other bands reconstructed from the $-16$\,d colours. The dashed line marks the $I$-band maximum.}
\label{fig:bolo_lc}
\end{figure}

Figure~\ref{fig:lightcurve} shows the multi-band light curves without host subtraction, Fig.~\ref{fig:ogle_gp_fit} the Gaussian-process fit to the OGLE $I$-band light curve that gives the epochs of explosion and maximum, and Fig.~\ref{fig:bolo_lc} the blackbody bolometric light curve with its temperature and radius (Sect.~\ref{subsec:lightcurve}).

\FloatBarrier

\section{Host galaxy SED fit} \label{app:host}

\begin{figure}
    \centering
    \includegraphics[width=0.99\linewidth]{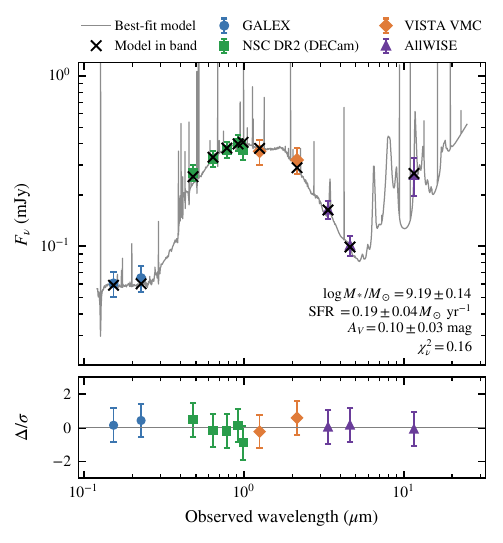}
    \caption{Pre-explosion SED of the host of SN 2019qem with the best-fit \texttt{CIGALE} model; the parameters in the lower right are from this fit. The lower panel shows the residuals in units of the total uncertainty.}
    \label{fig:host_sed}
\end{figure}

Figure~\ref{fig:host_sed} shows the pre-explosion SED of the host with the best-fit \texttt{CIGALE} model, and Table~\ref{tab:cigale} lists the parameter grid of the fit (Sect.~\ref{subsec:host}).

\begin{table}[ht]
\caption{\texttt{CIGALE} modules and parameter grid used for the host SED fit (Sect.~\ref{subsec:host}).}
\label{tab:cigale}
\centering
\footnotesize
\setlength{\tabcolsep}{4pt}
\begin{tabular}{ll}
\hline\hline
Parameter & Values \\
\hline
\multicolumn{2}{l}{Delayed-$\tau$ star-formation history with a recent burst} \\
\quad $\tau_{\rm main}$ (Gyr) & 0.1, 0.3, 1, 2, 4, 8 \\
\quad Main population age (Gyr) & 0.5, 1, 2, 3, 5, 8, 12 \\
\quad $\tau_{\rm burst}$, burst age (Myr) & 50, 50 \\
\quad Burst mass fraction & 0, 0.02, 0.10 \\
\multicolumn{2}{l}{Stellar populations \citep{2003MNRAS.344.1000B}} \\
\quad Initial mass function & \citet{2003PASP..115..763C} \\
\quad Metallicity $Z$ & 0.0004, 0.004, 0.008, 0.02, 0.05 \\
\quad Separation age (Myr) & 10 \\
\multicolumn{2}{l}{Nebular emission} \\
\quad $\log U$ & $-2.0$ \\
\quad Gas metallicity & 0.004, 0.008 \\
\quad $n_{\rm e}$ (cm$^{-3}$) & 100 \\
\quad $f_{\rm esc}$, $f_{\rm dust}$ & 0, 0 \\
\multicolumn{2}{l}{Modified \citet{2000ApJ...533..682C} attenuation} \\
\quad $E(B-V)_{\rm lines}$ (mag) & 0, 0.05, 0.1, 0.2, 0.4, 0.7 \\
\quad $E(B-V)_{\rm stars}/E(B-V)_{\rm lines}$ & 0.44 \\
\quad UV bump, power-law slope & 0, 0 \\
\multicolumn{2}{l}{Dust emission \citep{2014ApJ...780..172D}} \\
\quad $q_{\rm PAH}$ & 1.12, 2.50 \\
\quad $U_{\rm min}$ & 1, 5, 25 \\
\quad $\alpha$, $\gamma$ & 2.0, 0.02 \\
\hline
\end{tabular}
\tablefoot{The photometry covers GALEX FUV and NUV, DECam $grizY$ (NSC DR2), VISTA $J$ and $K_{\rm s}$, and WISE $W1$--$W3$, with an additional 10 per cent uncertainty added in quadrature. The redshift is fixed at $z=0.0427$, and the parameters are Bayesian estimates from the \texttt{CIGALE} likelihood-weighted analysis.}
\end{table}

\section{Light-curve model parameters} \label{app:params}

\begin{figure*}
\centering
\includegraphics[width=0.9\textwidth]{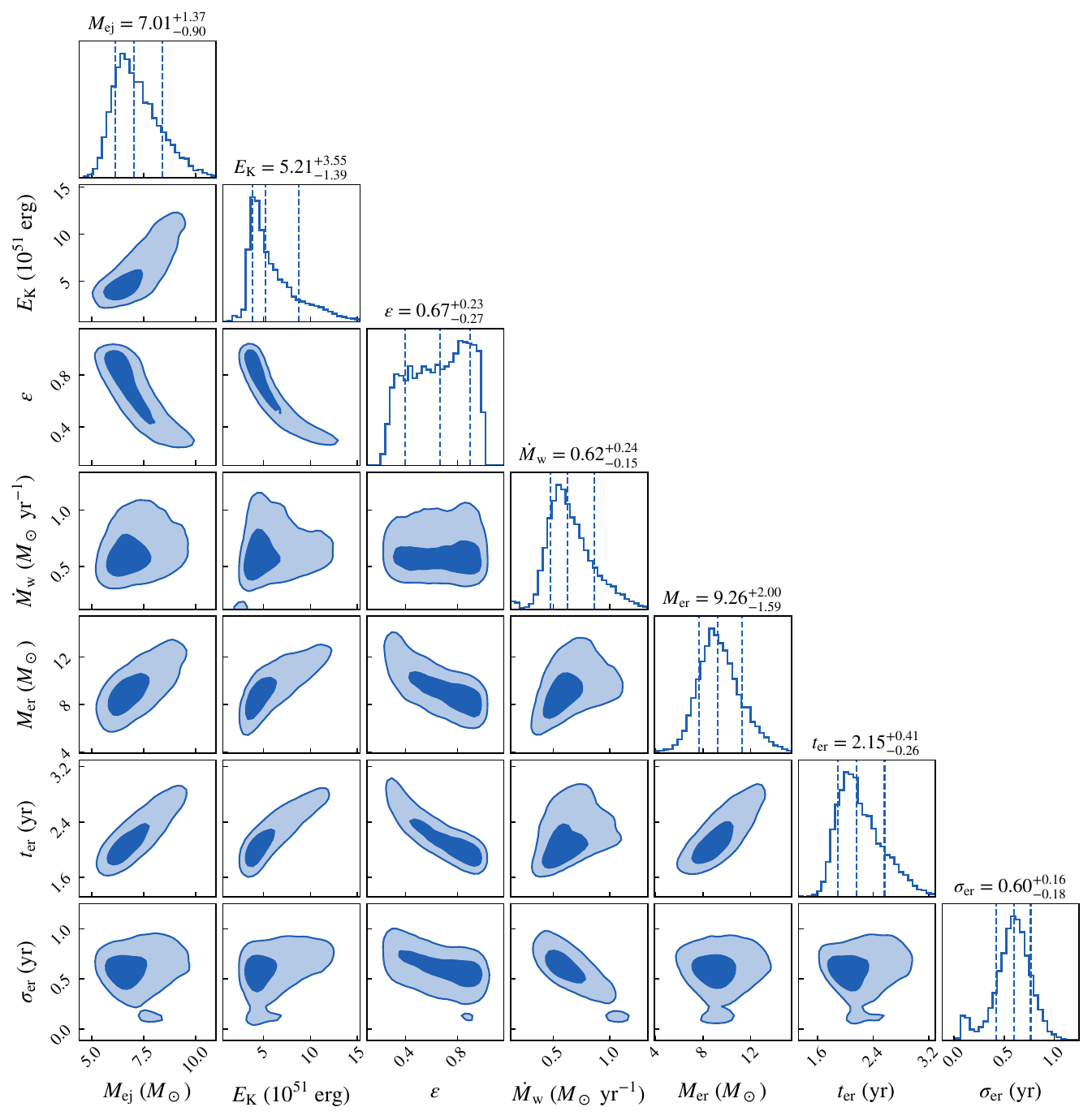}
\caption{Posterior of the physical parameters of the preferred model, namely the ejecta mass $M_{\rm ej}$, explosion energy $E_{\rm K}$, conversion efficiency $\epsilon$, wind mass-loss rate $\dot M_{\rm w}$, eruption mass $M_{\rm er}$, and the eruption epoch $t_{\rm er}$ and width $\sigma_{\rm er}$ in years before explosion. Contours enclose 39 and 86 per cent of the probability, and the titles give the median and the 16th and 84th percentiles. $\dot M_{\rm w}$, $t_{\rm er}$ and $\sigma_{\rm er}$ assume $v_w = 100$\,km\,s$^{-1}$; 1.8 per cent of the samples, with $t_{\rm er} < 1$\,yr, lie outside the axes. For skewed posteriors such as that of $E_{\rm K}$ the median lies above the peak of the distribution; the product $E_{\rm K}\epsilon = (3.5\pm0.3)\times10^{51}$\,erg is much better constrained.}
\label{fig:redback_corner}
\end{figure*}

Table~\ref{tab:redback} gives the priors and posteriors of the preferred \texttt{REDBACK} model, and Fig.~\ref{fig:redback_corner} shows its posterior distribution (Sect.~\ref{sec:discussion}).

\begin{table}[ht]
\caption{Priors and posteriors of the preferred \texttt{REDBACK} model (\texttt{gausswind\_bpl}).}
\label{tab:redback}
\centering
\small
\begin{tabular}{lll}
\hline\hline
Parameter & Prior & Posterior \\
\hline
$t_{\rm exp}$ (d from maximum) & $\mathcal{U}(-37.0, -29.6)$ & $-29.63^{+0.01}_{-0.02}$ \\
$M_{\rm ej}$ ($M_\odot$) & $\log\mathcal{U}(3, 40)$ & $7.0^{+1.4}_{-0.9}$ \\
$E_{\rm K}$ ($10^{51}$\,erg) & $\log\mathcal{U}(1, 20)$ & $5.2^{+3.6}_{-1.4}$ \\
$\epsilon$ & $\mathcal{U}(0.1, 1)$ & $0.67^{+0.23}_{-0.27}$ \\
$M_{\rm Ni}$ ($M_\odot$) & $\log\mathcal{U}(0.001, 0.1)$ & $<0.065$ (95\%) \\
$\dot M_{\rm w}$ ($M_\odot$\,yr$^{-1}$) & $\log\mathcal{U}(10^{-4}, 5)$ & $0.62^{+0.24}_{-0.15}$ \\
$M_{\rm er}$ ($M_\odot$) & $\log\mathcal{U}(0.01, 30)$ & $9.3^{+2.0}_{-1.6}$ \\
$t_{\rm er}$ (yr before explosion) & $\log\mathcal{U}(0.1, 100)$ & $2.15^{+0.41}_{-0.26}$ \\
$\sigma_{\rm er}$ (yr) & $\log\mathcal{U}(0.05, 30)$ & $0.60^{+0.16}_{-0.18}$ \\
$\sigma_{\rm sys}$ ($10^{41}$\,erg\,s$^{-1}$) & $\log\mathcal{U}$ & $8.6^{+1.3}_{-1.1}$ \\
\hline
\end{tabular}
\tablefoot{$\mathcal{U}$ and $\log\mathcal{U}$ denote uniform and log-uniform priors. The explosion-epoch prior spans the interval allowed by the OGLE photometry (Sect.~\ref{subsec:lightcurve}), and $\sigma_{\rm sys}$ is a scatter term added in quadrature to the luminosity errors, with a prior from $10^{-3}$ to 2 times the median luminosity. Fixed parameters are $v_w = 100$\,km\,s$^{-1}$, inner and outer ejecta density slopes $\delta = 1$ and $n = 11.5$, a maximum ejecta velocity of five times the transition velocity, and opacities $\kappa = 0.34$\,cm$^2$\,g$^{-1}$ and $\kappa_\gamma = 0.03$\,cm$^2$\,g$^{-1}$. The sampler used 2000 live points. The single power-law CSM model used the same explosion priors with $\rho\propto r^{\eta}$, $\eta\sim\mathcal{U}(-4, 0)$, $M_{\rm CSM}\sim\log\mathcal{U}(0.1, 50)\,M_\odot$, $\log(r_{\rm in}/{\rm cm})\sim\mathcal{U}(13, 15.5)$, and $r_{\rm out} = 10^{17}$\,cm. Posteriors are medians with 16th and 84th percentiles.}
\end{table}

\section{Spectroscopic log and \texorpdfstring{H$\alpha$}{H-alpha} fits} \label{app:spec}

Table~\ref{tab:log_spectra} lists the spectroscopic observations (Sect.~\ref{sec:observations}), and Table~\ref{tab:halpha_fits} gives the three-component fits to H$\alpha$ (Sect.~\ref{subsec:line_evo}).

\begin{table*}
    \centering
    \caption{Log of spectroscopic observations of SN 2019qem.}
    \footnotesize
    \renewcommand{\arraystretch}{0.9}
    \begin{tabular}{c c c c c c c}
    \toprule
    Date & MJD & Phase$^{\text{a}}$ & Instrument & Grism + slit & Spectral range & $\Delta\lambda$ \\
    && (days) &&&($\text{\AA}$)&($\text{\AA}$) \\
    \midrule
    2019-09-20 & 58\,746.50 & $-22$  & RSS-SALT   & PG0300+1.5$^{\prime\prime}$    & 3800 -- 8800  & 3.0 \\
    2019-09-25 & 58\,751.69 & $-17$  & FLOYDS-FTS & 2$^{\prime\prime}$             & 3294 -- 10174 & 2.3 \\
    2019-10-07 & 58\,763.50 & $-5$   & RSS-SALT   & PG0900+1.5$^{\prime\prime}$    & 4323 -- 7384  & 1.0 \\
    2019-10-08 & 58\,764.59 & $-4$   & FLOYDS-FTS & 2$^{\prime\prime}$             & 3297 -- 10180 & 2.3 \\
    2019-10-16 & 58\,772.65 & $+4$   & FLOYDS-FTS & 2$^{\prime\prime}$             & 3293 -- 10171 & 2.3 \\
    2019-10-26 & 58\,782.22 & $+13$  & EFOSC2-NTT & Gr11+Gr16+1$^{\prime\prime}$   & 3344 -- 10000 & 4.1 \\
    2019-11-04 & 58\,791.33 & $+21$  & EFOSC2-NTT & Gr11+Gr16+1$^{\prime\prime}$   & 3340 -- 10000 & 4.1 \\
    2019-11-20 & 58\,807.33 & $+37$  & EFOSC2-NTT & Gr11+Gr16+1$^{\prime\prime}$   & 3343 -- 10002 & 4.1 \\
    2019-12-06 & 58\,823.30 & $+52$  & EFOSC2-NTT & Gr11+Gr16+1$^{\prime\prime}$   & 3343 -- 10003 & 4.1 \\
    2019-12-20 & 58\,837.15 & $+65$  & EFOSC2-NTT & Gr11+Gr16+1.5$^{\prime\prime}$ & 3349 -- 10002 & 4.1 \\
    2020-01-02 & 58\,850.28 & $+78$  & EFOSC2-NTT & Gr11+Gr16+1$^{\prime\prime}$   & 3341 -- 10001 & 4.1 \\
    2020-01-16 & 58\,864.19 & $+91$  & EFOSC2-NTT & Gr11+Gr16+1$^{\prime\prime}$   & 3340 -- 10002 & 4.1 \\
    2020-02-02 & 58\,881.20 & $+108$ & EFOSC2-NTT & Gr11+Gr16+1$^{\prime\prime}$   & 3349 -- 10001 & 4.1 \\
    2020-02-14 & 58\,893.12 & $+119$ & EFOSC2-NTT & Gr11+Gr16+1$^{\prime\prime}$   & 3345 -- 10004 & 4.1 \\
    2020-03-16 & 58\,924.15 & $+149$ & EFOSC2-NTT & Gr11+Gr16+1$^{\prime\prime}$   & 3343 -- 10003 & 4.1 \\
    2020-09-19 & 59\,111.30 & $+328$ & FORS2-VLT  & 300V+1.6$^{\prime\prime}$      & 4608 -- 8644  & 3.3 \\
    2020-10-23 & 59\,145.27 & $+361$ & EFOSC2-NTT & Gr11+Gr16+1$^{\prime\prime}$   & 3348 -- 9990  & 4.1 \\
    2020-11-05 & 59\,158.31 & $+373$ & EFOSC2-NTT & Gr11+Gr16+1$^{\prime\prime}$   & 3351 -- 9992  & 4.1 \\
    2020-11-22 & 59\,175.20 & $+390$ & EFOSC2-NTT & Gr11+Gr16+1$^{\prime\prime}$   & 3347 -- 9993  & 4.1 \\
    2020-12-16 & 59\,199.10 & $+412$ & EFOSC2-NTT & Gr11+Gr16+1$^{\prime\prime}$   & 3348 -- 9994  & 4.1 \\
    2021-01-19 & 59\,233.05 & $+445$ & EFOSC2-NTT & Gr11+Gr16+1$^{\prime\prime}$   & 3347 -- 9992  & 4.1 \\
    2021-03-04 & 59\,277.13 & $+487$ & EFOSC2-NTT & Gr11+Gr16+1$^{\prime\prime}$   & 3345 -- 9990  & 4.1 \\
    2021-05-02 & 59\,336.98 & $+545$ & EFOSC2-NTT & Gr13+1$^{\prime\prime}$        & 3648 -- 9247  & 5.5 \\
    2021-07-30 & 59\,425.32 & $+629$ & EFOSC2-NTT & Gr13+1$^{\prime\prime}$        & 3642 -- 9242  & 5.5 \\
    2021-10-06 & 59\,493.28 & $+695$ & EFOSC2-NTT & Gr13+1$^{\prime\prime}$        & 3648 -- 9250  & 5.5 \\
    2021-10-10 & 59\,497.50 & $+699$ & RSS-SALT   & PG0900+1.5$^{\prime\prime}$    & 4319 -- 7384  & 1.0 \\
    \bottomrule
    \end{tabular}
    \begin{flushleft}
    $^{\text{a}}$Rest-frame phases relative to $I$-band maximum light, MJD~$=58\,769.0$, corrected for time dilation by a factor $(1+z)$ with $z=0.0427$.
    \end{flushleft}
    \label{tab:log_spectra}
\end{table*}

\begin{table*}
\caption{Three-component fits to H$\alpha$. The Gaussian components (G1-G3, ordered by width) follow \citet{2021MNRAS.506.4715R}; the last two columns compare the Gaussian and Lorentzian fits.}
\label{tab:halpha_fits}
\centering
\footnotesize
\renewcommand{\arraystretch}{0.9}
\begin{tabular}{rrrrrrrcrrc}
\toprule
Phase & \multicolumn{2}{c}{G1} & \multicolumn{2}{c}{G2} & \multicolumn{2}{c}{G3} & Flux fractions & $F_{{\rm H}\alpha}$ & $\Delta$BIC & Pref. \\
(d) & $v_0$ & FWHM & $v_0$ & FWHM & $v_0$ & FWHM & G1/G2/G3 & ($10^{-14}$) & (G$-$L) & \\
\midrule
$-22$ & $-52$ & 880\tablefootmark{u} & $-413$ & 2580 & $-1275$ & 14115 & 0.37/0.29/0.34  & 4.37& $+58$ & L \\
$-17$ & $-153$ & 967 & $-591$ & 2672 & $-4268$ & 6143 & 0.52/0.35/0.12  & 3.52& $-7$ & G \\
$-5$ & $-6$ & 312\tablefootmark{u} & $+20$ & 1558 & $-1440$ & 3609 & 0.30/0.54/0.16  & 4.49& $-773$ & G \\
$-4$ & $+115$ & 925 & $-432$ & 2517 & $-1835$ & 12331 & 0.51/0.23/0.26  & 5.19& $-49$ & G \\
$+4$ & $+64$ & 715\tablefootmark{u} & $-11$ & 2217 & $-2376$ & 13651 & 0.30/0.33/0.37  & 6.56& $-12$ & G \\
$+13$ & $+46$ & 758 & $-35$ & 2264 & $-406$ & 10001 & 0.32/0.30/0.38  & 6.19& $+12$ & L \\
$+21$ & $-2$ & 765 & $-104$ & 2143 & $-833$ & 9603 & 0.27/0.23/0.50  & 6.69& $-27$ & G \\
$+37$ & $-84$ & 404\tablefootmark{u} & $-31$ & 1643 & $-782$ & 10095 & 0.05/0.24/0.71  & 8.88& $-62$ & G \\
$+52$ & $-142$ & 365\tablefootmark{u} & $-104$ & 1871 & $-585$ & 10231 & 0.04/0.16/0.80  & 9.57& $-110$ & G \\
$+65$ & $+181$ & 1160 & $-697$ & 2442 & $-505$ & 9954 & 0.19/0.08/0.73  & 10.21& $-69$ & G \\
$+78$ & $+7$ & 1435 & $-1382$ & 5527 & $-601$ & 13584 & 0.17/0.35/0.48  & 11.46& $+249$ & L \\
$+91$ & $+44$ & 1118 & $-1000$ & 4915 & $-908$ & 12723 & 0.16/0.33/0.51  & 12.34& $+503$ & L \\
$+108$ & $+72$ & 1481 & $-988$ & 5888 & $-995$ & 16515 & 0.17/0.47/0.36  & 13.67& $+156$ & L \\
$+119$ & $-85$ & 1469 & $-955$ & 5910 & $-958$ & 16632 & 0.16/0.50/0.34  & 14.86& $+294$ & L \\
$+149$ & $-44$ & 1477 & $-1197$ & 3977 & $-677$ & 11501 & 0.18/0.28/0.54  & 15.95& $+393$ & L \\
$+328$ & $-175$ & 1901 & $-868$ & 5291 & $-3288$ & 20906 & 0.28/0.54/0.18  & 16.16& $+391$ & L \\
$+361$ & $-146$ & 1823 & $-854$ & 5114 & $-3742$ & 20852 & 0.31/0.52/0.17  & 14.36& $+286$ & L \\
$+373$ & $-147$ & 1803 & $-889$ & 4929 & $-3186$ & 19439 & 0.32/0.50/0.18  & 14.16& $+289$ & L \\
$+390$ & $-192$ & 1767 & $-956$ & 4718 & $-3057$ & 18578 & 0.33/0.49/0.18  & 12.61& $+422$ & L \\
$+412$ & $-143$ & 1760 & $-976$ & 4259 & $-1987$ & 14970 & 0.35/0.45/0.20  & 12.00& $+293$ & L \\
$+445$ & $-149$ & 1733 & $-1125$ & 3586 & $-1207$ & 11648 & 0.39/0.36/0.25  & 9.46& $+15$ & L \\
$+487$ & $-191$ & 1638 & $-1095$ & 3156 & $-1555$ & 10206 & 0.42/0.33/0.25  & 6.53& $-117$ & G \\
$+545$ & $-125$ & 1685 & $-1201$ & 3330 & $-3021$ & 17567 & 0.53/0.28/0.19  & 2.83& $-262$ & G \\
$+629$\tablefootmark{a} & $-20$ & 1556 & $-800$ & 1973 & $-2239$ & 6682 & 0.52/0.28/0.21  & 1.79& $-243$ & G \\
$+695$\tablefootmark{b} & $-90$ & 1201 & $-604$ & 1944 & $-2503$ & 5832 & 0.27/0.55/0.18  & 0.66& $-198$ & G \\
$+699$\tablefootmark{c} & $+3$ & 243\tablefootmark{u} & $-75$ & 2136 & $-2396$ & 4610 & 0.36/0.30/0.34  & 0.14& $-1271$ & G \\
\bottomrule
\end{tabular}
\tablefoot{Velocities ($v_0$) and FWHMs are in km\,s$^{-1}$; velocities are relative to H$\alpha$ at $z=0.0427$. $F_{{\rm H}\alpha}$ is the total H$\alpha$ flux of the three-Gaussian fit in units of $10^{-14}$\,erg\,s$^{-1}$\,cm$^{-2}$; its absolute scale follows the flux calibration of each spectrum, which is unreliable at $+699$\,d (Sect.~\ref{subec:spectra}). FWHMs are observed, not corrected for the instrumental resolution. $\Delta{\rm BIC}>0$ favours the Lorentzian model (L), $\Delta{\rm BIC}<0$ the Gaussian model (G); because neither model fits most epochs within the errors (reduced $\chi^2>1$), the $\Delta$BIC values overstate the significance of the preference. Before $+78$\,d the red wing between about $+1200$ and $+4000$\,km\,s$^{-1}$ is masked.
\tablefoottext{u}{Unresolved (FWHM $\leq1.2\times$ the instrumental resolution).}
\tablefoottext{a}{Components degenerate; only the total profile is used.}
\tablefoottext{b}{First of three exposures.}
\tablefoottext{c}{Host dominated: G1 is host H$\alpha$, and host [N\,{\sc ii}] is not modelled.}}
\end{table*}

\clearpage
\end{appendix}

\end{document}